\documentclass[aps,twocolumn,superscriptaddress,amsmath,longbibliography,nofootinbib]{revtex4-1}
\usepackage{graphicx,color,hyperref,xcolor}
\usepackage{soul}
\usepackage{amsfonts}
\usepackage{amssymb}
\usepackage{amsmath}
\usepackage{amsthm}
\usepackage{bm}
\usepackage{braket}

\makeatletter
\newcommand*{\rom}[1]{\expandafter\@slowromancap\romannumeral #1@}
\makeatother

\begin{document}

\title{Microscopic Examination of SRF-quality Nb Films through Local Nonlinear Microwave Response}

\author{Chung-Yang Wang}
\affiliation{Quantum Materials Center, Department of Physics, University of Maryland, College Park, Maryland 20742, USA}

\author{Carlota Pereira}
\affiliation{CERN, European Organization for Nuclear Research, 1211 Geneva 23, Switzerland}

\author{Stewart Leith}
\affiliation{CERN, European Organization for Nuclear Research, 1211 Geneva 23, Switzerland}

\author{Guillaume Rosaz}
\affiliation{CERN, European Organization for Nuclear Research, 1211 Geneva 23, Switzerland}

\author{Steven M. Anlage}
\affiliation{Quantum Materials Center, Department of Physics, University of Maryland, College Park, Maryland 20742, USA}

\begin{abstract}

The performance of superconducting radio-frequency (SRF) cavities is sometimes limited by local defects. To investigate the RF properties of these local defects, especially those that nucleate RF magnetic vortices, a near-field magnetic microwave microscope is employed. Local third harmonic response ($P_{3f}$) and its temperature-dependence and RF power-dependence are measured for one Nb/Cu film grown by Direct Current Magnetron Sputtering (DCMS) and six Nb/Cu films grown by High Power Impulse Magnetron Sputtering (HiPIMS) with systematic variation of deposition conditions. Five out of the six HiPIMS Nb/Cu films show a strong third harmonic response that is likely coming from RF vortex nucleation due to a low-$T_c$ surface defect with a transition temperature between 6.3 K and 6.8 K, suggesting that this defect is a generic feature of air-exposed HiPIMS Nb/Cu films. A phenomenological model of surface defect grain boundaries hosting a low-$T_{c}$ impurity phase is introduced and studied with Time-Dependent Ginzburg-Landau (TDGL) simulations of probe/sample interaction to better understand the measured third harmonic response. The simulation results show that the third harmonic response of RF vortex nucleation caused by surface defects exhibits the same general features as the data, including peaks in third harmonic response with temperature, and their shift and broadening with higher microwave amplitude. We find that the parameters of the phenomenological model (the density of surface defects that nucleate RF vortices and the depth an RF vortex travels through these surface defects) vary systematically with film deposition conditions. From the point of view of these two properties, the Nb/Cu film that is most effective at reducing the nucleation of RF vortices associated with surface defects can be identified.

\end{abstract}

\maketitle



\section{Introduction}
\label{sec:Introduction}

In high-energy physics, there is continued interest in building next-generation particle accelerators (for example, the International Linear Collider, ILC) using bulk Nb superconducting radio-frequency (SRF) cavities \cite{padamsee201750,bambade2019international}. For the ILC, around 10000 SRF cavities will be built.

The quality of an SRF cavity is typically quantified by its quality factor (Q-factor) as a function of the accelerating gradient for the particle beam. Real-world materials are not perfect. The Q-factors of SRF cavities are usually below their theoretical predictions. In particular, as the accelerating gradient, and hence the RF magnetic field on the Nb surfaces, becomes strong, the Q-factor drops significantly (this is called the Q-slope) \cite{ciovati2006review,gurevich2006multiscale,gurevich2012superconducting,gurevich2017theory}. Such a Q-slope phenomenon limits the RF field supported by the SRF cavities, which then limits the performance of the particle accelerator. Besides the Q-slope phenomenon, quenches are also frequently observed in many SRF cavities \cite{champion2009quench,bao2019quench,posen2015radio}. One reason for a quench is that a superconductor is locally heated up to exceed its critical temperature and loses superconductivity. Both the Q-slope and defect-nucleated quenches indicate that the performance of SRF cavities is limited by breakdown events below the theoretically predicted intrinsic critical field of the superconductor \cite{gurevich2006multiscale,gurevich2012superconducting,padamsee201750}. These breakdowns are sometimes caused by uncontrolled local defects \cite{kneisel2015review,antoine2019influence,weingarten2023field}. Candidates of defects in SRF cavities include oxides \cite{yoon2008atomic,proslier2008tunneling,romanenko2017understanding,semione2019niobium,semione2021temperature}, impurities \cite{russo2007quality,kharitonov2012surface}, grain boundaries \cite{carlson2021analysis,lee2007flux,polyanskii2011magneto,koszegi2017magneto,wang2018investigation,wang2022investigation,lee2020grain}, dislocations \cite{wang2022investigation,romanenko2010role,bieler2010physical}, surface roughness \cite{wang2022effects,ries2020superconducting}, etc. To make high Q-factor SRF cavities that operate to high accelerating gradients, it is necessary to understand these defects, in particular their influence on the RF properties of SRF cavities. Therefore, there is a need to understand in detail the RF properties of these local defects.

In SRF material science, various kinds of techniques have been developed to characterize SRF cavities and SRF materials. For example, researchers routinely measure the Q-factor \cite{dhakal2014enhancement} and residual resistance \cite{gonnella2014cool} of SRF cavities. However, it is costly and time-consuming to fabricate and measure an entire cavity. As a result, many measurements are performed on coupon samples of SRF materials, including measurements of RF quench field \cite{posen2015radio,kleindienst2015commissioning,keckert2021characterizing} and surface resistance \cite{kleindienst2015commissioning,arzeo2022enhanced,keckert2021mitigation,keckert2021characterizing}.

Another quantity of interest is the vortex penetration field because SRF cavities are expected to operate best in the Meissner state (vortex-free) to avoid dissipation due to vortex motion \cite{james2012superconducting,posen2015radio,ries2020superconducting,carlson2021analysis,antoine2019optimization}. Superconductors show strong nonlinearity in the presence of vortices and show relatively weak nonlinearity in the vortex-free Meissner state. The nonlinear electrodynamic response arises when properties of the superconductor (such as the superfluid density) become time-dependent during the RF cycle. One manifestation of nonlinearity is that the superconductor creates response currents to the stimulation at frequencies other than the driving frequency. Utilizing the connection between vortices and nonlinearity, the vortex penetration field can be determined by measuring the third harmonic response of a superconductor subjected to a time-harmonic magnetic field \cite{lamura2009first}. In particular, the vortex penetration field of thin films and multilayer structures have been studied with such alternating current (AC) (kHz regime) third harmonic response magnetometry \cite{antoine2010characterization,antoine2011characterization,antoine2013study,katyan2015characterization,aburas2017local,antoine2019optimization,ito2019lower,ito2020vortex}.

The techniques described above (Q-factor, residual resistance, RF quench field, surface resistance, vortex penetration field, etc.) help physicists to characterize the global properties of SRF materials. However, none of them can directly study the local RF properties of SRF materials.

Motivated by the need to study RF properties of local defects, we successfully built and operated a near-field magnetic microwave microscope using a scanned loop (the original version) \cite{lee2000magnetic,lee2003spatially,lee2005doping,lee2005microwave,mircea2009phase} as well as a magnetic writer from a magnetic recording hard-disk drive (the microwave microscope adopted in this work) \cite{tai2011nonlinear,tai2012nanoscale,tai2014modeling,tai2014near,tai2015nanoscale,oripov2019high} to measure locally-generated third harmonic response. The spatial resolution of the local probe of the microwave microscope is in the sub-micron scale, and the excitation frequency is in the range of several GHz.

Our microwave microscope offers a complementary view of the material properties that limit SRF cavity performance. The quantities of interest in a finished SRF accelerator cavity are the quality factor Q, the surface resistance of the material making up the walls of the cavity, and the changes in Q as the accelerating gradient of the cavity is increased. It is desirable to maintain a high Q-factor up to the point where the superconductor reaches its critical surface magnetic field. Our microscope measures local electrodynamic properties of small samples of materials that make up SRF cavities. It utilizes nonlinear response that results from strong and inhomogeneously imposed RF magnetic fields as a surrogate for the high-gradient conditions experienced by the materials inside an SRF cavity. The microscope applies surface magnetic fields quite different from those in an SRF cavity, and imposes unique electromagnetic stresses to the material under study. Note that the ohmic losses of superconducting materials in the microwave range at temperatures substantially below $T_c$ are well below the sensitivity limit of any existing microwave microscope, hence these properties are not studied.

Bulk Nb is the standard choice for fabricating SRF cavities. The main reason is that Nb has the highest critical temperature ($T_c = 9.3$ K) and the highest first critical field ($B_{c1} = 180$ mT) of all the elemental metals at ambient pressure. Besides bulk Nb, there are some candidate alternative materials for SRF applications \cite{valente2016superconducting}, including Nb film on Cu \cite{arbet2001superconducting,russo2007quality,james2012superconducting,ries2020superconducting,sublet2015developments,arzeo2022enhanced,roach2012niobium,rosaz2022niobium,ghaemi2023growth}, Nb\textsubscript{3}Sn on bulk Nb substrate \cite{lee2020grain,posen2015proof,posen2015radio,posen2017nb3sn,trenikhina2017performance,ilyina2019development,carlson2021analysis}, multilayer structure (superconductor-insulator-superconductor structures, for instance) \cite{gurevich2006enhancement,kubo2014radio,kubo2016multilayer,wang2022effects,gurevich2017theory,antoine2010characterization,antoine2019optimization}, etc. The potential benefits of using materials other than bulk Nb would be a higher $T_c$ and a potentially higher critical field $B_c$. Here we focus on Nb films on Cu.

The development of the deposition of Nb films onto Cu cavities has a long history \cite{calatroni200620}. In particular, the first Nb/Cu cavities were produced at CERN in the early 1980s \cite{benvenuti1984niobium}. Motivations for Nb thin film technology for SRF applications include better thermal stability (Nb/Cu cavities allow operation at 4 K, rather than 2K, because of the superior thermal conductivity of Cu) and reducing material cost (high purity Nb costs around 40 times more than Cu). The performance of bulk Nb cavities is approaching the intrinsic limit of the material. On the contrary, Nb/Cu cavities typically suffer from serious Q-slope problems \cite{aull2015understanding,palmieri2015thermal}, which limits their use in high accelerating fields. Solving the Q-slope problem in Nb/Cu cavities is essential for making them competitive for use in high-field accelerators.

In this work, we use our near-field magnetic microwave microscope to study the local third harmonic response (on sub-micron scales, at several GHz) of SRF-quality Nb/Cu films produced at CERN. Our objective is not to provide a comprehensive characterization of all types of surface defects present in these Nb/Cu films. Surface defects that do not nucleate RF vortices fall outside the scope of this study. Instead, our focus is on locally measuring the third harmonic response to explore the surface defects that do nucleate RF vortices. Specifically, we aim to extract the properties of RF vortices linked to surface defects through experiments and simulations investigating the third harmonic response and its dependence on temperature and RF field amplitude. Subsequently, we conduct a qualitative comparison of these Nb/Cu films based on our findings. This allows us to identify the Nb/Cu film that is most effective at reducing the nucleation of RF vortices associated with surface defects.

The outline of this paper is as follows: In Sec. \ref{sec:Setup}, we describe the experimental setup. In Sec. \ref{sec:Results}, we show the experimental results of these Nb/Cu films and focus on surface defect signals. In Sec. \ref{sec:Discussion}, we perform Time-Dependent Ginzburg-Landau (TDGL) simulations to better understand the experimental results. Through these simulations, we compare the Nb/Cu films and determine which film is most effective at reducing the nucleation of RF vortices associated with surface defects. In Sec. \ref{sec:Conclusion}, we summarize the results on the Nb/Cu films.


\section{Experimental setup}
\label{sec:Setup}

\begin{figure}
\includegraphics[width=0.45\textwidth]{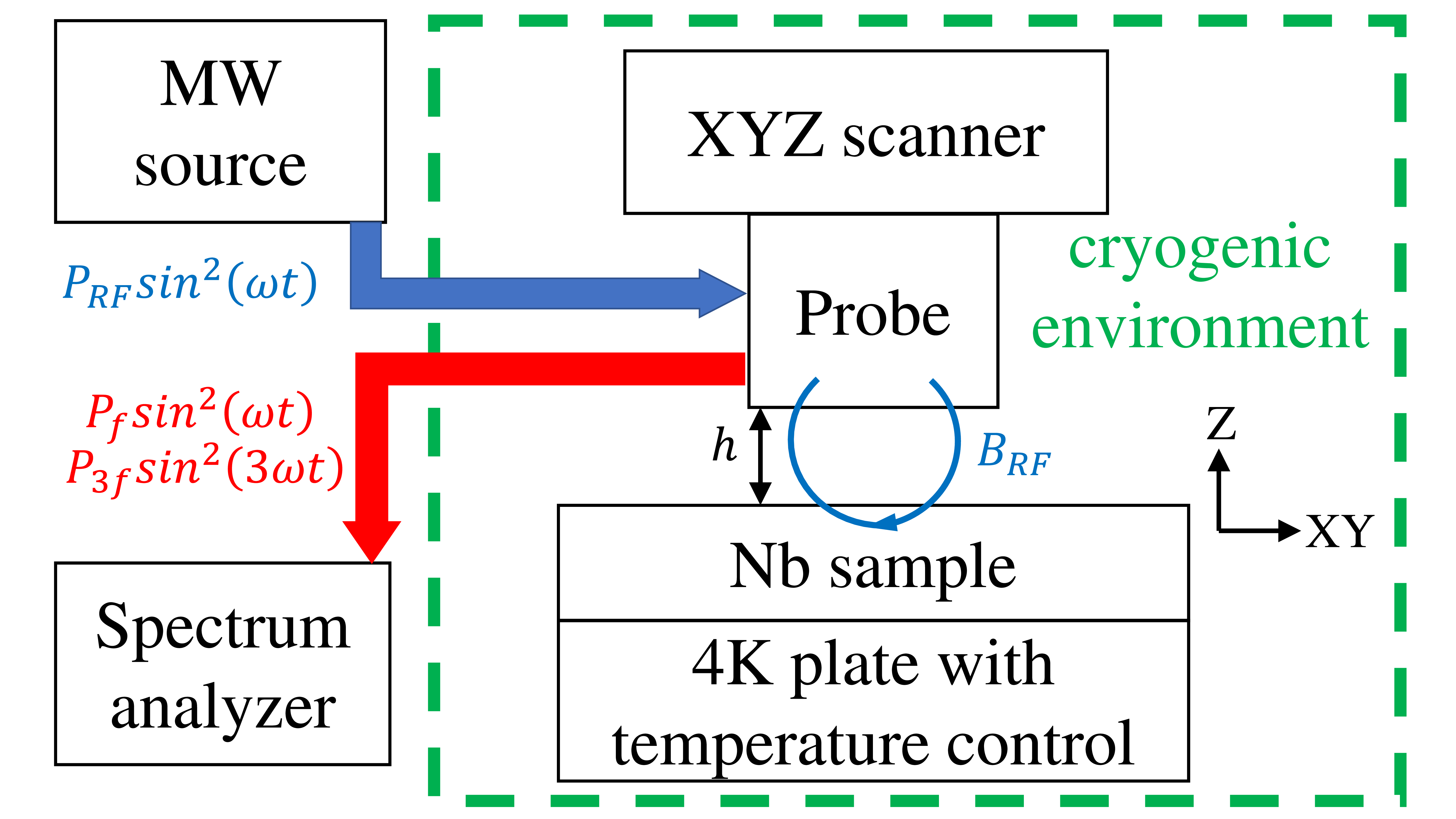}
\caption{\label{fig:Setup}Schematic of experiment setup. The microwave source (MW source) sends a signal of $P_{RF}sin^{2}(\omega t)$ to the magnetic writer (Probe), which then generates $B_{RF}sin(\omega t)$ acting on the sample surface locally. The sample response is collected by the same magnetic writer, and the third Fourier component $P_{3f}sin^{2}(3 \omega t)$ is studied. The dashed line box represents the cryostat.}
\end{figure}

The setup of our near-field magnetic microwave microscope (identical to that described in Ref. \cite{oripov2019high}) is described in this section. A schematic of the setup is shown in Fig. \ref{fig:Setup}.

The heart of our microwave microscope is the magnetic writer head (provided by Seagate Technology) that is used in conventional hard-disk drives. The central part of the magnetic writer head is basically a solenoid that generates a localized RF magnetic field. The solenoid is in the sub-micron scale, which sets the spatial resolution of our microscope. In the setup, a Seagate magnetic writer head is attached to a cryogenic XYZ positioner (sub-micron spatial resolution) and used in a scanning probe microscope fashion. The probe is in contact with the sample during the third harmonic measurement. However, the surfaces of the probe and sample are not perfectly flat, resulting in a finite probe-sample separation $h$, which influences the spatial resolution of the microscope and the peak microwave magnetic field experienced by the sample $B_{pk}$. The height $h$ is estimated to be less than 1 micron, and the spatial resolution is estimated to be in the sub-micron scale, and $B_{pk}$ is estimated to fall within the range of tens of mT for a 0 dBm input microwave power \cite{oripov2019high}.

The microwave source signal $P_{RF}sin^{2}(\omega t)$ is sent to the probe (magnetic writer head) by its built-in and highly-engineered transmission line. The probe then produces a local (sub-micron scale) RF magnetic field $B_{RF}sin(\omega t)$ acting on the sample surface. The superconducting sample then generates a screening current on the surface in an effort to maintain the Meissner state. This screening current generates a response magnetic field that is coupled back to the same probe, creating a propagating signal whose third harmonic component $P_{3f}sin^{2}(3 \omega t)$ is measured by a spectrum analyzer at room temperature. The third harmonic response $P_{3f}$ is measured because it arises from both the nonlinear Meissner effect \cite{groll2010measurement,makita2022nonlinear}, and when a vortex penetrates the sample surface and forms a vortex semi-loop \cite{gurevich2008dynamics} (as discussed in Sec. \ref{sec:Introduction}), either due to an intrinsic mechanism or local defects (weak spot for a vortex to penetrate).

Measurements of superconductor nonlinear response show tremendous dynamic range, often more than 20 dB \cite{tai2011nonlinear,tai2012nanoscale,tai2014near,tai2015nanoscale,oripov2019high}. The excellent instrumental nonlinear background of our measurements ($\sim$ -155 dBm) allows for very sensitive measurements of superconductor nonlinearity and its variation with temperature, driving RF power, location, and probe-sample separation.  Note that measurements are recorded in dBm and later converted to linear power for further study.

To improve the signal-to-noise ratio, microwave filters are installed as follows. Low-pass filters are installed between the microwave source and the probe to block the unwanted harmonic signals generated by the microwave source. High-pass filters are installed between the probe and the spectrum analyzer to block the fundamental input frequency signal from reaching the spectrum analyzer and producing unwanted nonlinear signals. 

Measurements are performed with a variety of fixed input frequencies between 1.1 GHz and 2.2 GHz, while varying temperature and applied RF field amplitude. No external DC magnetic field is applied. The measured residual DC field near the sample at low temperatures is around 35 $\mu$T, as measured by a cryogenic 3-axis magnetometer.

The base temperature for a sample in the cryostat is around 3.5 K. The sample and the thermometer are both directly mounted on the cold plate of the cryogen-free refrigerator to ensure good thermalization.


\section{Experimental results}
\label{sec:Results}

\begin{figure}
\includegraphics[width=0.45\textwidth]{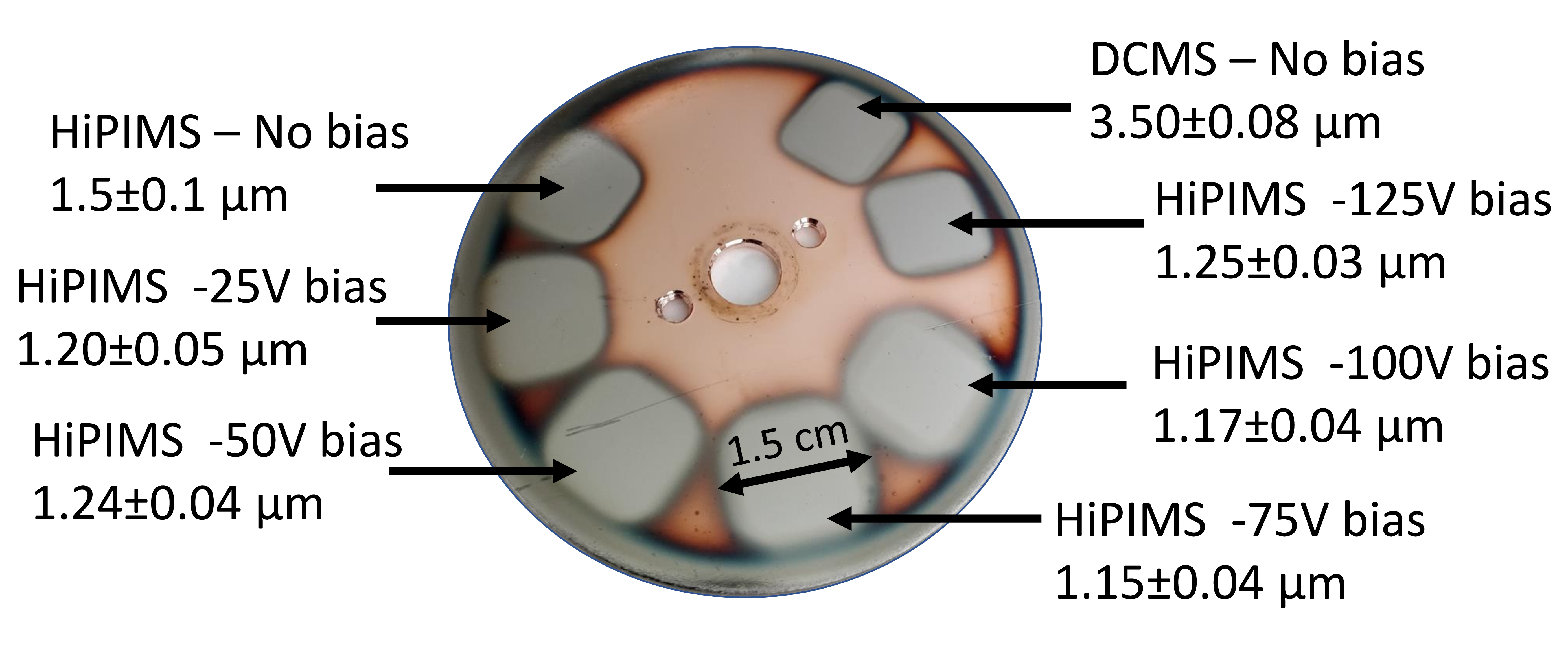}
\caption{\label{fig:SamplePhoto}Photo of the seven Nb films on one common Cu substrate. The samples are prepared at CERN.}
\end{figure}

In this work, we study seven Nb films deposited on one common Cu substrate, as shown in Fig. \ref{fig:SamplePhoto}. One of the samples is prepared by Direct Current Magnetron Sputtering (DCMS) with zero bias, and the sample thickness is around 3.5 $\mu$m. The other six samples are prepared by High Power Impulse Magnetron Sputtering (HiPIMS), with bias from 0 V to 125 V, and the sample thickness ranges from 1.15 $\mu$m to 1.5 $\mu$m. The preparation of the seven Nb/Cu films is discussed in Appendix \ref{sec:SamplePreparation}.

Our previous research investigated the third harmonic response of both bulk Nb and Nb/Cu films \cite{oripov2019high}. However, maintaining a consistent probe-sample separation proved challenging due to the non-flat surfaces of the samples, which provided motivation for the samples created for the present work. In contrast, the Nb/Cu films examined in this study exhibit remarkable flatness, ensuring consistent probe-sample separation. Furthermore, HiPIMS Nb/Cu cavities fabricated at CERN show high repeatability \cite{vegacid:srf2023-weixa02}, and the HiPIMS Nb/Cu films examined in this study adopt the same fabrication recipe. These two features enable us to conduct more meaningful comparisons between these samples.

In the following, the HiPIMS 25 V bias Nb/Cu sample is discussed in detail, and the results of all the seven Nb/Cu samples are summarized and compared in Table \ref{tbl:ResultSummary} and in Table \ref{tbl:SampleComparison}.

\begin{figure}
\includegraphics[width=0.45\textwidth]{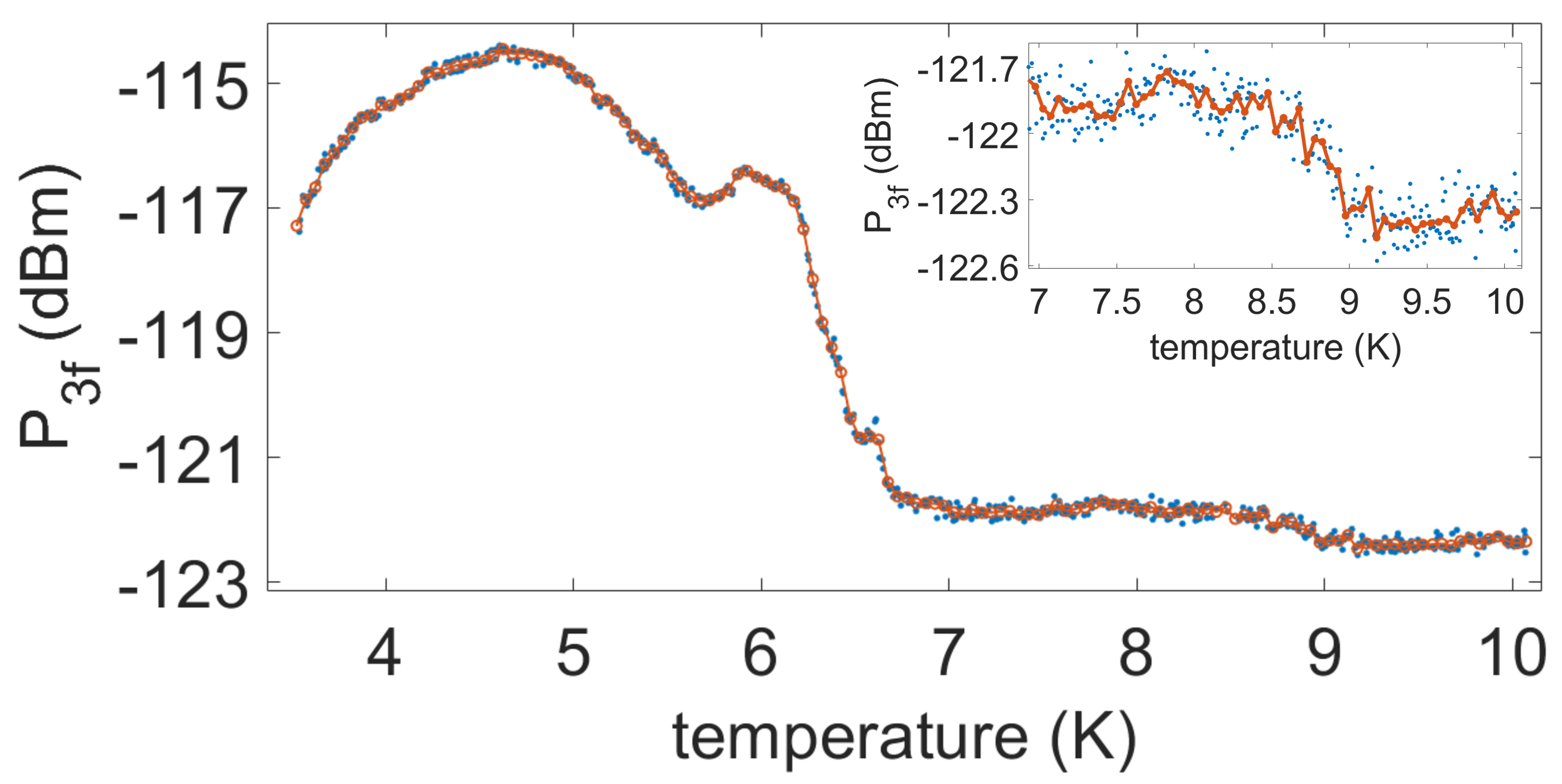}
\caption{\label{fig:P3fRaw}Representative data for $P_{3f}$ as a function of temperature for the HiPIMS 25 V bias Nb/Cu sample. The input frequency is 1.86 GHz and the input power is 2 dBm. The blue dots are the raw data, and the red curve is the $P_{3f}$ averaged over 0.05 K range bins. Inset: enlargement of the figure above 7 K.}
\end{figure}

Fig. \ref{fig:P3fRaw} shows the representative data for the third harmonic response power $P_{3f}$ as a function of temperature at a fixed location on the HiPIMS 25 V bias Nb/Cu sample. A representative measurement protocol is as follows. The sample is warmed up to 10 K (above $T_{c}$), and then the microwave source is turned on with fixed input frequency ($\omega/{2\pi}$=1.86 GHz) and input power ($P_{RF}$=+2 dBm), and then $P_{3f}$ is measured as the sample is gradually cooled down to 3.5 K. In other words, the surface of the sample experiences a fixed RF field $B_{RF}sin(\omega t)$ in a sub-micron scale area during the cooldown process from 10 K to 3.5 K.

The measured $P_{3f}(T)$ in Fig. \ref{fig:P3fRaw} can be decomposed into three segments: the region above 9.1 K, the region below 6.7 K, and the region in between. The magnetic writer head (the probe of our microwave microscope) itself has temperature-independent nonlinearity. The signal above 9.1 K comes from this probe background and is indeed temperature-independent. A transition around 9.1 K can be seen in the inset of Fig. \ref{fig:P3fRaw}. This transition comes from the intrinsic nonlinear response of the Nb film. The strongest $P_{3f}$ signal here shows up below 6.7 K. In the following, such a $P_{3f}$ onset temperature is called the $P_{3f}$ transition temperature and is denoted as $T_{c}^{P_{3f}}$. Compared to the signal around 9.1 K, the onset around 6.7 K is dramatic. Such a signal suggests that some mechanism shows up at and below 6.7 K that produces strong nonlinearity. The mechanisms leading to strong $P_{3f}$ below 6.7 K are extrinsic and are likely due to surface defects. Note that $P_{3f}$ below 6.7 K is much stronger than the intrinsic Nb signal around 9.1 K, suggesting that our local $P_{3f}$ measurement is sensitive to surface defects.

\begin{figure}
\includegraphics[width=0.45\textwidth]{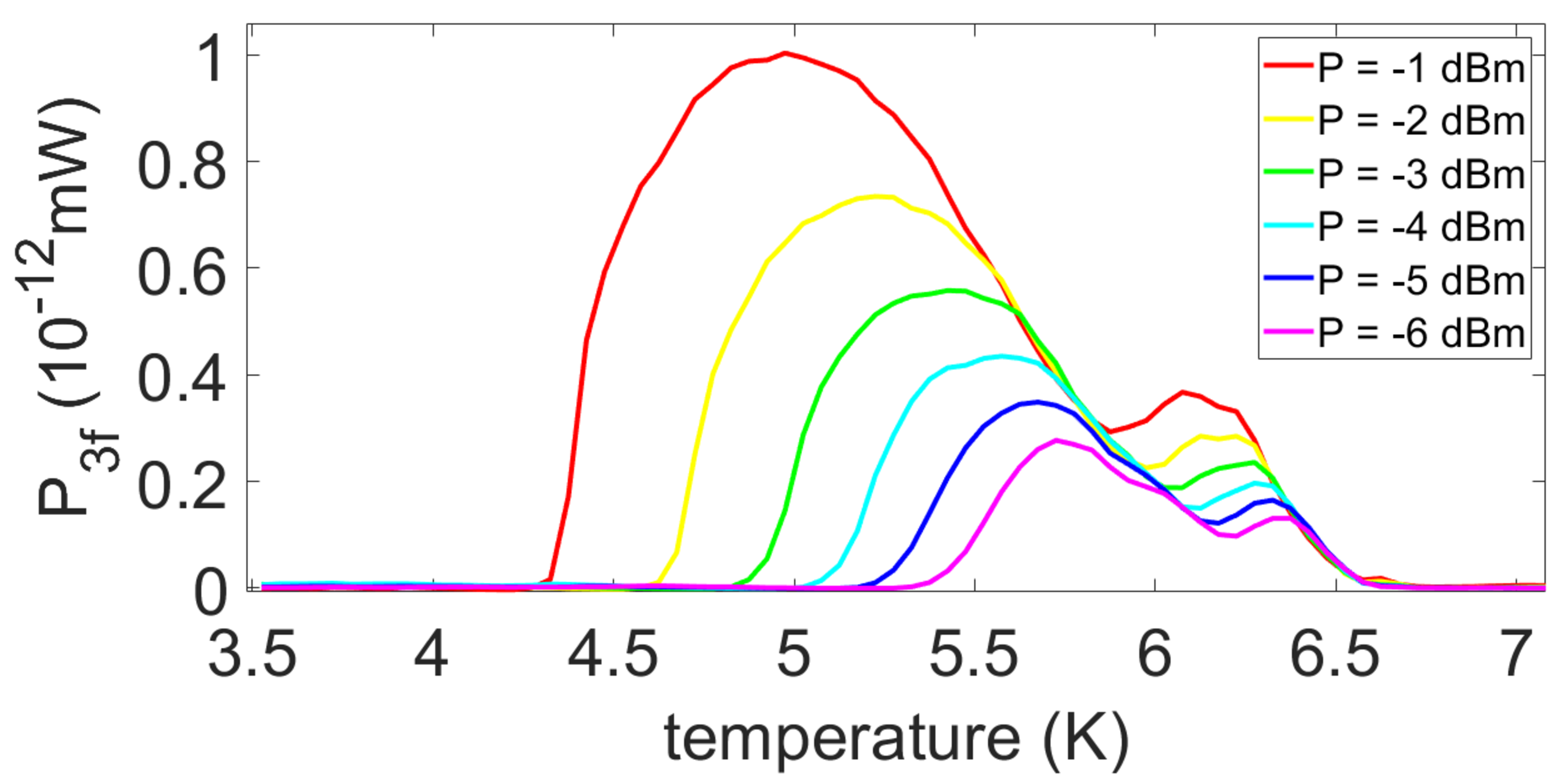}
\caption{\label{fig:P3fMultiplePowers}Measured linear power scale $P_{3f}(T)$ from strong (red) to weak (purple) input power for the HiPIMS 25 V bias Nb/Cu sample. The input frequency is 1.86 GHz. Compared to the raw data shown in Fig. \ref{fig:P3fRaw}, here the probe background is subtracted. Note that input power (in dBm scale) is proportional to $B_{RF}^2$, and hence $B_{RF}^{red}=1.78B_{RF}^{purple}$.}
\end{figure}

Since the main objective of this work is investigating RF properties of surface defects, the strong $P_{3f}$ below 6.7 K is the main focus in the following.

To further study the nature of $P_{3f}$ below 6.7 K, $P_{3f}(T)$ for various input powers $P_{RF}$ (and hence various applied RF field amplitudes $B_{RF}$) are measured. For each measurement, the sample is warmed up to 10 K and then cooled down to 3.5 K while experiencing an applied RF field with fixed input frequency and input power. Since the measured $P_{3f}$ is the combination of probe background and sample contribution, the probe background (probe background is obtained by averaging the magnitude of $P_{3f}(T)$ between 9.5 K and 10 K) is subtracted from the total signal to isolate the sample signal. The process is repeated six times, each time with a different input power. The results of the six $P_{3f}(T)$ of different input powers/RF field amplitudes are shown in the linear format in Fig. \ref{fig:P3fMultiplePowers}.

\begin{figure}
\includegraphics[width=0.45\textwidth]{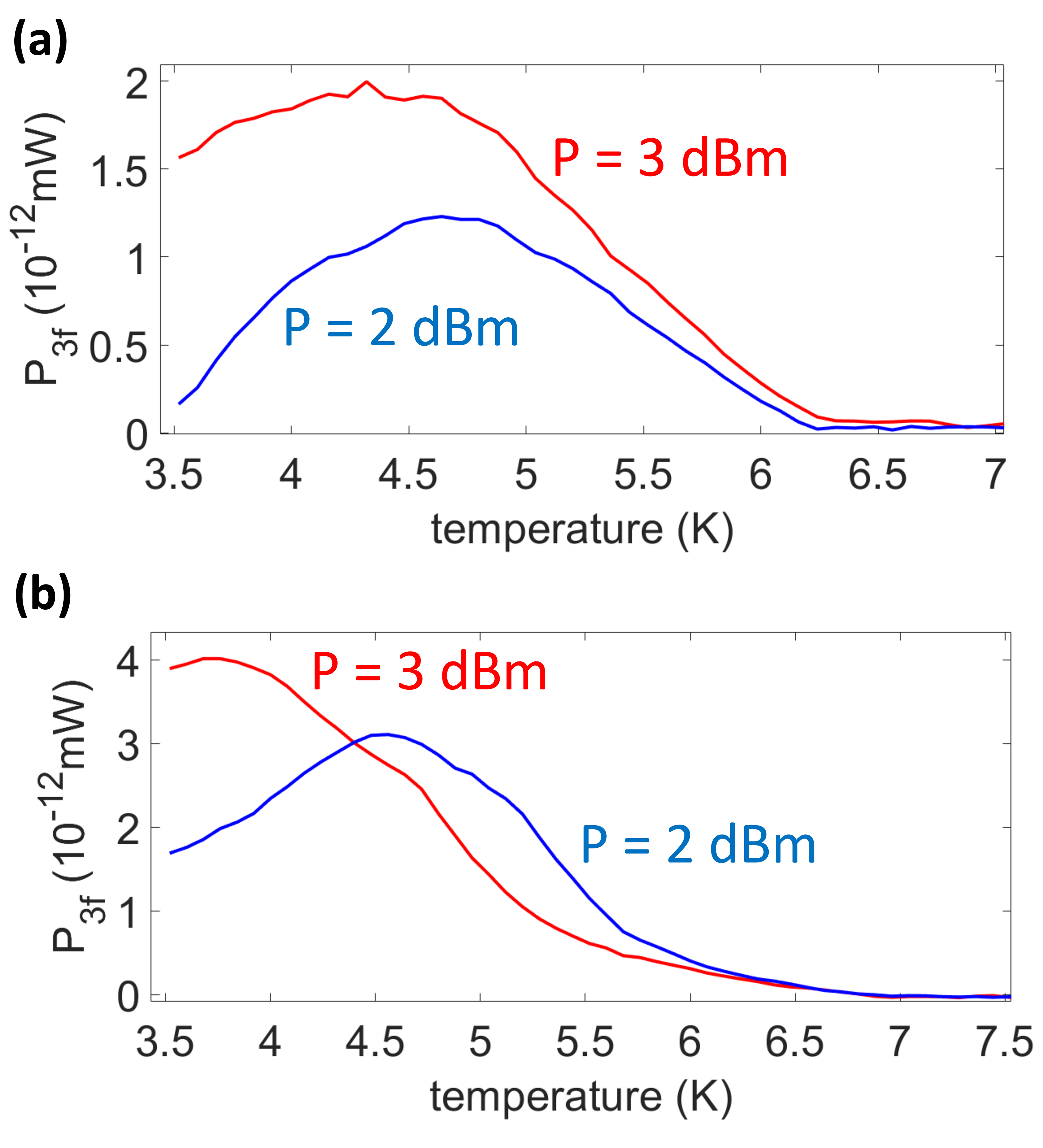}
\caption{\label{fig:75Vand125V} Measured $P_{3f}(T)$ for two input powers for the HiPIMS 75 V bias Nb/Cu sample with an input frequency of 1.98 GHz (a) and for the HiPIMS 125 V bias Nb/Cu sample with an input frequency of 1.66 GHz (b).}
\end{figure}

In Fig. \ref{fig:P3fMultiplePowers}, all the six $P_{3f}(T)$ exhibit a two-peak feature, and their $T_{c}^{P_{3f}}$ are consistently around 6.7 K. For both peaks, as the RF field amplitude increases (purple to red in Fig. \ref{fig:P3fMultiplePowers}), the $P_{3f}(T)$ maximum increases; in addition, the $P_{3f}(T)$ maximum and the $P_{3f}(T)$ low-temperature onset both show up at a lower temperature. These features of $P_{3f}(T)$ are listed in Table \ref{tbl:FourKeyFeatures}. We will see that the four features of $P_{3f}(T)$ listed in Table \ref{tbl:FourKeyFeatures} are the key features of all the nonlinear data in the sense that they show up in all the Nb/Cu films measurement results (see Fig. \ref{fig:P3fMultiplePowers} and Fig. \ref{fig:75Vand125V}) and also in numerical simulations of superconductor nonlinear response (see Fig. \ref{fig:TDGLforNb3}, Fig. \ref{fig:OneGBResult}, and Fig. \ref{fig:TDGLheight}).

\begin{table}
\begin{center}
\includegraphics[width=0.45\textwidth]{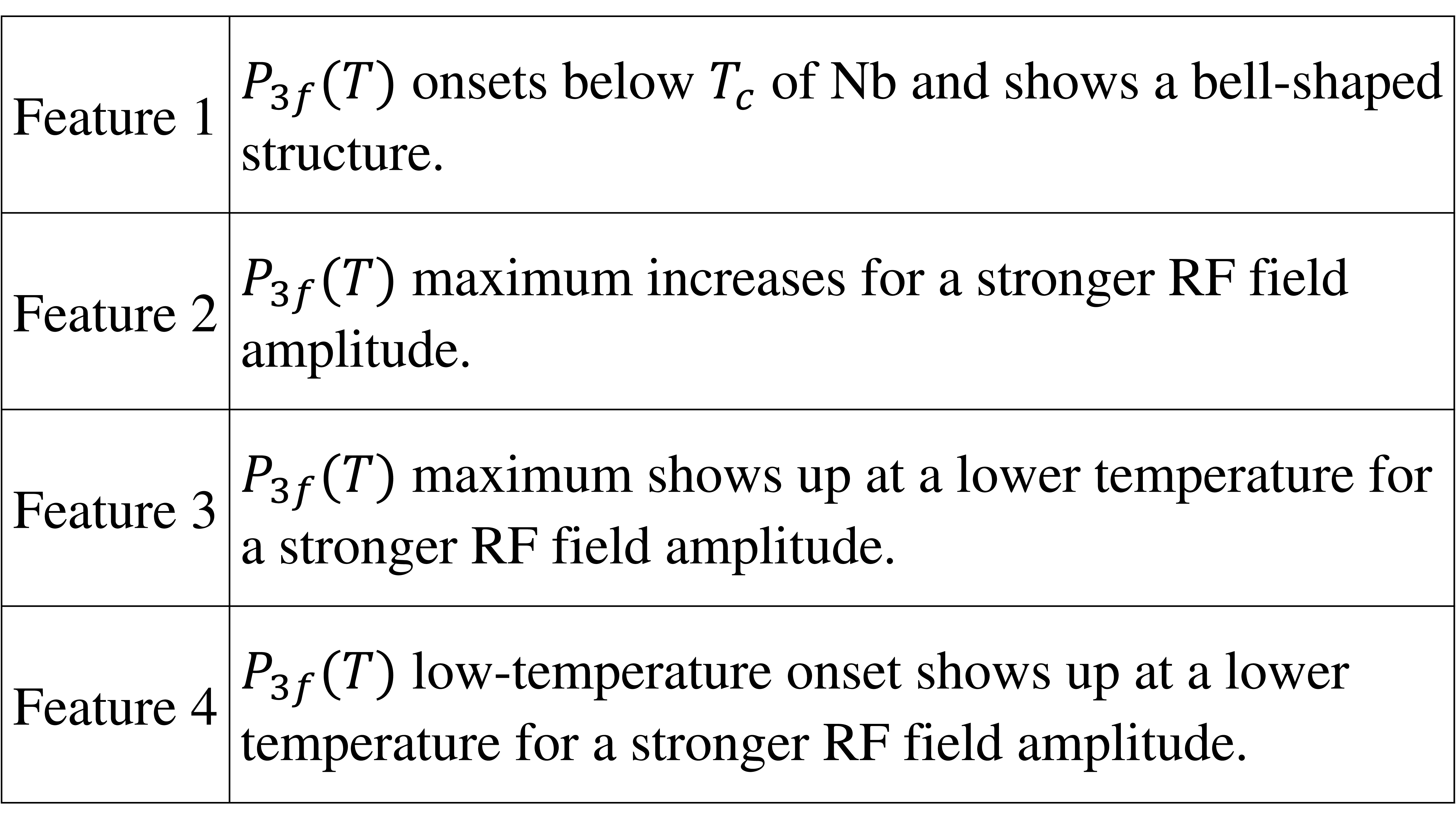}
\caption{The four key features of $P_{3f}(T)$ in our measurements of Nb/Cu films.}
\label{tbl:FourKeyFeatures}
\end{center}
\end{table}

\begin{figure}
\includegraphics[width=0.45\textwidth]{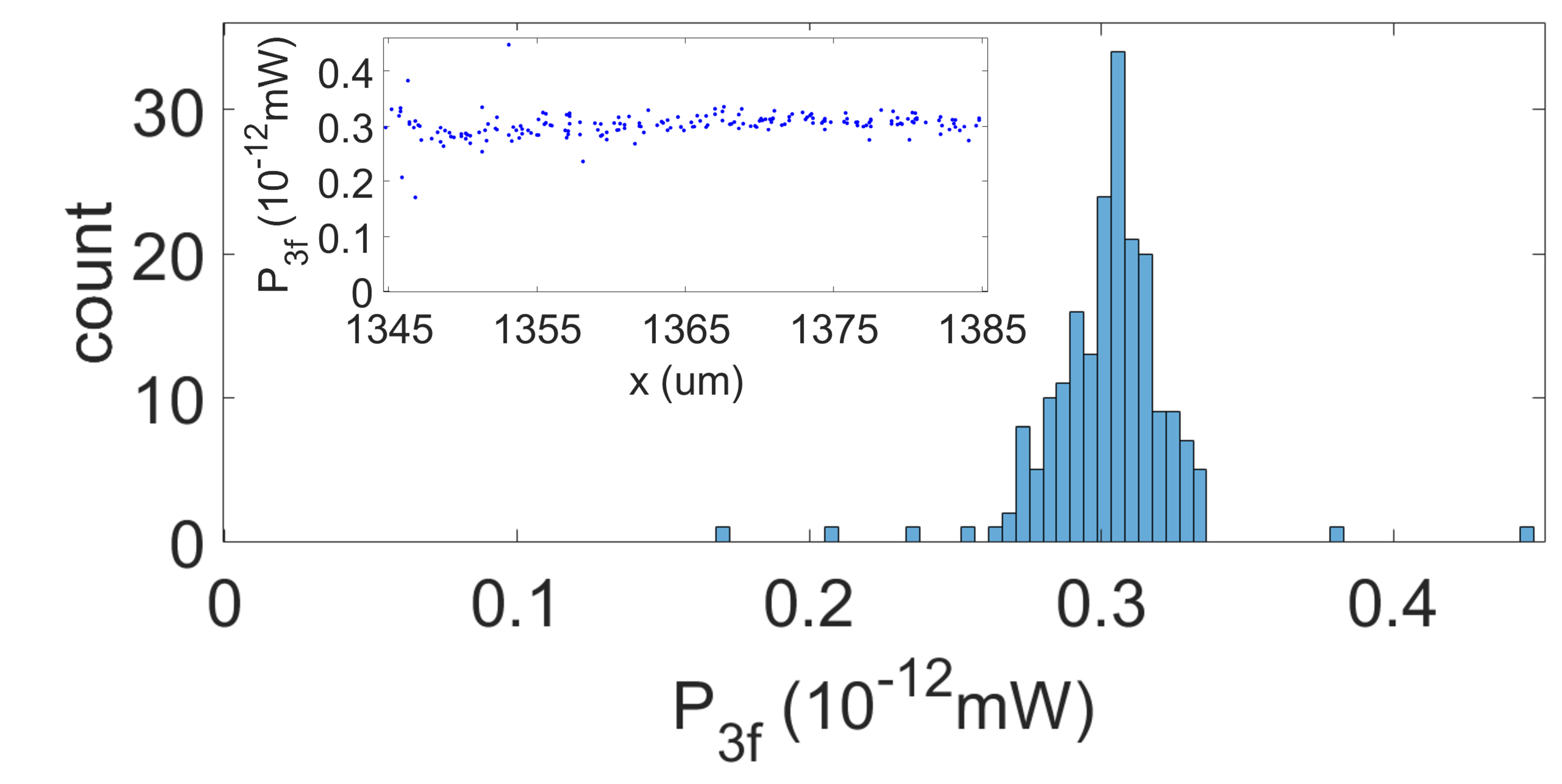}
\caption{\label{fig:P3fScan}Statistics of one-dimensional scan for $P_{3f}(T=5.7 \quad \textrm{K}, \quad P=-5 \quad \textrm{dBm}, \quad f=1.86 \quad \textrm{GHz})$ for the HiPIMS 25 V bias Nb/Cu sample. The coefficient of variation (the ratio of the standard deviation to the mean) is 0.08. Inset: spatial distribution of this one-dimensional scan.}
\end{figure}

So far the measurements are taken at one single location on the surface of the sample. How about the situation at other locations? In particular, does the defect signal (the strong $P_{3f}$ below 6.7 K) show up at other locations as well? To answer this question, we measure $P_{3f}(T)$ with P = -2 dBm at four distinct locations on the sample surface and obtain consistent results. The separation between any two of these locations exceeds 10 $\mu$m. In addition, a one-dimensional (1D) scan is performed for a 40 $\mu$m range with a step size of 0.2 $\mu$m. Temperature and input power are chosen to be T = 5.7 K and P = -5 dBm, and our goal is to determine the homogeneity of the signal from the lower temperature peak of the blue curve in Fig. \ref{fig:P3fMultiplePowers}. The 1D scan result is shown in Fig. \ref{fig:P3fScan}. According to this scanning result, the $P_{3f}$ signal from the defect is quite consistent and uniform at the $\mu$m-scale (coefficient of variation = 0.08). One possible explanation of such uniformity is that the size of one single defect is at the nm-scale and the defect spacing is smaller than the resolution of the microscope (sub-micron scale), and hence the $P_{3f}$ signal of our sub-micron scale measurements comes from the contributions of multiple defects. As an example, the average grain size is around 390 nm for the HiPIMS Nb/Cu films studied in Ref. \cite{arzeo2022enhanced} (page 3 of Ref. \cite{arzeo2022enhanced}). Suppose the HiPIMS Nb/Cu films studied in this work have a similar structure to those in Ref. \cite{arzeo2022enhanced}, and suppose the surface defects responsible for the $P_{3f}$ signal in Fig. \ref{fig:P3fScan} are grain boundaries. In that case, a microwave microscope with a resolution significantly better than 400 nm is required to resolve $P_{3f}$ signals from two distinct grain boundaries.

\begin{table}
\begin{center}
\includegraphics[width=0.45\textwidth]{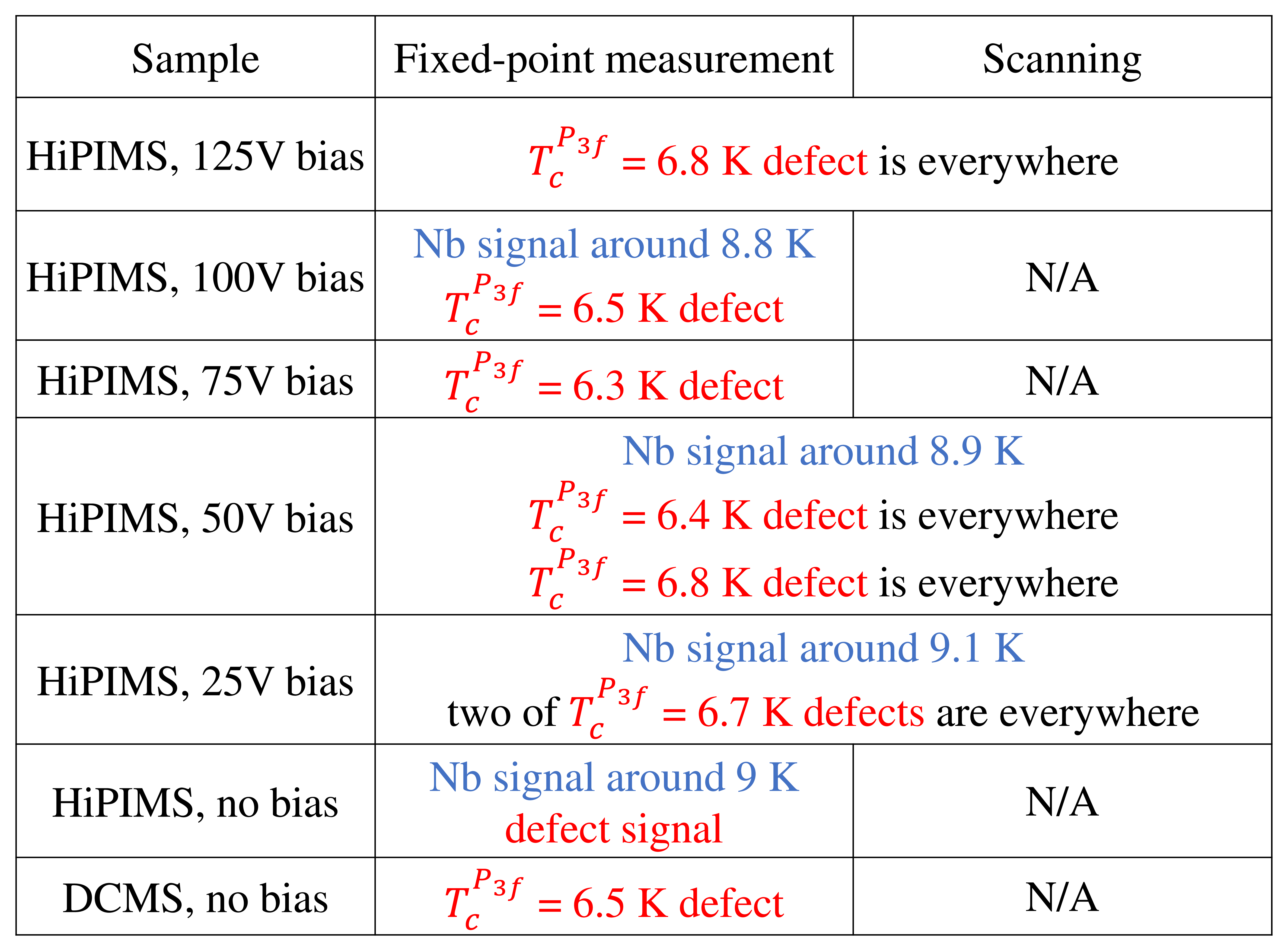}
\caption{Summary of the defect signal (red) and the Nb signal (blue) of the seven Nb/Cu samples. See Appendix \ref{sec:NoBiasSample} for the discussion of the HiPIMS 0 V bias sample.}
\label{tbl:ResultSummary}
\end{center}
\end{table}

So far only the HiPIMS 25 V bias Nb/Cu sample has been discussed. Temperature dependence and input power dependence of $P_{3f}$ are studied for the other six Nb/Cu samples in the same manner as the HiPIMS 25 V bias Nb/Cu sample. The HiPIMS 25 V bias Nb/Cu sample $P_{3f}(T)$ shows the two-peak feature, suggesting that the measured $P_{3f}(T)$ is the superposition of two defect signals. On the other hand, for some of the Nb/Cu samples (75 V bias, 100 V bias, 125 V bias), the $P_{3f}(T)$ defect signal shows a single-peak structure, suggesting that the measured $P_{3f}(T)$ captures only one defect signal. Fig. \ref{fig:75Vand125V} shows the representative $P_{3f}(T)$ for the 75 V bias sample (Fig. \ref{fig:75Vand125V} (a)) and the 125 V bias sample (Fig. \ref{fig:75Vand125V} (b)). For both Fig. \ref{fig:75Vand125V} (a) and (b), $P_{3f}(T)$ displays Features 1, 2, and 3 (Table \ref{tbl:FourKeyFeatures}), suggesting that these features are quite universal. Although we have no access to the temperature regime below 3.5 K, Feature 4 (Table \ref{tbl:FourKeyFeatures}) is likely to be present based on $P_{3f}(T)$ above 3.5 K.

The results of the defect signal and the Nb signal of all seven Nb/Cu samples are summarized in Table \ref{tbl:ResultSummary}. For the six HiPIMS Nb/Cu samples, it is quite universal that $P_{3f}$ measurements reveal the intrinsic Nb signal around 9 K and the extrinsic defect signal at low temperatures. Specifically, the defect signal with $T_{c}^{P_{3f}}$ between 6.3 K and 6.8 K is observed for five out of the six HiPIMS Nb/Cu samples, suggesting that such a defect is a generic feature for these HiPIMS Nb/Cu samples. Moreover, the defect signals are always much stronger than the intrinsic Nb signal around 9 K, like the situation shown in Fig. \ref{fig:P3fRaw}. Scanning measurements are performed for three HiPIMS samples (25 V bias sample, 50 V bias sample and 125 V bias sample), and the results show that the defect signals are quite uniform at the $\mu$m-scale for all three samples. Data for the HiPIMS 50 V bias sample, the HiPIMS 100 V bias sample and the DCMS sample are shown in Appendix \ref{sec:ExtraData}. Data for the HiPIMS 0 V bias sample is qualitatively different from all the other six samples and is discussed in Appendix \ref{sec:NoBiasSample}.

In all the measurements discussed so far, $P_{3f}(T)$ is measured as a function of decreasing temperature starting above $T_c$ (from 10 K to 3.5 K). One may wonder whether this procedure produces trapped vortices when a Nb film cools down through its $T_c$. Indeed, it is known that cooling a Nb film across its $T_c$ in the presence of a DC magnetic field would produce trapped vortices. In our experiment, however, there is a nominally zero DC magnetic field. The Nb film cools down through $T_c$ in the presence of an RF field instead of a DC field. The RF vortices are transient, appearing and disappearing twice per RF cycle. Compared to the dynamics of an RF vortex (on the order of nanoseconds), the temperature change during a typical cool-down is in the extreme adiabatic limit. To the best of our knowledge, there is no unambiguous evidence suggesting that an RF vortex can be converted to a DC vortex and trapped as a Nb film cools down through $T_c$. One way to verify whether RF vortices are trapped in our setup is to compare $P_{3f}(T)$ taken in a warm-up process to $P_{3f}(T)$ taken in a cool-down process for a sample showing a defect signal (such a study is presented in Appendix \ref{sec:Hysteresis}). As shown in Fig. \ref{fig:Hysteresis}, $P_{3f}(T)$ does not exhibit a clear hysteresis, which justifies our interpretation about the absence of trapped DC vortices as the Nb film cools down through $T_c$ in the presence of only an RF field.


\section{Discussion}
\label{sec:Discussion}

The objective of this section is to better understand the third harmonic response associated with RF vortices nucleating at surface defects through numerical simulations, and then using these insights to analyze the experimental results further. In the following, we first introduce our simulation framework (Sec. \ref{sec:TDGLIntroduction}). As a warmup, we consider the case of a defect-free bulk Nb ($T_c = 9.3$ K) (Sec. \ref{sec:BulkNbTDGL}) for developing central concepts relevant to RF vortex nucleation and nonlinear response. Next, we propose a phenomenological surface defect toy model (Sec. \ref{sec:GBModel}) whose $P_{3f}$ shares common features with the experimental results (the four key features in Table \ref{tbl:FourKeyFeatures}). Specifically, the temperature dependence and the RF field amplitude dependence of $P_{3f}$ can be qualitatively explained by the two parameters (``how deep an RF vortex semi-loop penetrates into a sample through a surface defect" and ``the number of surface defects that nucleate RF vortices in each half of the RF cycle") of the toy model (Sec. \ref{sec:GBHeight} and Sec. \ref{sec:TwoGBmodel}). Finally, we conduct a qualitative comparison of these Nb/Cu films based on these two parameters (Sec. \ref{sec:Comparison}).


\subsection{Introduction to numerical simulations}
\label{sec:TDGLIntroduction}

In our measurements, the applied field is a localized RF magnetic field instead of a uniform DC magnetic field. In addition, the configuration of the RF magnetic field produced by the probe is non-uniform but is similar to the field produced by a point dipole. In other words, the situation is different from the case of ``a constant DC magnetic field parallel to the sample surface". The probe also produces a more aggressive magnetic field than the one present in an SRF cavity. Instead of being nominally parallel to the surface, the RF magnetic field has a substantial normal component to the surface. As a result, it is required to study the third harmonic response more carefully.

The Time-Dependent Ginzburg-Landau (TDGL) model is widely used for studying vortex behavior in superconductors \cite{oripov2020time,wang2022effects,carlson2021analysis,pack2020vortex,kato1999charging,lara2015microwave,dobrovolskiy2020moving,hernandez2008dissipation}. In particular, vortices (see Sec. \ref{sec:BulkNbTDGL}) and the proximity effect (see Sec. \ref{sec:GBModel}) are incorporated naturally in TDGL simulations and hence TDGL is a good tool for SRF material science. To better understand our data, numerical simulations of the TDGL equations are performed. The full TDGL equations must be solved in this case because the superconductor is subjected to a time-dependent and inhomogeneous RF magnetic field. We do not assume or impose any spatial symmetries in the model, and solve Maxwell's equations for the dipole in free space above the superconductor, as well as inside the superconductor \cite{oripov2020time}. The TDGL equations solved are the same as those discussed in Ref. \cite{oripov2020time}. The equations, boundary conditions and material parameters of the TDGL simulations can be found in Appendix \ref{sec:TDGLparameters} and in Table \ref{tbl:TDGLparameters}.

In the simulations, the smooth and flat superconducting sample occupies the $z<0$ region, and the magnetic writer probe is approximated to be a pointlike magnetic dipole with a sinusoidal time-dependent magnetic moment $(M_{dp}sin(\omega t),0,0)$ (namely an RF magnetic dipole pointing in the x direction) whose frequency is $\omega/{2\pi}$=1.7 GHz. The RF dipole is located at $(0,0,h_{dp})$ with $h_{dp}$=400 nm. That is, the RF magnetic dipole is above the superconducting sample and parallel to the surface of the sample. Since the magnetic field produced by the RF dipole is non-uniform, the peak RF magnetic field amplitude experienced by the superconductor is specified and is denoted as $B_{pk}$. Here $(x,y,z)=(0,0,0)$ is the location in the sample that experiences the strongest field, and hence $B_{pk}$ is the RF field amplitude at $(x,y,z)=(0,0,0)$. Because the RF field is localized, nontrivial dynamics of the sample (vortex nucleation, for example) would show up only in the region that is underneath the RF magnetic dipole (namely near $(x,y,z)=(0,0,0)$), while the region that is far away from the RF magnetic dipole would be in the vortex-free Meissner state. 

In the simulations, TDGL is used to calculate the time evolution of the order parameter and the vector potential as the superconducting sample is stimulated by the time-dependent RF field produced by the horizontal point dipole above it. (There is no DC magnetic field in the simulations.) With the order parameter and the vector potential, the screening current and hence the magnetic field associated with that screening current (response of the sample) can be calculated. The response magnetic field is calculated at the location of the point dipole and it is assumed that this time-varying magnetic field induces a voltage wave that propagates up to the spectrum analyzer at room temperature. The quantity $\sqrt{P_{3f}}$ is proportional to the third harmonic Fourier component of the magnetic field (generated by the screening current) at the dipole location.

Most TDGL treatments assume a 2D sample and fields that are uniform in the third dimension. This oversimplifies the problem. It creates ``artificial features that extend uniformly" in the third dimension and thus creates infinitely long vortices in the superconductor. Our approach (multi-domain 3D simulation) does not make such unrealistic assumptions. In addition, our experiment and model examine the properties of finite-sized magnetic vortex semi-loops, which are thought to be the generic types of RF vortex excitations created at the surface of SRF cavities \cite{gurevich2008dynamics}.


\subsection{Calculated bulk Nb nonlinear response}
\label{sec:BulkNbTDGL}

\begin{figure}
\includegraphics[width=0.41\textwidth]{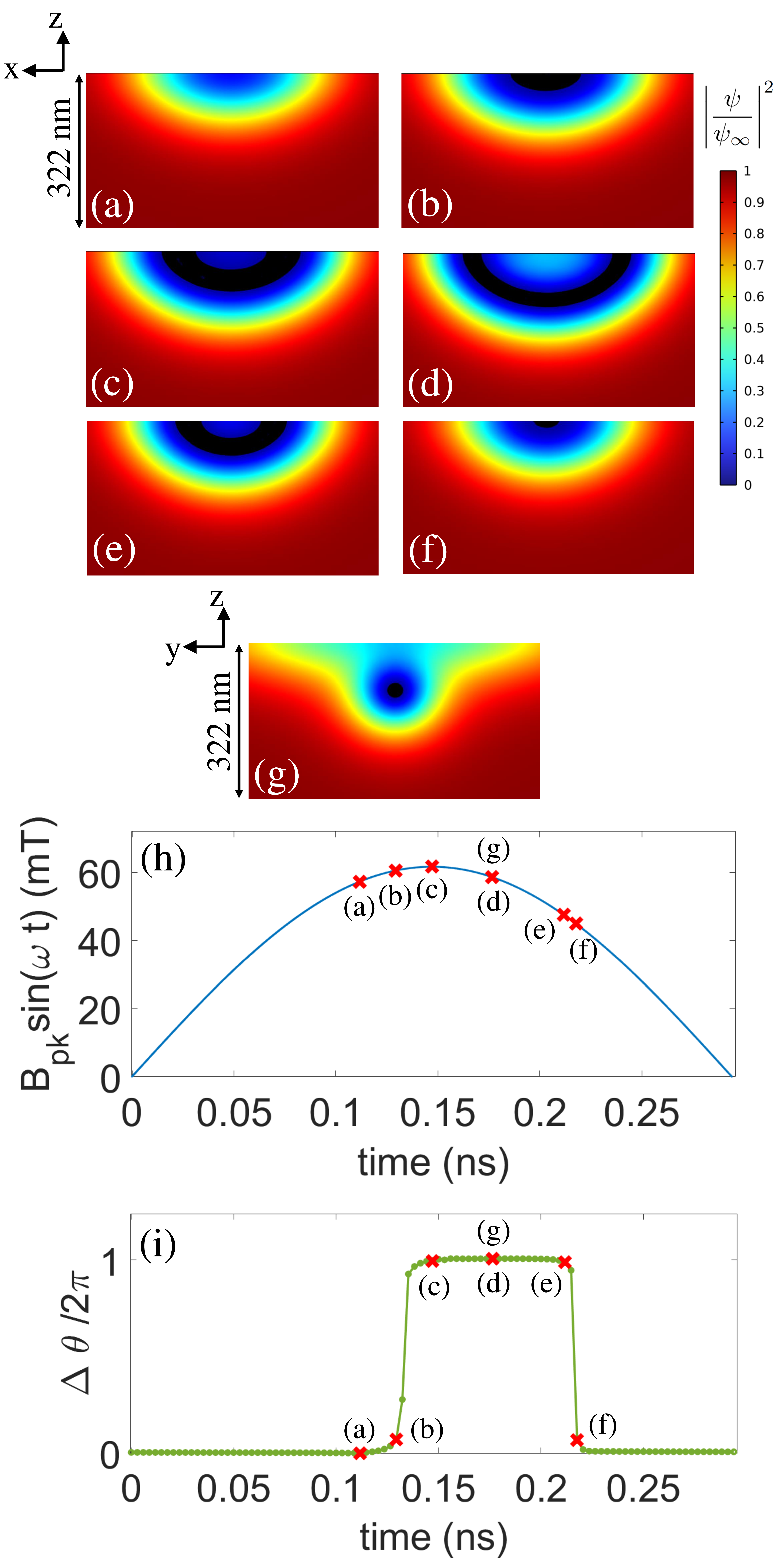}
\caption{\label{fig:VortexDynamics} Time-domain analysis for the dynamics of RF vortex semi-loops in bulk Nb. A time-dependent magnetic moment (RF dipole) is 400 nm above the superconductor surface, pointing in the x-direction, and producing $B_{pk} = 61.6$ mT at T=8.23 K. (a)-(f) show the square of the normalized order parameter ($|\psi / \psi_{\infty}|^2$) on the XZ plane cross-section at different times during the first half of the RF period. The black region is where $|\psi / \psi_{\infty}|^2 < 0.03$. (g) shows $|\psi / \psi_{\infty}|^2$ on the YZ plane cross-section at the same moment as (d). (h) RF field at $(x,y,z)=(0,0,0)$ versus time during the first half of the RF period. (i) Phase change for a closed contour (on the YZ plane) that is large enough to enclose the entire nontrivial region. Red crosses in (h) and (i) correspond to snapshots (a)-(g).}
\end{figure}

\begin{figure}
\includegraphics[width=0.45\textwidth]{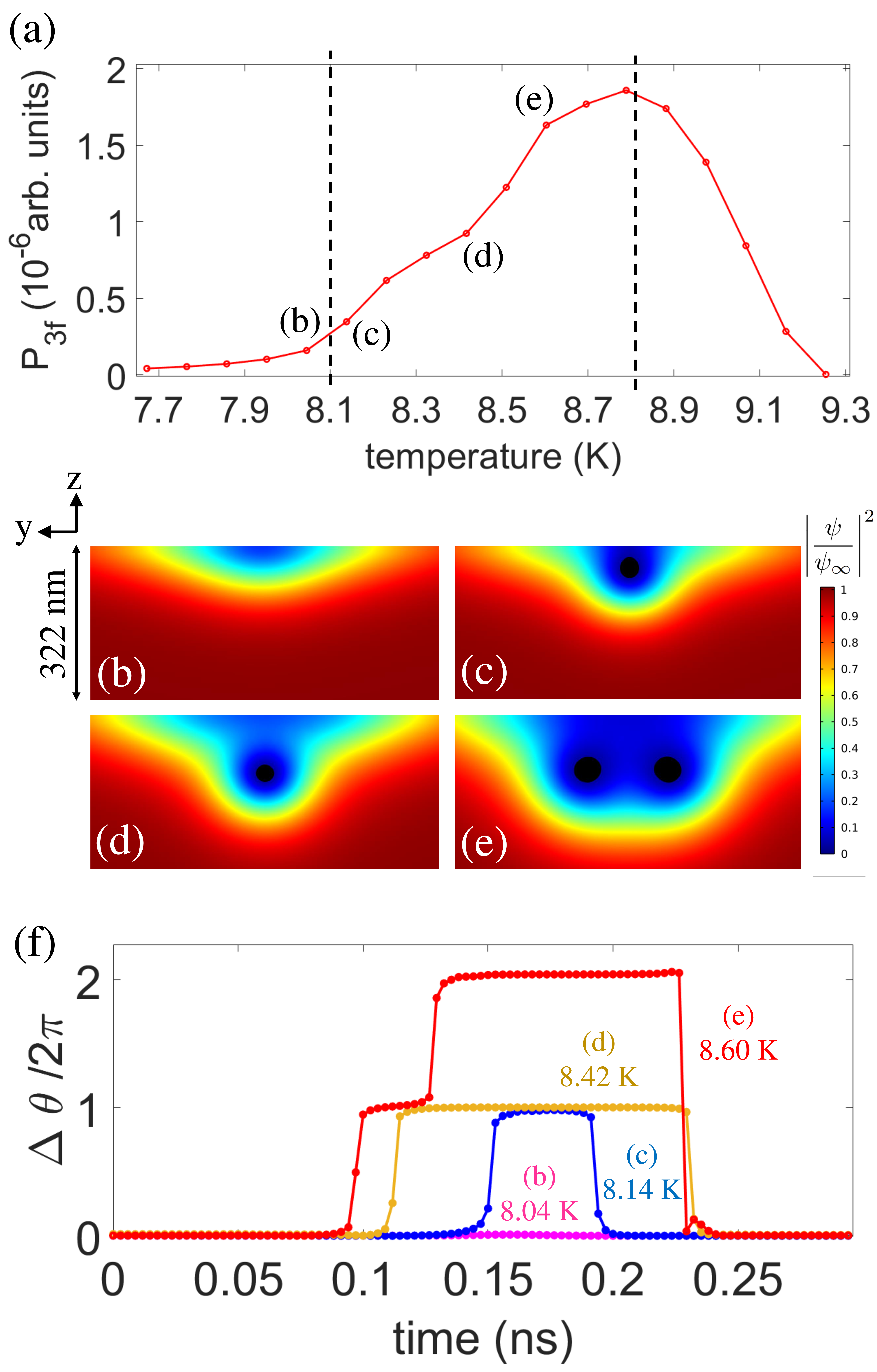}
\caption{\label{fig:TDGLforNbSingle} (a) TDGL simulation result of $P_{3f}(T)$ for a fixed RF field amplitude ($B_{pk} = 61.6$ mT) imposed by a point dipole source for bulk Nb. From left to right, $P_{3f}$ is weak at low temperatures (below 8.1 K), then increases with temperature (between 8.1 K and 8.8 K), and drops with the temperature at high temperatures (above 8.8 K). (b)-(e) show $|\psi / \psi_{\infty}|^2$ on the YZ plane cross-section at 8.04 K, 8.14 K, 8.42 K, and 8.60 K, respectively. The black region is where $|\psi / \psi_{\infty}|^2 < 0.03$. These snapshots are taken at $\omega t=0.6 \pi$. (f) shows $\Delta \theta / 2 \pi$ versus time for the four temperatures.}
\end{figure}

\begin{figure}
\includegraphics[width=0.5\textwidth]{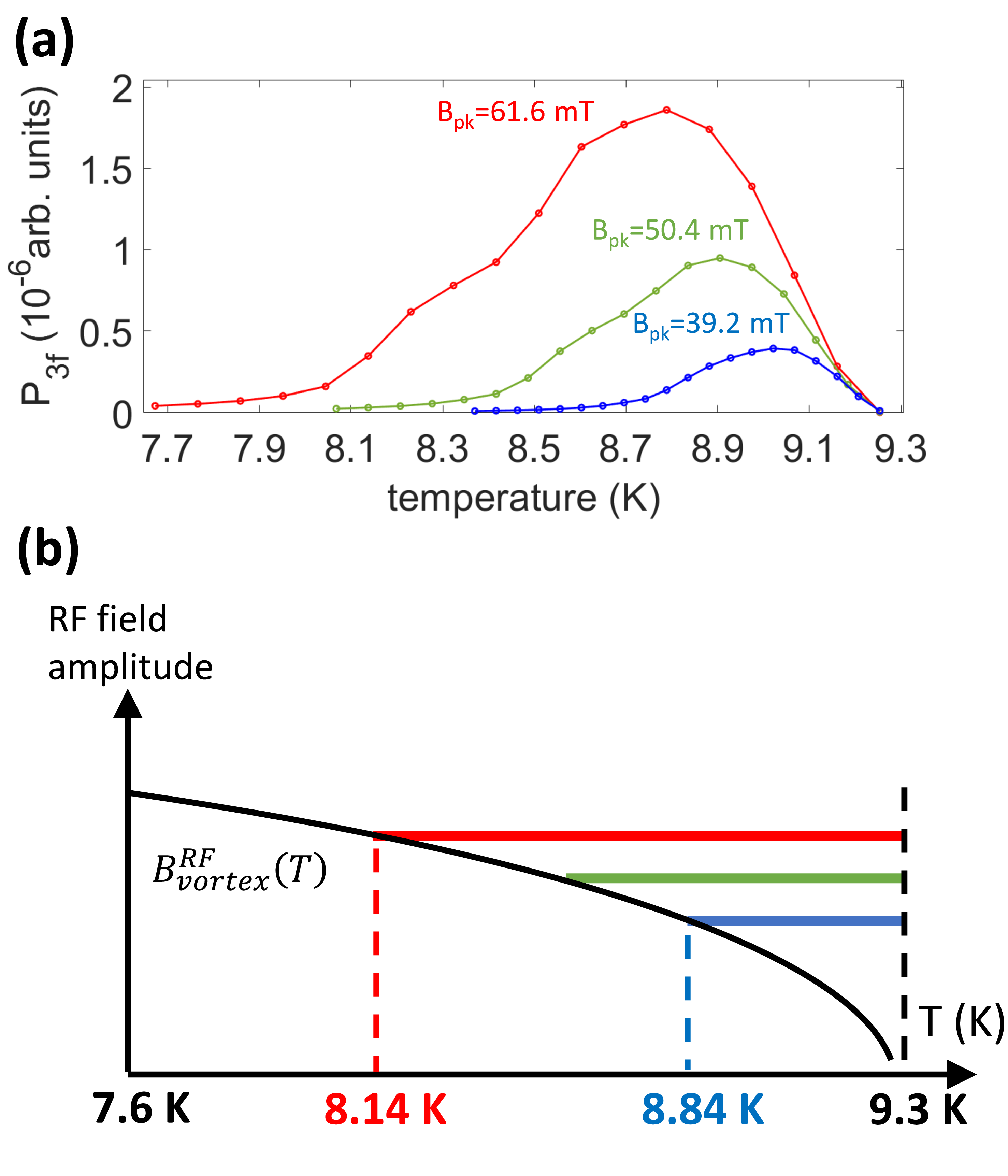}
\caption{\label{fig:TDGLforNb3}(a) TDGL simulation result of $P_{3f}(T)$ for three different RF field amplitudes for bulk Nb. (b) Schematic of the vortex penetration field $B_{vortex}^{RF}(T)$ (the black curve) and the temperature range of $P_{3f}(T)$ bell-shaped structure for the three RF field amplitudes (the three colorful horizontal lines). The y-axis is the RF field amplitude. Note that (a) and (b) share the same horizontal axis and the same color-coded RF field amplitudes.}
\end{figure}

Unlike a DC vortex whose behavior shows no time dependence, an RF vortex shows nontrivial dynamics, and should be examined in a time-domain manner. Here we demonstrate the time-domain analysis (focusing on the dynamics of RF vortices) for a specific RF field amplitude ($B_{pk} = 61.6$ mT) and a specific temperature (8.23 K). (Material parameters of Nb are used in the defect-free bulk Nb simulations. See Appendix \ref{sec:TDGLparameters} and Table \ref{tbl:TDGLparameters}.)

The dynamics of RF vortex semi-loops for bulk Nb during the first half of an RF cycle (frequency=1.7 GHz, period=$5.88\times10^{-10}$s) is shown in Fig. \ref{fig:VortexDynamics}, for a fixed RF field amplitude ($B_{pk} = 61.6$ mT) and a fixed temperature (8.23 K). Fig. \ref{fig:VortexDynamics} (a)-(g) show the space and time dependence of the square of the normalized order parameter ($|\psi / \psi_{\infty}|^2$) (here 1 means full superconductivity and 0 means no superconductivity); the black region is where $|\psi / \psi_{\infty}|^2 < 0.03$. Since the order parameter is suppressed significantly at the center of a vortex core, a vortex can be visualized by tracking the black region. Here $\psi_{\infty} = \psi_{\infty}(T)$ is the value of the order parameter deep inside bulk Nb at temperature T.

In the early stage of the RF cycle, there is no RF vortex (Fig. \ref{fig:VortexDynamics} (a) and (b)), and then an RF vortex semi-loop that is parallel to the direction of the RF dipole (which points in the x direction) shows up (Fig. \ref{fig:VortexDynamics} (c), (d), (e) and (g)). The RF vortex semi-loop disappears later in the RF cycle (Fig. \ref{fig:VortexDynamics} (f)). 

Besides examining the spatial distribution of the order parameter, another signature of vortices is the phase of the order parameter. Because Ginzburg-Landau theory is based on the existence of a single-valued complex superconducting order parameter ($\psi = |\psi| e^{i \theta}$), the phase $\theta$ must change by integral multiples of $2\pi$ in making a closed contour (see equation (4.45) in \cite{tinkham2004introduction}), namely

\begin{equation} \label{eqPhaseIntegral}
    \oint ds \cdot \nabla \theta = 2 \pi N,
\end{equation}

\noindent where N is a positive or negative integer, or zero. The integral $\frac{1}{2\pi} \oint ds \cdot \nabla \theta$ is quantized, and corresponds to the number of vortices enclosed by the closed contour. 

Fig. \ref{fig:VortexDynamics} (i) shows the value of the integral $\frac{1}{2\pi} \oint ds \cdot \nabla \theta$ (namely $\Delta \theta / 2 \pi$) as a function of time. The contour is on the YZ plane and is large enough to enclose the entire nontrivial region. Note that Fig. \ref{fig:VortexDynamics} (h) and Fig. \ref{fig:VortexDynamics} (i) share a common horizontal axis. Based on Fig. \ref{fig:VortexDynamics} (i), there are no vortices at the moments of (a), (b) and (f), and there is one vortex at the moments of (c), (d) and (e), which agrees with the order parameter analysis (Fig. \ref{fig:VortexDynamics} (a)-(g)).

The time-domain analysis described here (space and time dependence of the order parameter (Fig. \ref{fig:VortexDynamics} (a)-(g)) and $\Delta \theta / 2 \pi$ (Fig. \ref{fig:VortexDynamics} (i))) is applied to all TDGL simulations whenever we check whether or not there are RF vortex semi-loops.

Equipped with the picture of RF vortex nucleation, now let's move on to the resulting $P_{3f}$. The simulation result of $P_{3f}(T)$ for a fixed RF field amplitude ($B_{pk} = 61.6$ mT) for bulk Nb is shown in Fig. \ref{fig:TDGLforNbSingle} (a). The bell-shaped structure $P_{3f}(T)$ in Fig. \ref{fig:TDGLforNbSingle} (a) can be decomposed into three segments (separated by the two dashed vertical black lines) and can be understood with the vortex penetration field $B_{vortex}^{RF}(T)$ (see Fig. \ref{fig:TDGLforNb3}) and the strength of superconductivity. An RF vortex semi-loop shows up when $B_{pk}>B_{vortex}^{RF}(T)$. Below 8.1 K, $B_{vortex}^{RF}(T)>B_{pk}$ and hence the entire bulk Nb is in the vortex-free Meissner state (see Fig. \ref{fig:TDGLforNbSingle} (b) and the purple curve in (f)), whose nonlinear response is weak. As temperature increases, $B_{vortex}^{RF}(T)$ decreases and hence $B_{pk}$ would be greater than $B_{vortex}^{RF}(T)$ at a certain temperature depending on the strength of the RF stimulus. In this simulation ($B_{pk} = 61.6$ mT), from the full time-domain simulation (see the discussion for Fig. \ref{fig:VortexDynamics}) one finds that there is one RF vortex semi-loop (underneath the RF magnetic dipole) that penetrates the surface of the bulk Nb when the temperature is around 8.14 K (see Fig. \ref{fig:TDGLforNbSingle} (c) and the blue curve in (f)). Roughly speaking, this implies that $B_{vortex}^{RF}(T=8.14 \textrm{K}) \approx 61.6$ mT. As temperature increases, $B_{vortex}^{RF}(T)$ drops and vortex nucleation is favorable, and indeed the second RF vortex semi-loop shows up around 8.6 K (see Fig. \ref{fig:TDGLforNbSingle} (e) and the red curve in (f)) and thus $P_{3f}$ increases with temperature between 8.1 K and 8.8 K. Besides examining the order parameter (Fig. \ref{fig:TDGLforNbSingle} (b)-(e)), Fig. \ref{fig:TDGLforNbSingle} (f) also shows how the vortex number and duration change with temperature.

The nonlinear response of the superconductor is determined not only by the number of vortices (as described above in the language of $B_{vortex}^{RF}(T)$) but also by the strength of superconductivity. As the temperature approaches the transition temperature of a superconductor, its superconductivity and hence nonlinear response becomes weak. Such a temperature dependence leads to the decreasing tail of $P_{3f}(T)$ above 8.8 K in Fig. \ref{fig:TDGLforNbSingle} (a).

Fig. \ref{fig:TDGLforNb3}(a) summarizes the simulation results of $P_{3f}(T)$ for three different RF field amplitudes for bulk Nb. For all three RF field amplitudes, $P_{3f}$ is weak at low temperatures ($B_{pk}<B_{vortex}^{RF}(T)$, vortex-free Meissner state), arises at high temperatures ($B_{pk}>B_{vortex}^{RF}(T)$, RF vortex semi-loops), and then drops with temperature as the temperature is near the critical temperature. For the red curve, the first vortex semi-loop shows up around 8.14 K, which implies $B_{vortex}^{RF}(T=8.14 
 \textrm{K})\approx 61.6 \textrm{mT}$; for the blue curve, the first vortex semi-loop shows up around 8.84 K, which implies $B_{vortex}^{RF}(T=8.84 
 \textrm{K})\approx 39.2 \textrm{mT}$. Fig. \ref{fig:TDGLforNb3}(b) illustrates $B_{vortex}^{RF}(T)$ and the RF field amplitude dependence of the temperature range of $P_{3f}(T)$ bell-shaped structure. In this RF field amplitude-temperature phase diagram, the vortex-free Meissner state occupies the region below $B_{vortex}^{RF}(T)$ and RF vortex semi-loops show up in the region above $B_{vortex}^{RF}(T)$. It is clear that the $P_{3f}(T)$ bell-shaped structure extends to lower temperatures as the RF field amplitude becomes stronger (from the blue to the green to the red in Fig. \ref{fig:TDGLforNb3}(a) and (b)) because of the temperature dependence of $B_{vortex}^{RF}(T)$, and this explains Feature 4 in Table \ref{tbl:FourKeyFeatures}. Note that the four key features of $P_{3f}(T)$ (Table \ref{tbl:FourKeyFeatures}) are clearly observed in Fig. \ref{fig:TDGLforNb3}(a) (except that the onset temperature of $P_{3f}(T)$ is equal to but not below 9.3 K since this simulation is for a defect-free bulk Nb), suggesting that these key features are signatures of RF vortex nucleation.

Equipped with the intuition of $P_{3f}$ and RF vortex semi-loops and their temperature dependence and RF field amplitude dependence for the defect-free bulk Nb, now let's move on to the case with surface defects, which is the main focus of this paper.


\subsection{Simulation of vortex nucleation in grain boundaries}
\label{sec:GBModel}

\begin{figure}
\includegraphics[width=0.45\textwidth]{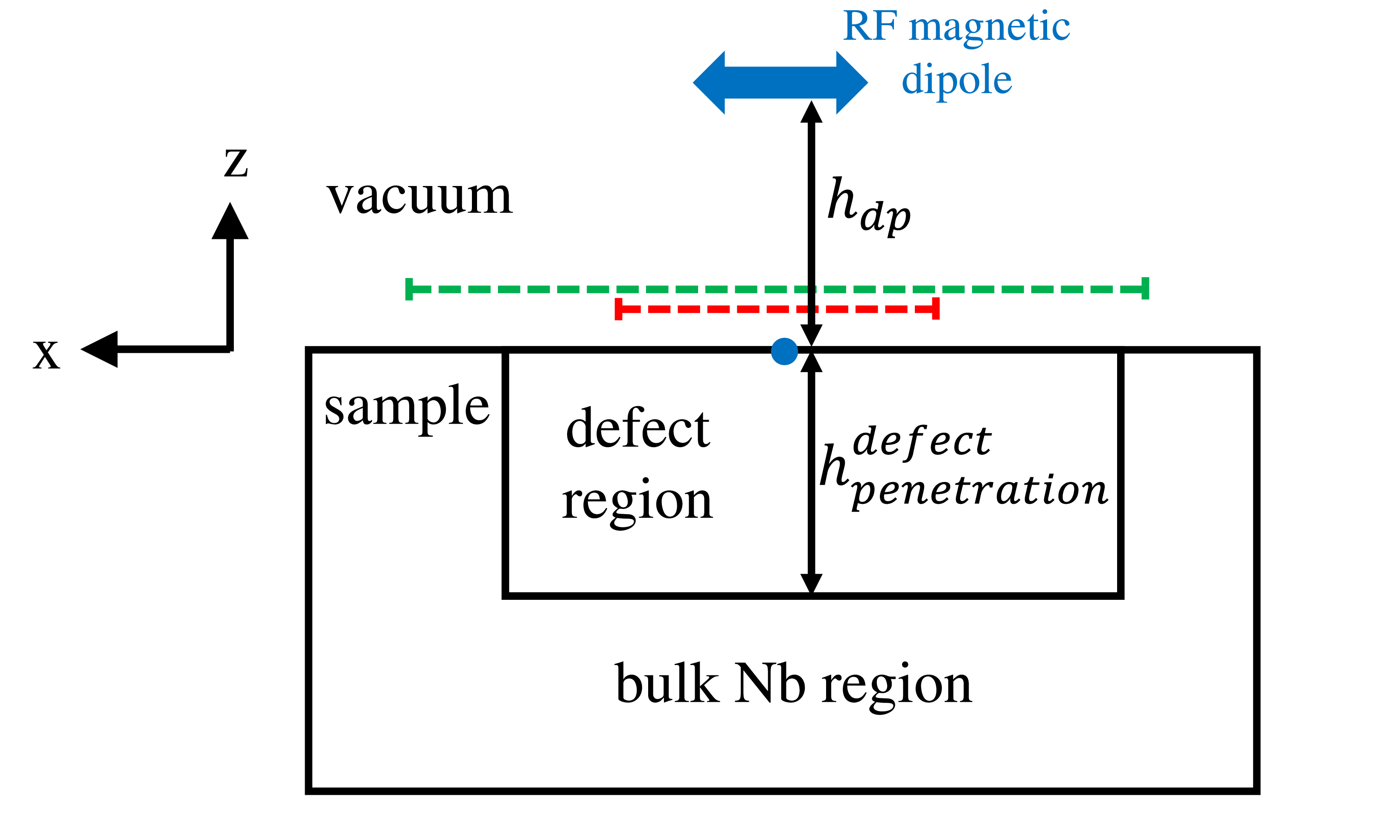}
\caption{\label{fig:OneGBSideView}Sketch of the side view (XZ plane) of the grain boundary model. The origin is marked by a blue dot. The RF dipole is represented by the blue arrow. The schematic is not to scale ($h_{dp}$=400 nm and $h_\textrm{penetration}^\textrm{defect}$=200 nm). Since the physics around the origin plays a dominant role, the region far away from the origin is set to be defect-free bulk Nb to reduce computational time. In addition, the region below $z=-h_\textrm{penetration}^\textrm{defect}$ is also set to be bulk Nb for simplicity. Note that the model is \underline{not} cylindrically symmetric. The top view of the region indicated by the red dashed line is shown in Fig. \ref{fig:OneGBTopView} (b), and the top view of the region indicated by the green dashed line is shown in Fig. \ref{fig:OneGBComplete}.}
\end{figure}

\begin{figure}
\includegraphics[width=0.45\textwidth]{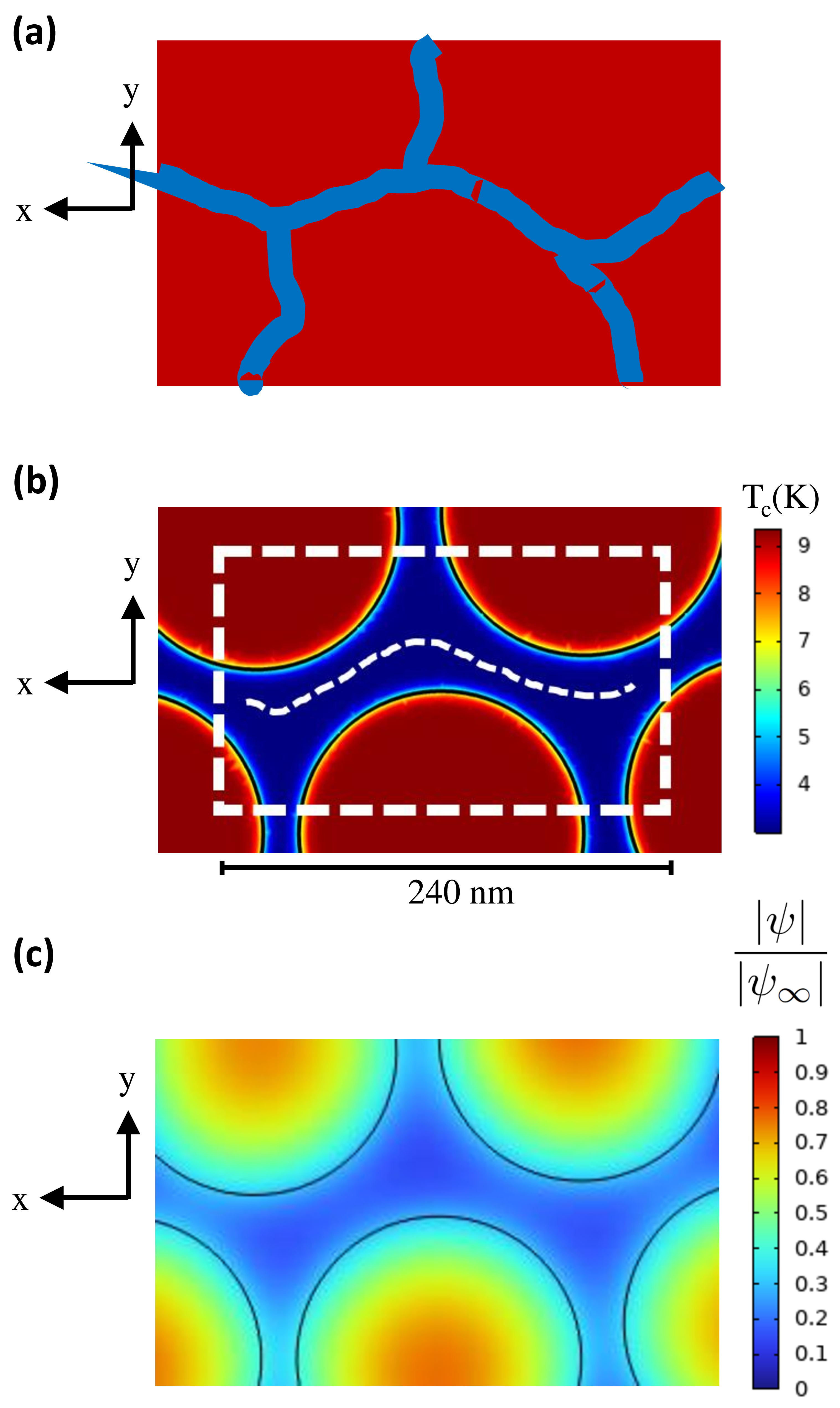}
\caption{\label{fig:OneGBTopView}A top view schematic of the grain boundary model. (a) Illustration of Nb grains and grain boundaries, with red being Nb grains and blue being grain boundaries filled with impurity phases. The realization of this illustration used in the TDGL simulations is shown in (b). (b) The distribution of critical temperature on the XY plane around the origin (corresponds to the region indicated by the red dashed line in Fig. \ref{fig:OneGBSideView}), with red being Nb and blue being the low-$T_c$ impurity phase with $T_c^\textrm{impurity}=3$ K. Here $(x,y,z)=(0,0,0)$ is at the center. The RF dipole is located at $(x,y,z)=(0,0,400 \textrm{nm})$ and points in the x direction. The white dashed curve indicates the grain boundary that is roughly parallel to the x direction. The model defect region is indicated by the white dashed rectangle. (c) A snapshot of the distribution of normalized order parameter $|\psi / \psi_{\infty}|$ obtained by a TDGL simulation with the temperature being 5.2 K (higher than $T_c^\textrm{impurity}$) and $B_{pk} = 50.9$ mT. This snapshot is taken at the end of an RF cycle, namely when the RF field drops to zero ($\omega t=2 \pi$ and hence $B_{RF}sin(\omega t)=0$). The normalized order parameter of the dark blue region is around 0.18.}
\end{figure}

Quantitatively comparing the details of the measured $P_{3f}$ among different Nb/Cu films is challenging for two main reasons. Firstly, the specifics of $P_{3f}$ depend on the probe-sample separation, which is difficult to reproduce. Secondly, different Nb/Cu films studied in this work might possess different types of surface defects that nucleate RF vortices. For example, grain boundaries may be the primary sources of $P_{3f}$ for one Nb/Cu film, while dislocations could be the main sources of $P_{3f}$ for another Nb/Cu film.

Although the details of RF vortex nucleation (and the resulting $P_{3f}$) by various surface defects are complicated and depend on details, we can analyze the dynamics of RF vortex nucleation in a phenomenological way and extract qualitative information. On the phenomenological level, the dynamics of RF vortices penetrating a sample surface through surface defects can be quantified by means of two key aspects: how many surface defects that nucleate RF vortices ($\rho^\textrm{defect}$) exist, and how deep do RF vortices travel into a sample through surface defects in half an RF cycle ($h_\textrm{penetration}^\textrm{defect}$)?

Here we illustrate the concept of $h_\textrm{penetration}^\textrm{defect}$ by considering RF vortex nucleation at grain boundaries in Nb films. For a given grain boundary that nucleates an RF vortex, the depth the RF vortex travels into the Nb film through the grain boundary in half an RF cycle ($h_\textrm{penetration}^\textrm{defect}$) is determined by multiple factors, including the width of the grain boundary, the angle between the grain boundary and the Nb film surface, the material properties of the impurity phases in the grain boundary, etc. Instead of considering the details of all possible microstructures, we adopt a phenomenological approach, where $h_\textrm{penetration}^\textrm{defect}$ serves as the phenomenological characterization of the depth the RF vortex travels into the Nb film through the grain boundary in half an RF cycle.

Nb is known to contain low-$T_{c}$ impurity phases, such as the oxides of Nb \cite{yoon2008atomic,proslier2008tunneling,romanenko2017understanding,semione2019niobium,semione2021temperature,halbritter1987oxidation}. In such phases, oxygen forms a solid solution in Nb and produces materials with critical temperatures below the bulk $T_c$ of pure Nb (9.3 K) \cite{desorbo1963effect,koch1974effects,weingarten2023field}. Nb samples with higher oxygen content tend to have a lower critical temperature. For instance, $T_c$ drops to around 7.33 K for 2\% oxygen content, and drops to around 6.13 K for 3.5\% oxygen content \cite{koch1974effects}. Another class of low-$T_{c}$ impurity phases in Nb are the niobium hydrides \cite{wang2022investigation,ford2013first}, and some of these phases exhibit superconducting transitions around 1.3 K \cite{isagawa1980hydrogen}.

Motivated by the existence of these low-$T_{c}$ impurity phases, here we consider a phenomenological surface defect toy model in which the grain boundaries of Nb host the low-$T_c$ material (impurity phases). Such grain boundaries might serve as weak spots for vortex nucleation. As shown later in this section, the proximity effect is active in this model of the grain boundaries. Here we consider one possible toy model realization of ``Nb grain boundaries filled with low-$T_c$ material". Of course, the phenomenological toy model (a grain boundary model) considered here is just one possible scenario of surface defects that might be able to qualitatively explain the experimental results. RF vortices are more prone to nucleate at wide grain boundaries compared to narrow ones. Consequently, instead of characterizing a typical grain boundary in Nb/Cu films, the toy model is designed to characterize specifically those that are wide.

Fig. \ref{fig:OneGBSideView} shows the side view of the grain boundary model. By the very definition of $h_\textrm{penetration}^\textrm{defect}$, we constrain the RF vortices to only probe the region of $0>z>-h_\textrm{penetration}^\textrm{defect}$, and hence the details in the region of $z<-h_\textrm{penetration}^\textrm{defect}$ play a minor role. As a result, this region can be treated as bulk Nb as an approximation. (In summary, $z=0$: sample surface; $0>z>-h_\textrm{penetration}^\textrm{defect}$: defect whose XY cross-section is shown in Fig. \ref{fig:OneGBTopView} (b); $z<-h_\textrm{penetration}^\textrm{defect}$: bulk Nb.) Here $h_\textrm{penetration}^\textrm{defect}$ is set to be 200 nm.

The RF dipole is located at $(0,0,h_{dp})$ and hence vortex semi-loops first show up near $(x,y,z)=(0,0,0)$. Therefore, the physics around the origin plays a dominant role. As an approximation, surface defects (Nb grain boundaries filled with low-$T_c$ material) are introduced near the origin (illustrated in Fig. \ref{fig:OneGBSideView}), while the region far away from the origin is set to be defect-free bulk Nb to reduce computational time. As a result, only the region around the origin characterizes the grain boundary scenario accurately (the model defect region), and thus the screening current is collected only from this region when calculating $P_{3f}$.

The side view of the grain boundary model is shown in Fig. \ref{fig:OneGBSideView}, and the top view ($(x,y,z=0)$) is shown in Fig. \ref{fig:OneGBTopView}. Fig. \ref{fig:OneGBTopView} (a) is an illustration of an Nb surface containing Nb grains (red) and grain boundaries filled with low-$T_{c}$ impurity phases (blue). The realization used in the TDGL simulations is shown in Fig. \ref{fig:OneGBTopView} (b). Fig. \ref{fig:OneGBTopView} (b) is a top view of the sample critical temperature distribution around the origin of the grain boundary model: the red region means Nb with $T_{c}=9.3$ K, and the blue region means the low-$T_{c}$ impurity phase with $T_c^\textrm{impurity}=3$ K. (Parameters of the grain boundary model are given in Appendix \ref{sec:TDGLparameters} and in Table \ref{tbl:TDGLparameters}.) The origin is at the center of Fig. \ref{fig:OneGBTopView} (b). The geometry is intentionally asymmetric in both the x direction and the y direction to prevent symmetry-induced artifacts. A top view of the sample critical temperature distribution with a broader scope (containing the defect region together with the bulk Nb region) is shown in Appendix \ref{sec:OneGBComplete}.

An RF vortex semi-loop that nucleates in the sample tends to be parallel to the direction of the RF dipole, which points in the x direction. Therefore, as $B_{pk}>B_{vortex}^{RF}(T)$, RF vortex semi-loops show up in the grain boundaries that are roughly parallel to the x direction. Note that there is one grain boundary in Fig. \ref{fig:OneGBTopView} (b) that is roughly parallel to the x direction and it is marked by a white dashed curve. 

It is worth mentioning that the proximity effect shows up naturally in TDGL simulations. Fig. \ref{fig:OneGBTopView} (c) shows a snapshot of the distribution of normalized order parameter $|\psi / \psi_{\infty}|$ (here 1 means full superconductivity and 0 means no superconductivity) obtained by a TDGL simulation with the temperature being 5.2 K and $B_{pk} = 50.9$ mT. This snapshot is taken at the end of an RF cycle, namely when the RF field drops to zero ($\omega t=2 \pi$ and hence $B_{RF}sin(\omega t)=0$). Due to the proximity effect, the normalized order parameter of the dark blue region is around 0.18 but not zero, even though the temperature (5.2 K) is higher than $T_c^\textrm{impurity}$ (3 K). As a result, a grain boundary filled with a low-$T_{c}$ impurity phase can host RF vortices even for $T>T_c^\textrm{impurity}$.

\begin{figure}
\includegraphics[width=0.45\textwidth]{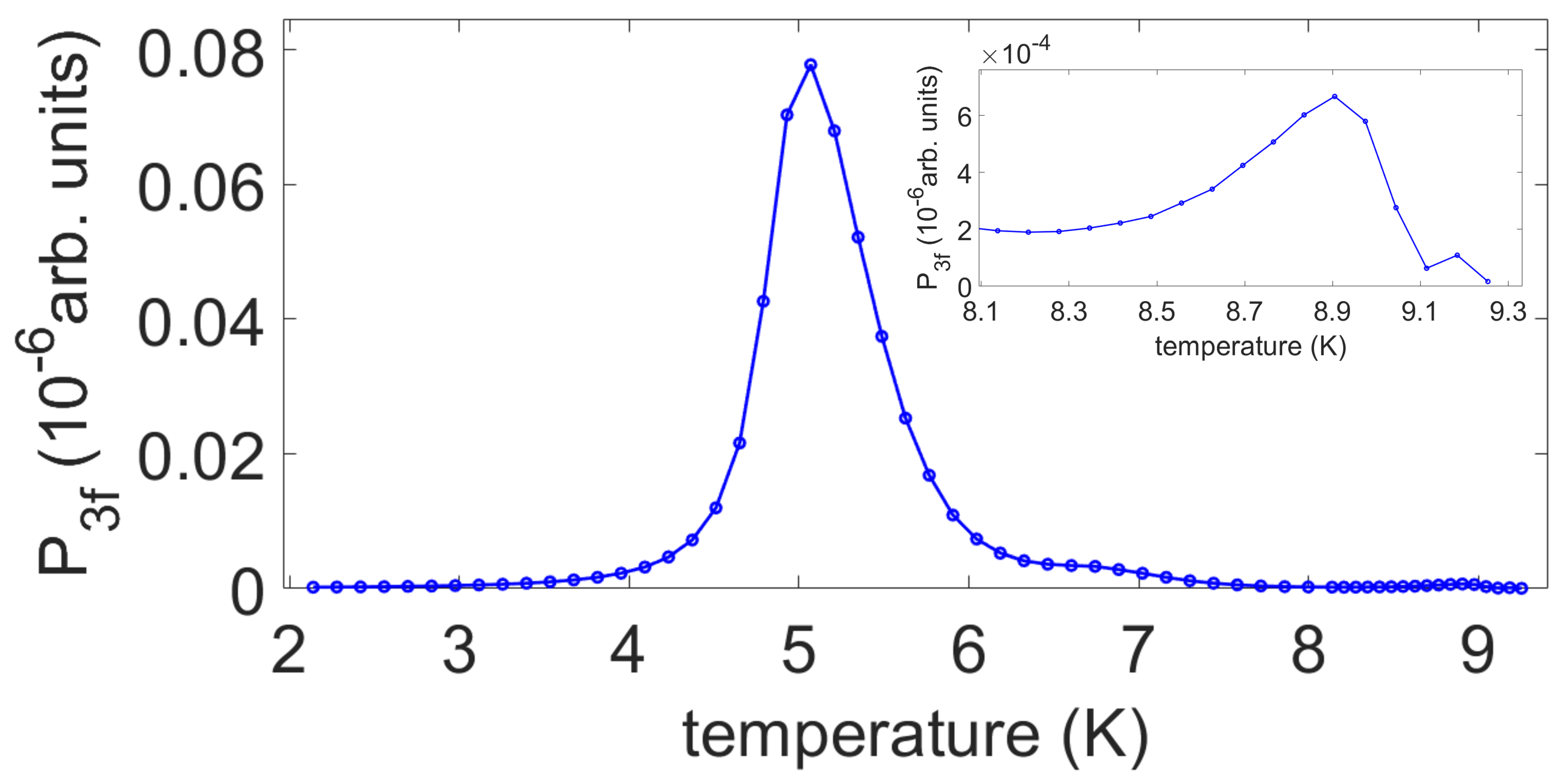}
\caption{\label{fig:OneGBBroadT}TDGL simulation result of $P_{3f}(T)$ for the grain boundary model shown in Fig. \ref{fig:OneGBSideView} and Fig. \ref{fig:OneGBTopView}. Here $h_\textrm{penetration}^\textrm{defect}=200$ nm and $B_{pk} = 50.9$ mT. Inset: enlargement of the figure above 8.1 K.}
\end{figure}

\begin{figure}
\includegraphics[width=0.45\textwidth]{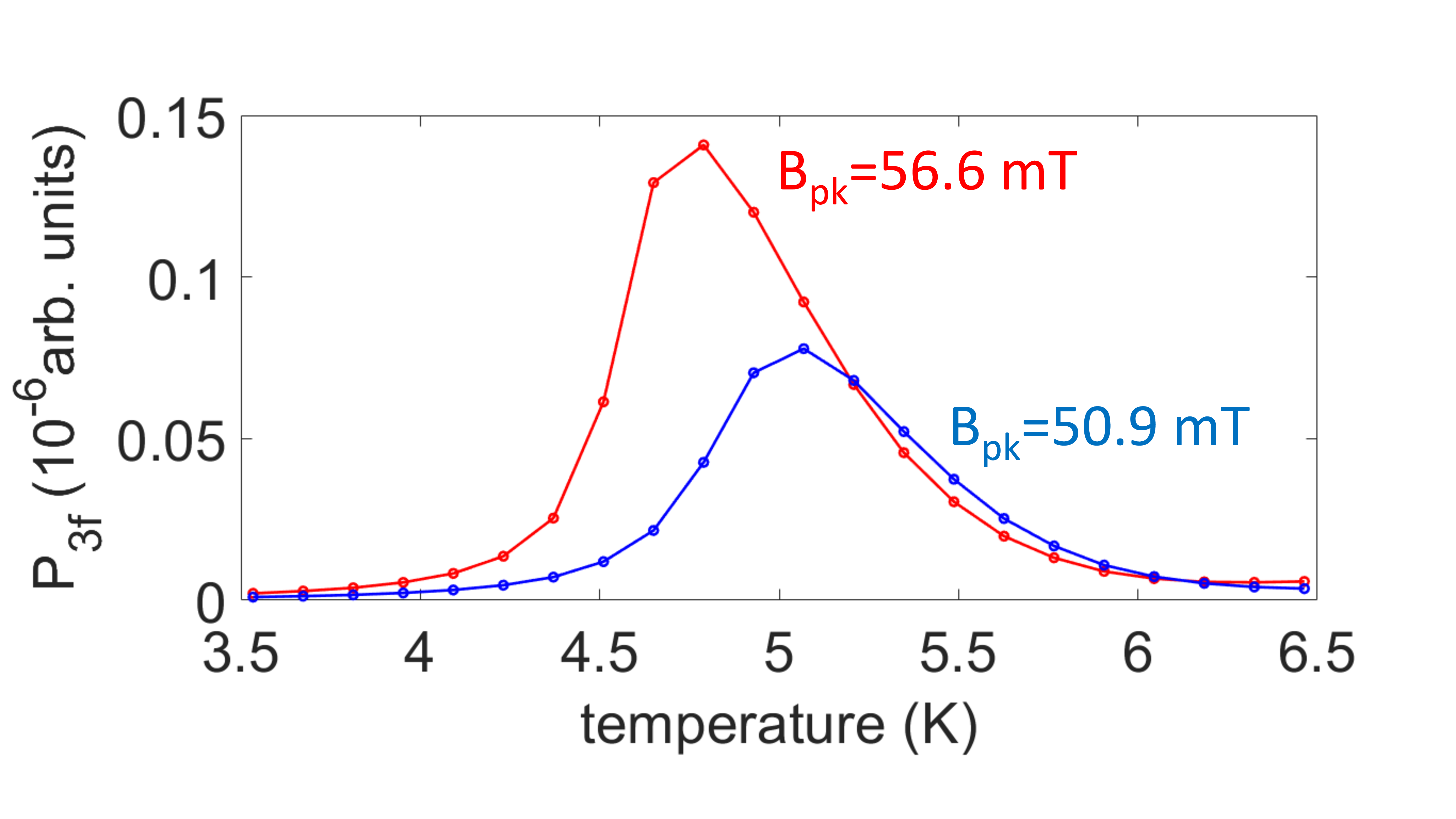}
\caption{\label{fig:OneGBResult}TDGL simulation result of $P_{3f}(T)$ for two different RF field amplitudes for the grain boundary model shown in Fig. \ref{fig:OneGBSideView} and Fig. \ref{fig:OneGBTopView}. Here $h_\textrm{penetration}^\textrm{defect}=200$ nm.}
\end{figure}

Fig. \ref{fig:OneGBBroadT} shows the simulation result of $P_{3f}(T)$ for the grain boundary model shown in Fig. \ref{fig:OneGBSideView} (side view of the model) and Fig. \ref{fig:OneGBTopView} (top view of the model). Compared to the $P_{3f}$ around 8.9 K, the $P_{3f}$ between 4.5 K and 6 K is much stronger. In other words, in the presence of surface defects, $P_{3f}$ generated by surface defects is much stronger than the intrinsic $P_{3f}$ of Nb. In the following, we focus on the $P_{3f}$ generated by surface defects.

Fig. \ref{fig:OneGBResult} shows the simulation result of $P_{3f}(T)$ for two different RF field amplitudes for the grain boundary model. At low temperatures, the sample is in the Meissner state and $P_{3f}$ is weak. As temperature increases, RF vortex semi-loops nucleate in the grain boundary marked by the white dashed curve in Fig. \ref{fig:OneGBTopView} (b) and result in strong $P_{3f}$. (The existence of RF vortex semi-loops is verified by examining the order parameter in a time-domain manner as described in Fig. \ref{fig:VortexDynamics}.) This can be interpreted as $B_{pk}>B_{vortex}^{RF}(T)$, where $B_{vortex}^{RF}(T)$ is the vortex penetration field of the region around that specific grain boundary. Note that RF vortex semi-loops show up in the grain boundary, indicating that the grain boundary serves as the weak spot for RF vortex nucleation.

Here the upper onset temperature of $P_{3f}$ is around 6 K. The grain boundary model is a mixture of the $T_c^\textrm{Nb}=9.3$ K Nb and the $T_c^\textrm{impurity}=3$ K impurity phase, and hence $T_{c}^{P_{3f}}$ is between 3 K and 9.3 K. Of course, the numerical value of $T_{c}^{P_{3f}}$ depends on how the Nb and the impurity phase are distributed. Our objective with this model is not to propose a specific microstructure of the sample, but to illustrate the generic nonlinear properties of a proximity-coupled defective region of the sample.

Similar to the case of bulk Nb (Fig. \ref{fig:TDGLforNb3}(a)), $P_{3f}(T)$ in Fig. \ref{fig:OneGBResult} also exhibits the four key features (Table \ref{tbl:FourKeyFeatures}). Therefore, the ideas illustrated in Fig. \ref{fig:TDGLforNb3}(b) could describe the physics of the grain boundary model, with $T_{c}^{P_{3f}}$ being 6 K but not 9.3 K and $B_{vortex}^{RF}(T)$ being the vortex penetration field of the region around the specific grain boundary, but not a bulk property. The fact that experimental data (Fig. \ref{fig:P3fMultiplePowers} and Fig. \ref{fig:75Vand125V}) and TDGL simulation of the grain boundary model (Fig. \ref{fig:OneGBResult}) both exhibit the four key features (Table \ref{tbl:FourKeyFeatures}) and a $T_{c}^{P_{3f}}$ below 9.3 K suggests that ``RF vortex semi-loops nucleate in grain boundaries hosting low-$T_{c}$ impurity phases" is indeed one of the possible mechanisms of the observed $P_{3f}$ signals for the HiPIMS Nb/Cu samples.


\subsection{Effect of $h_\textrm{penetration}^\textrm{defect}$ on $P_{3f}(T)$}
\label{sec:GBHeight}

\begin{figure}
\includegraphics[width=0.45\textwidth]{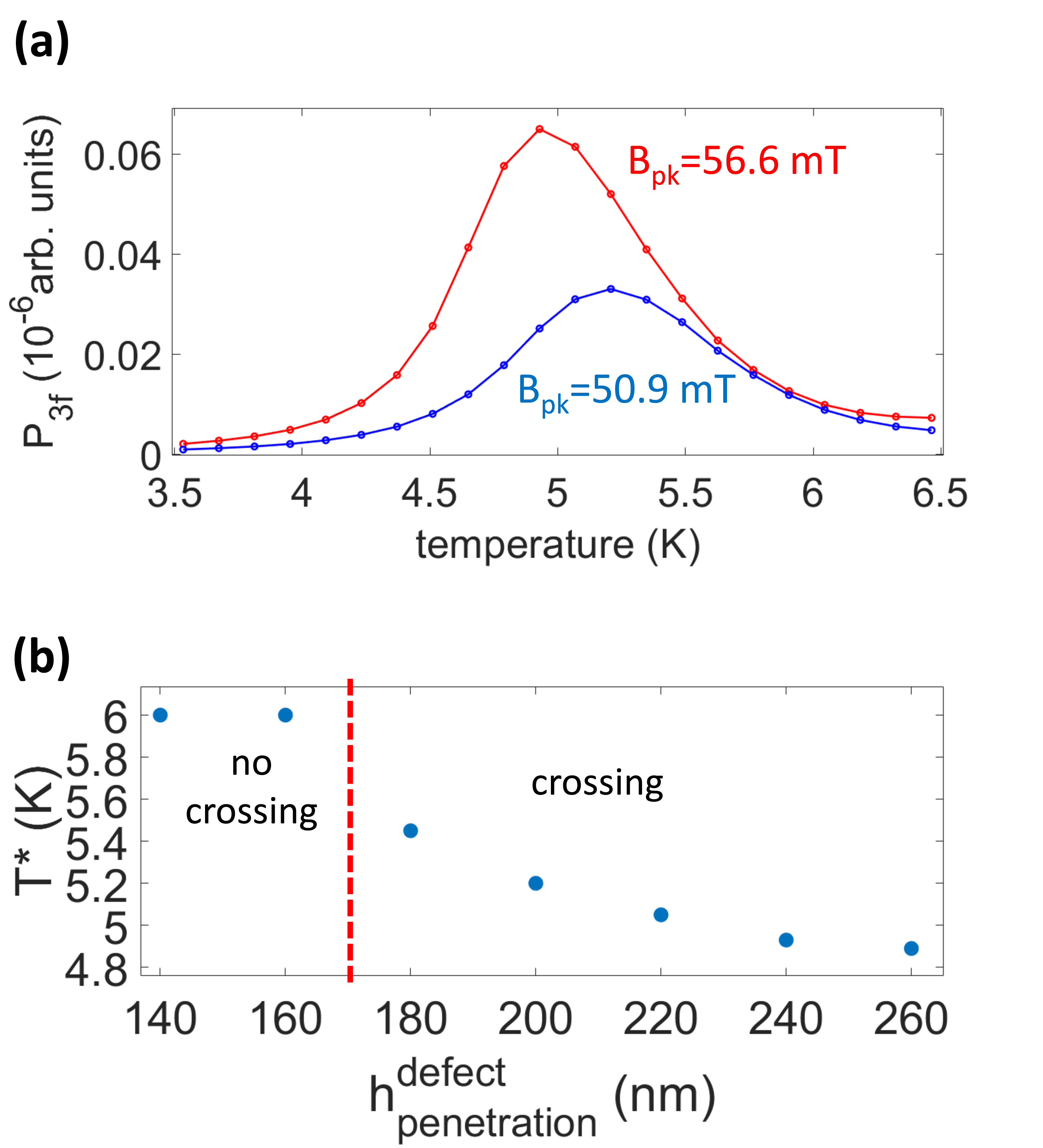}
\caption{\label{fig:TDGLheight} 
(a) TDGL simulation result of $P_{3f}(T)$ for two different RF field amplitudes for the grain boundary model shown in Fig. \ref{fig:OneGBSideView} and Fig. \ref{fig:OneGBTopView} for $h_\textrm{penetration}^\textrm{defect}=160$ nm. (b) Crossing temperature $T^{*}$ as a function of $h_\textrm{penetration}^\textrm{defect}$.
}
\end{figure}

In Sec. \ref{sec:GBModel}, $h_\textrm{penetration}^\textrm{defect}$ is set to be 200 nm. Here we consider the effect of varying $h_\textrm{penetration}^\textrm{defect}$, with everything else being the same as described in Sec. \ref{sec:GBModel}.

Fig. \ref{fig:TDGLheight} (a) shows the result for $h_\textrm{penetration}^\textrm{defect}=160$ nm. Fig. \ref{fig:TDGLheight} (a) is similar to the data in Fig. \ref{fig:75Vand125V} (a), in the sense that a stronger RF field amplitude leads to a stronger $P_{3f}$ for all temperatures; Fig. \ref{fig:OneGBResult} ($h_\textrm{penetration}^\textrm{defect}=200$ nm) is similar to the data in Fig. \ref{fig:75Vand125V} (b), in the sense that $P_{3f}(T)$ shows a ``crossing" effect: a stronger RF field amplitude leads to a weaker $P_{3f}$ (the red curve is below the blue curve) for temperatures close to $T_{c}^{P_{3f}}$. The temperature where the red curve ($P_{3f}(T)$ with a strong RF field amplitude) and the blue curve ($P_{3f}(T)$ with a weak RF field amplitude) cross is denoted as $T^{*}$. The numerical value of $T^{*}$ depends on the choice of the two RF field amplitudes (Here $B_{pk} =$ 56.6 mT and 50.9 mT). For Fig. \ref{fig:TDGLheight} (a), there is no crossing and hence $T^{*}=T_{c}^{P_{3f}}=6$ K. For Fig. \ref{fig:OneGBResult}, $T^{*}=5.2 \textrm{K} <T_{c}^{P_{3f}}=6 \textrm{K}$. Fig. \ref{fig:TDGLheight} (b) shows how the crossing temperature $T^{*}$ changes with $h_\textrm{penetration}^\textrm{defect}$. For a small $h_\textrm{penetration}^\textrm{defect}$, there is no crossing and hence $T^{*}=T_{c}^{P_{3f}}$. The crossing shows up when $h_\textrm{penetration}^\textrm{defect}$ is beyond a critical depth. As $h_\textrm{penetration}^\textrm{defect}$ becomes larger, the crossing effect becomes more significant ($T^{*}$ becomes smaller, which means that the temperature window that a stronger RF field amplitude leads to a weaker $P_{3f}$ becomes larger) and eventually tends to saturate. 

The crossing effect can be understood as follows. In experiments/simulations, $P_{3f}$ is collected by the magnetic writer/at the RF dipole location, which is at $z=h_{dp}$ (above the sample surface). For a weak RF field amplitude, the RF vortex semi-loop in the grain boundary stays close to the sample surface ($z=0$). As the RF field amplitude increases, the RF vortex semi-loop in the grain boundary is pushed toward the bottom of the grain boundary that it can penetrate ($z=-h_\textrm{penetration}^\textrm{defect}$)(see Appendix \ref{sec:VortexSnapshot} for a visualization of this effect), which means that the RF vortex semi-loop is farther away from the magnetic writer/the RF dipole location ($z=h_{dp}$), and hence the measured $P_{3f}$ becomes weaker. Such a phenomenon shows up only when the RF vortex semi-loop in the grain boundary can be pushed far away from the sample surface. For a small $h_\textrm{penetration}^\textrm{defect}$, the RF vortex semi-loop always stays just below the sample surface instead of penetrating deep into the sample, and hence $P_{3f}$ doesn't decrease as the RF field amplitude increases (no crossing effect).

Note that the results presented in Fig. \ref{fig:TDGLheight} are obtained using the grain boundary model, which is meant to be a phenomenological toy model. Therefore, the numerical values of $h_\textrm{penetration}^\textrm{defect}$ should not be interpreted literally. The key insight here is that there is no crossing effect for a small $h_\textrm{penetration}^\textrm{defect}$ (shallow penetration), and the crossing effect becomes apparent for a large $h_\textrm{penetration}^\textrm{defect}$ (deep penetration).


\subsection{Simulation of a defect model with two grain boundaries}
\label{sec:TwoGBmodel}

\begin{figure}
\includegraphics[width=0.45\textwidth]{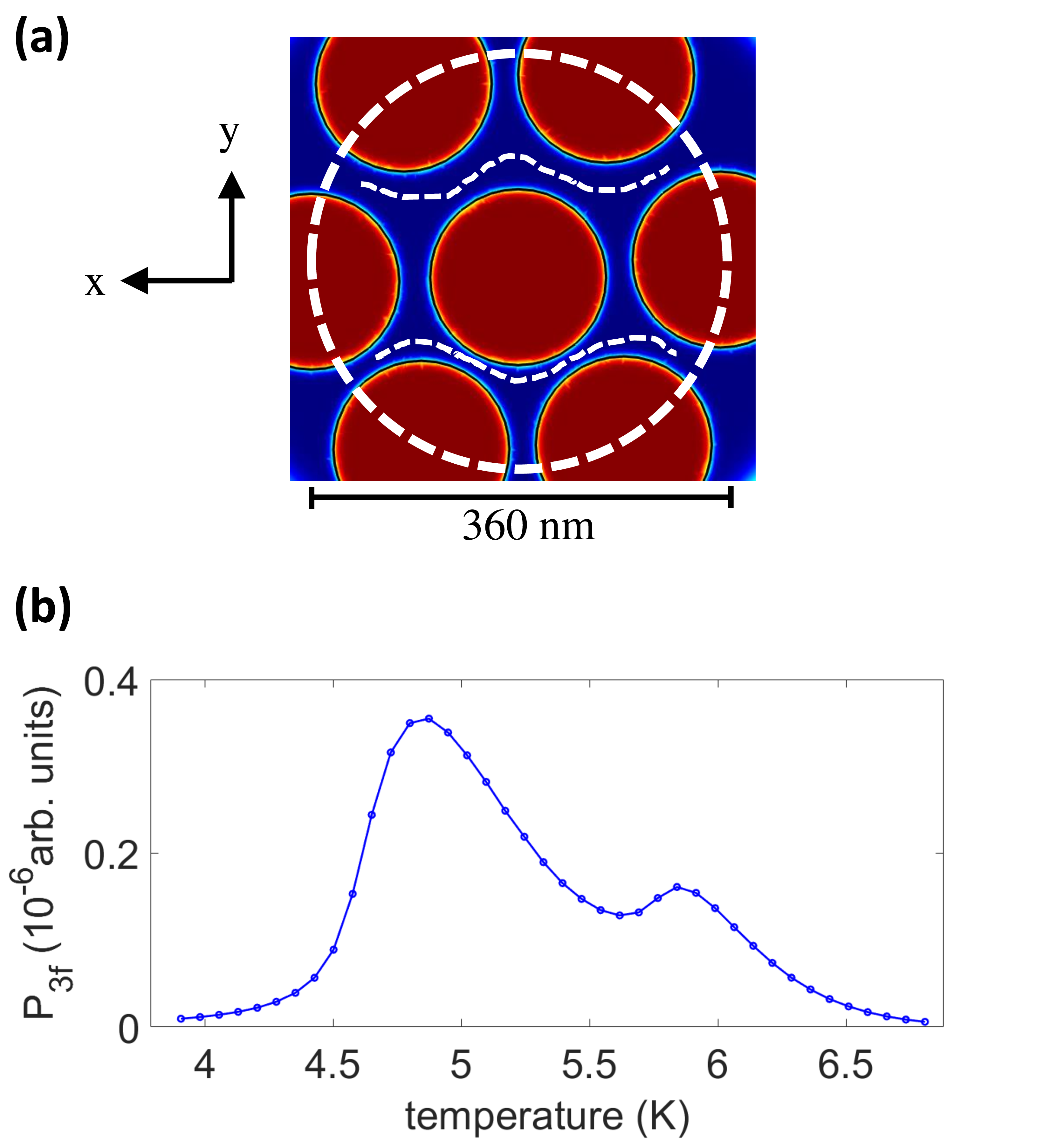}
\caption{\label{fig:TDGL2GB} A grain boundary model containing two grain boundaries that are roughly parallel to the x direction. (a) The distribution of critical temperature on the XY plane around the origin, with red being Nb and blue being the low-$T_c$ impurity phase with $T_c^\textrm{impurity}=3$ K. Here $(x,y,z)=(0,0,0)$ is at the center. The RF dipole is located at $(x,y,z)=(0,0,400 \textrm{nm})$, which is above the center of the image, and points in the x direction. The two white dashed curves indicate the grain boundaries (the top grain boundary and the bottom grain boundary) that are roughly parallel to the x direction. The model defect region is indicated by the white dashed circle. (b) TDGL simulation result of $P_{3f}(T)$ for the grain boundary model shown in (a). Here $h_\textrm{penetration}^\textrm{defect}=280$ nm, and $B_{pk} = 50.9$ mT.}
\end{figure}

In the grain boundary model discussed in Sec. \ref{sec:GBModel} and Sec. \ref{sec:GBHeight}, there is only one single grain boundary underneath and roughly parallel to the RF dipole, and thus RF vortex semi-loops nucleate in one single grain boundary and $P_{3f}(T)$ shows a single-peak feature. Such a scenario corresponds to the case where the density of the sample grain boundaries that nucleates RF vortices ($\rho^\textrm{defect}$) is low. For a sample with a high density of grain boundaries that nucleates RF vortices, it can be modeled as a grain boundary model that contains two grain boundaries underneath and roughly parallel to the RF dipole.

Here we consider a grain boundary model that contains two grain boundaries that are near the origin and roughly parallel to the x direction. The basic setting of the model is the same as described in Sec. \ref{sec:GBModel}. The only difference is how the Nb and the impurity phase are distributed horizontally, as shown in Fig. \ref{fig:TDGL2GB} (a). Fig. \ref{fig:TDGL2GB} (a) shows the top view of the critical temperature distribution and Fig. \ref{fig:TDGL2GB} (b) shows the TDGL simulation result of $P_{3f}(T)$ for this model. 

The two-peak feature of $P_{3f}(T)$ in Fig. \ref{fig:TDGL2GB} (b) can be understood as follows. At low temperatures, the sample is in the Meissner state and $P_{3f}$ is weak. As temperature increases, around 4.50 K an RF vortex semi-loop nucleates in the top grain boundary and results in the lower temperature peak, and then around 5.77 K another RF vortex semi-loop nucleates in the bottom grain boundary and results in the higher temperature peak. That is, RF vortex semi-loops nucleate in both grain boundaries and thus result in the two-peak feature of $P_{3f}(T)$. The nucleation of the two RF vortex semi-loops is verified by monitoring $\Delta \theta / 2 \pi$ (the same analysis as shown in Fig. \ref{fig:TDGLforNbSingle}). Note that the two-peak feature of the $P_{3f}(T)$ in Fig. \ref{fig:TDGL2GB} (b) (simulation) is also observed in Fig. \ref{fig:P3fMultiplePowers} (measurement).


\subsection{Nb/Cu samples comparison}
\label{sec:Comparison}

\begin{table}
\begin{center}
\includegraphics[width=0.45\textwidth]{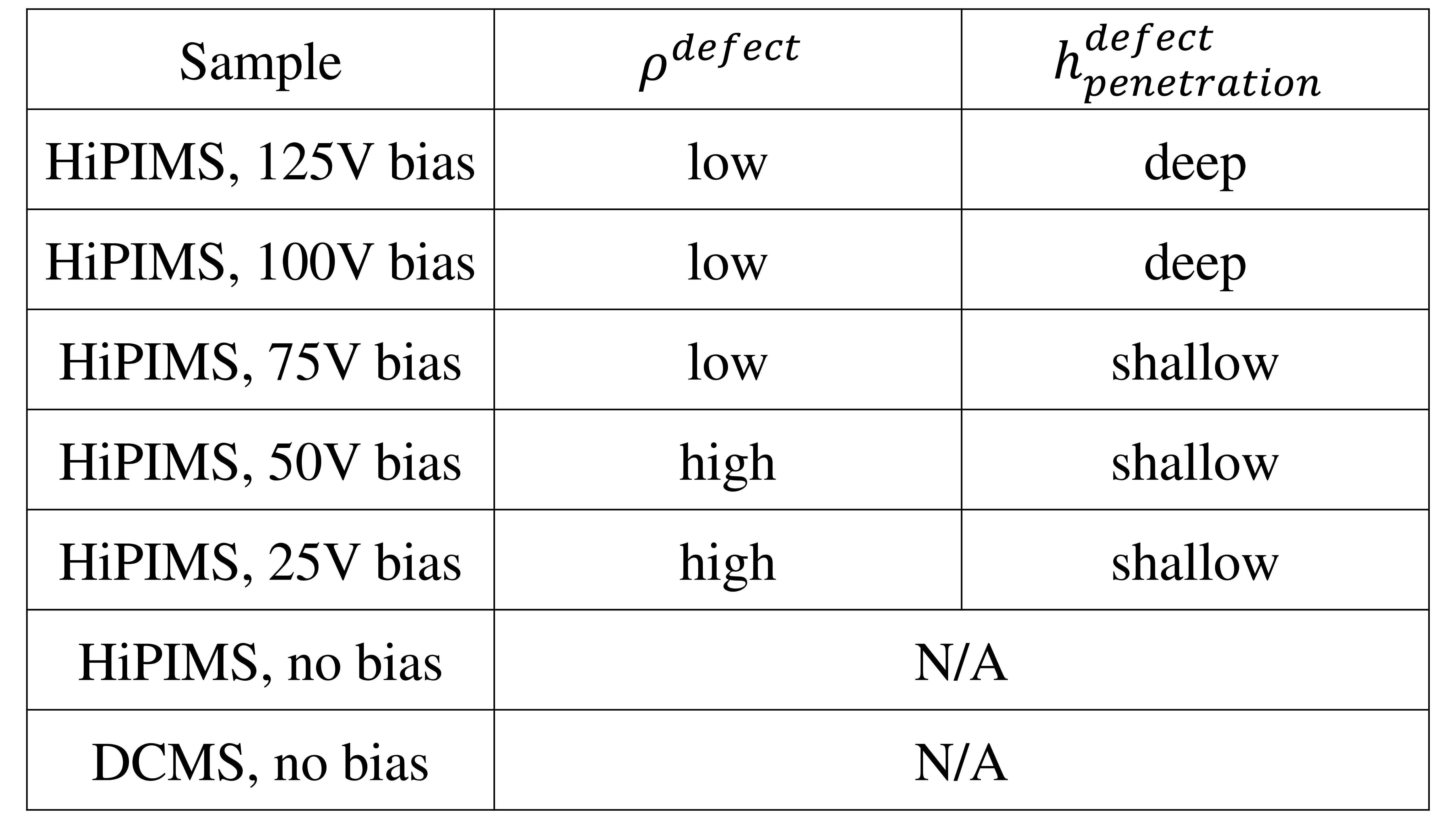}
\caption{Comparison of the seven Nb/Cu samples. The second column shows qualitative assignments of the density of surface defects that nucleate RF vortex semi-loops ($\rho^\textrm{defect}$) and the third column shows qualitative assignments of how deep an RF vortex semi-loop penetrates into a sample through a surface defect ($h_\textrm{penetration}^\textrm{defect}$).}
\label{tbl:SampleComparison}
\end{center}
\end{table}

Larger values of $P_{3f}$ are closely related to the presence of RF vortex semi-loops. By studying $P_{3f}$, we investigate surface defects that nucleate RF vortex semi-loops, which are the kinds of surface defects that are closely related to the RF performance of SRF cavities. In the following, surface defects refer to the kind of surface defects that nucleate RF vortex semi-loops. 

Equipped with the simulation results, the measured $P_{3f}$ can be analyzed further. First, the number of $P_{3f}(T)$ peaks can be related to the density of surface defects ($\rho^\textrm{defect}$). Single-peak $P_{3f}(T)$ corresponds to a low density of surface defects, and two-peak $P_{3f}(T)$ corresponds to a high density of surface defects. Second, the crossing effect of $P_{3f}(T)$ can be related to how deep an RF vortex semi-loop penetrates into a sample through a surface defect ($h_\textrm{penetration}^\textrm{defect}$). The absence of the crossing effect corresponds to a small $h_\textrm{penetration}^\textrm{defect}$ (shallow), and the presence of the crossing effect corresponds to a large $h_\textrm{penetration}^\textrm{defect}$ (deep).

The assignment of qualitative surface defect properties of all seven Nb/Cu samples is summarized in Table \ref{tbl:SampleComparison}. The second column shows an estimate for the density of surface defects that nucleate RF vortex semi-loops ($\rho^\textrm{defect}$) and the third column shows how deep an RF vortex semi-loop penetrates into a sample through a surface defect ($h_\textrm{penetration}^\textrm{defect}$). RF vortex penetration through surface defects is one of the main enemies of SRF applications. To achieve good SRF performance, the surface defect density should be low and the defect should be shallow. From the perspective of these two properties, among the five HiPIMS Nb/Cu samples with non-zero voltage bias, the HiPIMS 75 V bias Nb/Cu sample is the most effective in reducing the nucleation of RF vortices associated with surface defects.

In the numerical simulations of this work, surface defects that nucleate RF vortex semi-loops are modeled as grain boundaries. Since $P_{3f}$ is associated with the behavior of RF vortex semi-loops, it is likely that the ideas of $\rho^\textrm{defect}$ and $h_\textrm{penetration}^\textrm{defect}$ could apply to surface defects other than grain boundaries (dislocations, for example).

For the DCMS sample, the actual $P_{3f}$ value of the sample response remains unknown due to interference with the background signal and the lack of relative phase information (see Fig. \ref{fig:50Vand100V} (d) and related discussion in Appendix \ref{sec:ExtraData}). Consequently, the TDGL simulation results cannot be directly applied to the experimental data of the DCMS sample.


\section{Conclusion}
\label{sec:Conclusion}

In this work, we study seven Nb/Cu films that are candidates for SRF applications. Local $P_{3f}$ measurements reveal surface defects with $P_{3f}$ onset temperatures between 6.3 K and 6.8 K for five out of the six HiPIMS Nb/Cu samples, indicating that such defects are a generic feature of these air-exposed HiPIMS Nb/Cu films. The signal from the low-$T_c$ surface defect is much stronger than the intrinsic Nb response around 9 K, suggesting that our local $P_{3f}$ measurement is sensitive to surface defects. With the capability of $\mu$m-scale scanning, it is found that such a defect is quite uniform in space on the $\mu$m-scale.  

TDGL simulations are performed to analyze the experimental results further. In particular, the simulations suggest that the density of surface defects that nucleate RF vortices and how deep RF vortices travel through these surface defects can be extracted qualitatively from our local $P_{3f}$ measurements. From the point of view of these two properties, the HiPIMS 75 V bias Nb/Cu sample stands out as the most effective in reducing the nucleation of RF vortices associated with surface defects.


\section{Acknowledgement}
\label{sec:Acknowledgement}

The authors would like to thank Javier Guzman from Seagate Technology for providing magnetic write heads. C.Y.W. would like to thank Bakhrom Oripov and Jingnan Cai for helpful discussions. This work is funded by the U.S. Department of Energy/High Energy Physics through grant No. DESC0017931 and the Maryland Quantum Materials Center.

\appendix


\section{Sample preparation}
\label{sec:SamplePreparation}

\begin{table}
\begin{center}
\includegraphics[width=0.45\textwidth]{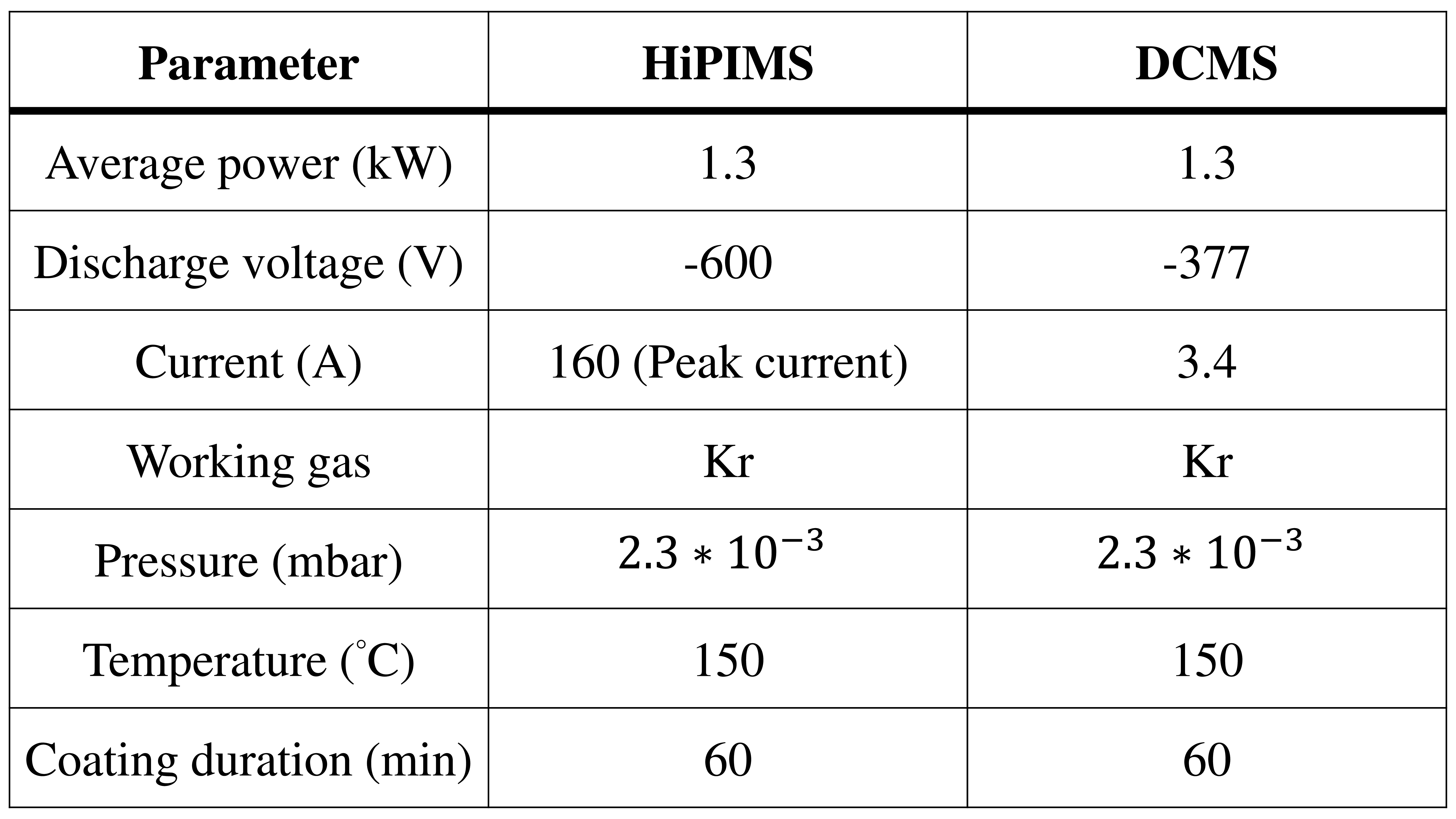}
\caption{Sample preparation parameters of the seven Nb/Cu films studied in this paper.}
\label{tbl:SampleParameter}
\end{center}
\end{table}

Here we discuss how the seven Nb/Cu films studied in this paper are prepared.

The substrate used for the Nb coatings is a 2 mm thick, oxygen-free electronic (OFE) copper disk measuring 75 mm in diameter. Prior to coating, the substrate disk is degreased using commercial detergent. The sample is then chemically polished using a mixture of sulfamic acid ($\textrm{H} _\textrm{3} \textrm{NSO} _\textrm{3}$, 5 g/L), hydrogen peroxide ($\textrm{H} _\textrm{2} \textrm{O} _\textrm{2}$, 5\% vol.), n-butanol (5\% vol.) and ammonium citrate (1 g/L) heated up at 72$^{0} \textrm{C}$ for 20 minutes. After polishing, the disk is rinsed with sulfamic acid to remove the build-up of native oxide and cleaned with de-ionized water and ultra-pure ethanol.

The Cu substrate is mounted on an ultra-high vacuum (UHV) stainless steel chamber equipped with a rotatable shutter to expose in turn the areas to be coated, and the chamber is then connected to a sputtering system. Both assemblies are performed inside an ISO5 cleanroom, and the sputtering apparatus is described in detail in \cite{rosaz2022niobium}. The entire system is transported to the coating bench where it is coupled to the pumping group and gas injection lines, and pumped down to about $1\times10^{-7}$ mbar. The pumping group and the sputtering system undergo a 48-hour bakeout at 200$^{0} \textrm{C}$, during which a 4-hour activation of the Non-Evaporable Getter (NEG) pump is performed. The temperature of the UHV chamber is maintained at 150$^{0} \textrm{C}$ until the start of the coating. After cooling down, the system reaches a base pressure around $9.3\times10^{-10}$ mbar. Ultra-pure krypton (99.998\%) is injected into the system until a process pressure of $2.3\times10^{-3}$ mbar is reached. The seven coatings are then performed according to the deposition parameters outlined in Table \ref{tbl:SampleParameter}. One of the samples is prepared by Direct Current Magnetron Sputtering (DCMS) with zero bias, and the coating thickness is around 3.5 $\mu$m. The other six samples are prepared by High Power Impulse Magnetron Sputtering (HiPIMS), with bias voltages ranging from 0 V to -125 V, with coating thicknesses ranging from 1.15 $\mu$m to 1.5 $\mu$m. 

During the coating process, the sample temperature was monitored with an infrared thermal sensor (OMEGA OS100-SOFT) and kept constant at 150$^{0} \textrm{C}$. The HiPIMS plasma discharge was maintained using a pulsed power supply (Huettinger TruPlasma HighPulse 4006) and the negative bias voltage was applied to the samples using a DC power supply (TruPlasma Bias 3018). The DCMS discharge was maintained using a Huettinger Truplasma 3005 power supply. The discharge and bias voltages and currents were monitored throughout the entire coating process using voltage (Tektronix P6015A) and current (Pearson current monitor 301×) probes whose signals are recorded by a digital oscilloscope (Picoscope 2000). After the coating, the samples were cooled down to room temperature, after which the chamber was vented with dry air. The Nb layer thickness is measured by X-ray fluorescence via the attenuation method.


\section{Data for two HiPIMS samples and the DCMS sample}
\label{sec:ExtraData}

\begin{figure}
\includegraphics[width=0.4\textwidth]{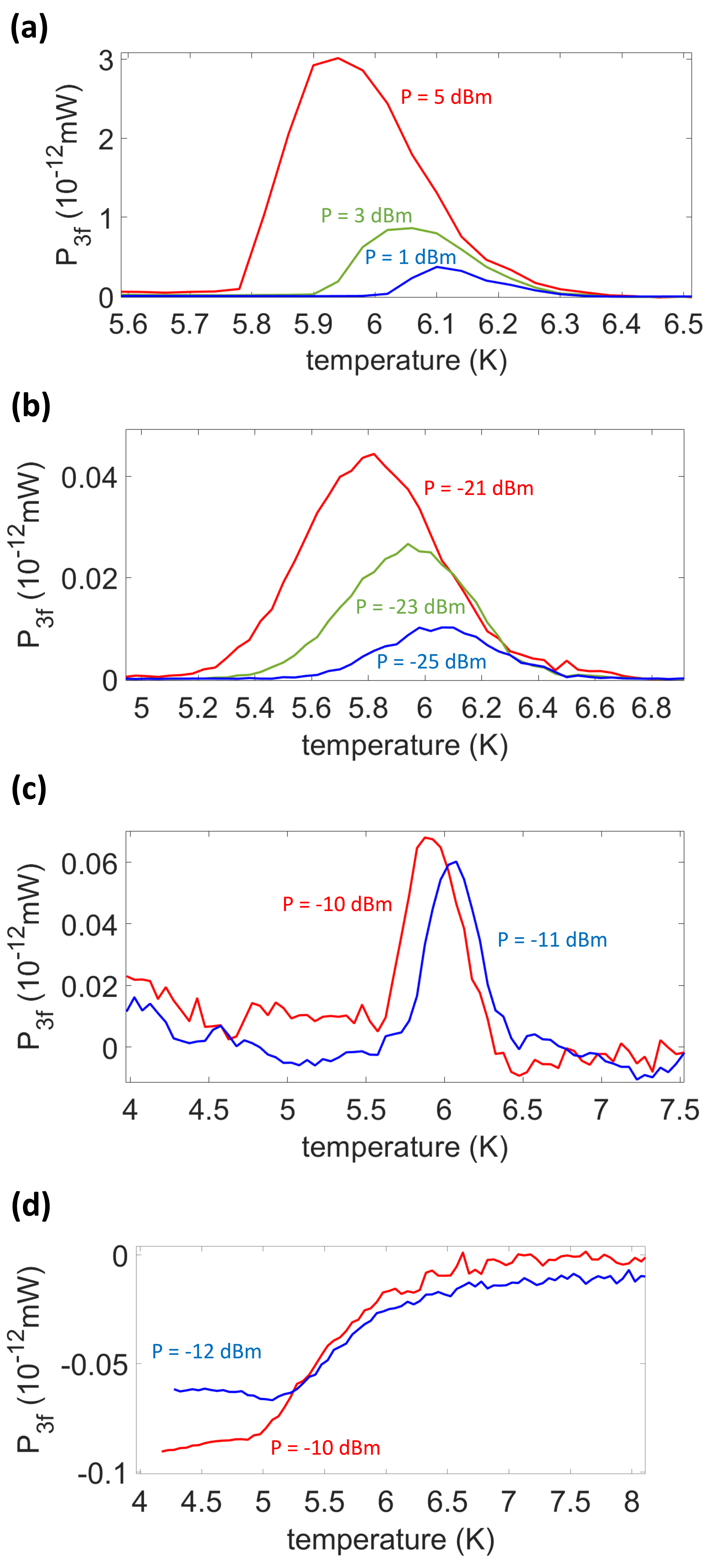}
\caption{\label{fig:50Vand100V} (a) and (b) show $P_{3f}(T)$ for the HiPIMS 50 V bias Nb/Cu sample with an input frequency of 2.16 GHz in a strong input power regime and in a weak input power regime, respectively. (c) shows $P_{3f}(T)$ for the HiPIMS 100 V bias Nb/Cu sample with an input frequency of 1.18 GHz. (d) shows $P_{3f}(T)$ for the DCMS Nb/Cu sample with an input frequency of 1.22 GHz. The negative value for $T<6.5$K is discussed in the second paragraph of Appendix \ref{sec:ExtraData}.}
\end{figure}

Here we show the representative data for $P_{3f}(T)$ generated by surface defects for the HiPIMS 50 V bias sample, the HiPIMS 100 V bias sample and the DCMS sample. For the HiPIMS 50 V bias sample, $P_{3f}(T)$ with $T_{c}^{P_{3f}}$ around 6.4 K is observed in a strong input power regime (Fig. \ref{fig:50Vand100V} (a)), and $P_{3f}(T)$ with $T_{c}^{P_{3f}}$ around 6.8 K is observed in a weak input power regime (Fig. \ref{fig:50Vand100V} (b)). For the HiPIMS 100 V bias sample, $P_{3f}(T)$ with $T_{c}^{P_{3f}}$ around 6.5 K is observed (Fig. \ref{fig:50Vand100V} (c)).

For the DCMS sample, $P_{3f}(T)$ with $T_{c}^{P_{3f}}$ around 6.5 K is observed (Fig. \ref{fig:50Vand100V} (d)). The measured $P_{3f}$ is the superposition of the probe background and the sample response. Consequently, the total signal may be weaker than the probe background if these two components are out of phase. In the absence of phase information, we simply subtract the probe background from the total signal in a scalar manner. This naive background subtraction can lead to negative $P_{3f}$ values, as observed in Fig. \ref{fig:50Vand100V} (d) for $T<6.5$K.

The phase information can be obtained by conducting measurements with a vector network analyzer \cite{mircea2009phase,tai2012nanoscale}, which provides both amplitude and phase data, instead of a spectrum analyzer that only measures amplitude. However, due to its superior noise floor (approximately -155 dBm), we chose to use a spectrum analyzer for this work, rather than a vector network analyzer, which has a noise floor of around -130 dBm.


\section{Data for the HiPIMS no bias sample}
\label{sec:NoBiasSample}

\begin{figure}
\includegraphics[width=0.4\textwidth]{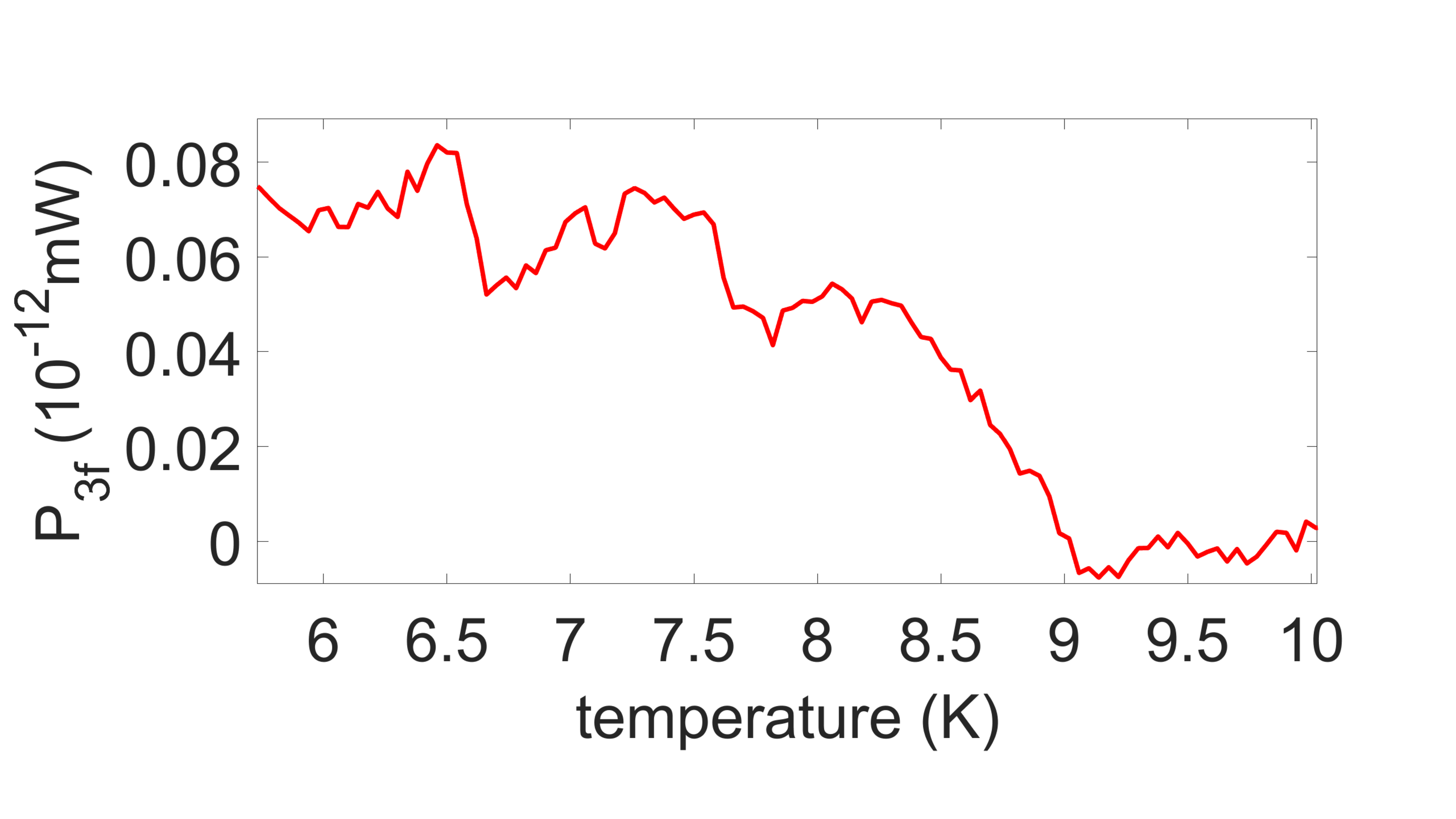}
\caption{\label{fig:0Vsample} Measured $P_{3f}(T)$ for the HiPIMS 0 V bias Nb/Cu sample. The input frequency is 1.98 GHz and the input power is 3 dBm.}
\end{figure}

Here we discuss the data of the HiPIMS 0 V bias Nb/Cu sample, as shown in Fig. \ref{fig:0Vsample}. For this sample, $P_{3f}(T)$ shows a clear transition around 9 K. In addition to the intrinsic Nb response, $P_{3f}(T)$ demonstrates a non-trivial temperature dependence around 7.6 K and 6.5 K, indicating the presence of surface defects.

The surface defect signal of the HiPIMS 0 V bias Nb/Cu sample is qualitatively different from all the other samples examined in this work in two aspects. First, $P_{3f}(T)$ shows a bell-shaped structure in all the other samples (see Fig. \ref{fig:P3fMultiplePowers}, Fig. \ref{fig:75Vand125V} and Fig. \ref{fig:50Vand100V}). Second, in all other samples that display both responses, the surface defect signal is much stronger than the intrinsic Nb response (see Fig. \ref{fig:P3fRaw}). Consequently, the HiPIMS 0 V bias Nb/Cu sample likely exhibits a unique mechanism generating nonlinearity, distinct from the mechanisms observed in all the other six samples. Notably, when deposited on complex-shaped substrates, the HiPIMS 0 V samples show significant porosity inside the film, which is quite different from all other HiPIMS films at non-zero bias \cite{rosaz2022niobium}. This porosity may give rise to a qualitatively different type of nonlinear response, which is beyond the scope of the present work.


\section{Check for hysteresis in $P_{3f}$ measurements}
\label{sec:Hysteresis}

\begin{figure}
\includegraphics[width=0.45\textwidth]{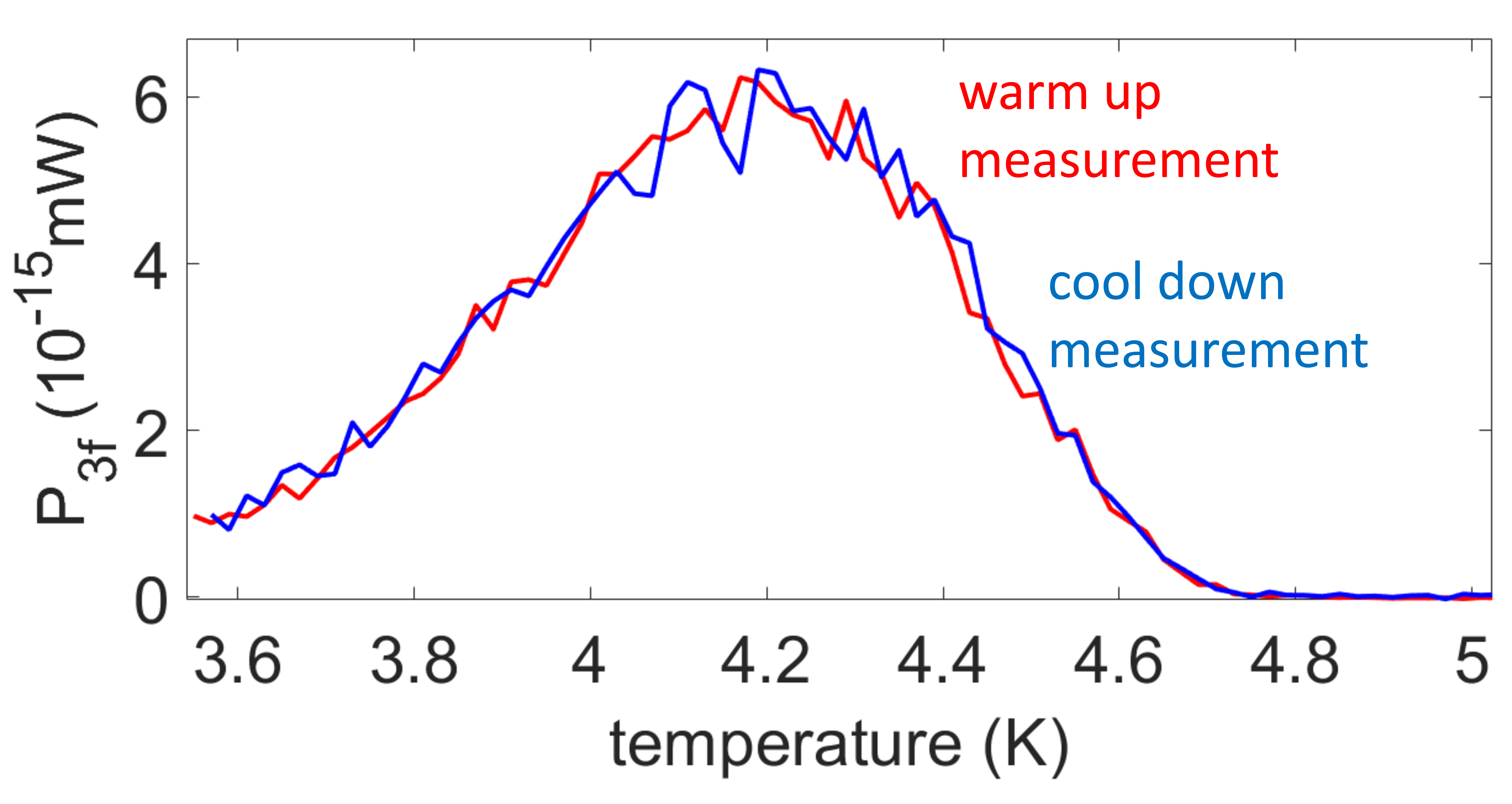}
\caption{\label{fig:Hysteresis}Check for hysteresis in $P_{3f}(T)$ measurement for the HiPIMS 125 V bias Nb/Cu sample. The input frequency is 1.88 GHz and the input power is -25 dBm.}
\end{figure}

Here we check whether or not $P_{3f}(T)$ measurements exhibit hysteresis. The HiPIMS 125 V bias Nb/Cu sample first undergoes a cool down from 10 K to 3.6 K with zero RF field. The microwave signal is turned on after the temperature stabilizes at 3.6 K. The sample then gradually warms up from 3.6 K to 10 K (warm-up $P_{3f}(T)$), and then cools down from 10 K to 3.6 K (cool-down $P_{3f}(T)$), with the microwave signal being turned on in this process. As shown in Fig. \ref{fig:Hysteresis}, such $P_{3f}(T)$ measurement does not exhibit a clear hysteresis.

The measurement of Fig. \ref{fig:Hysteresis} is performed one year and a half after the measurement of Fig. \ref{fig:75Vand125V} (b). It is quite likely that different regions of the sample surface are studied in these two measurements. This might explain why Fig. \ref{fig:Hysteresis} ($T_{c}^{P_{3f}}$=4.8 K) and Fig. \ref{fig:75Vand125V} (b) ($T_{c}^{P_{3f}}$=6.8 K) show different $P_{3f}$ signals.


\section{Equations and parameters for TDGL simulations}
\label{sec:TDGLparameters}

\begin{table}
\begin{center}
\includegraphics[width=0.45\textwidth]{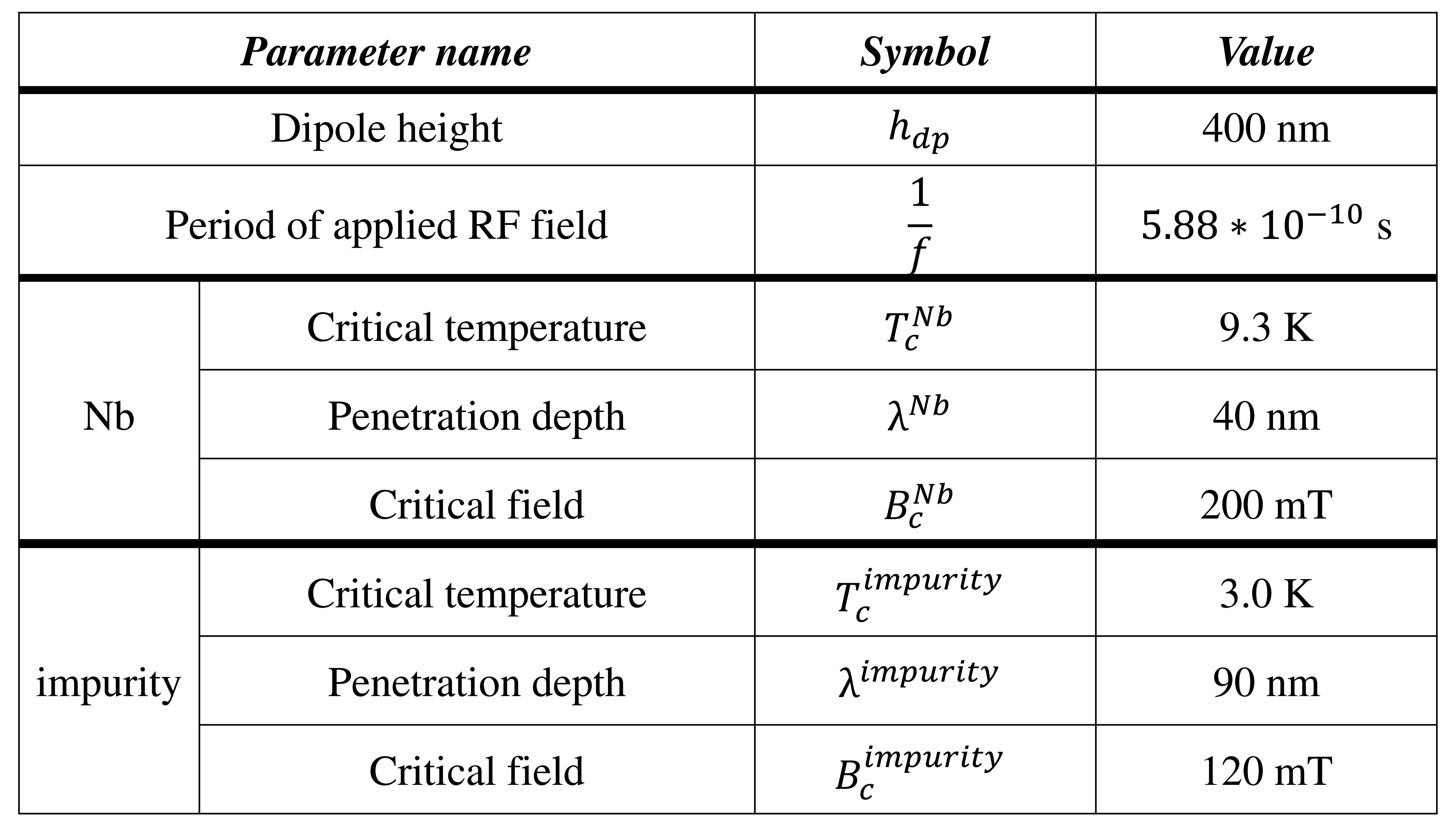}
\caption{Values of parameters used in TDGL simulations for bulk Nb (Sec. {\ref{sec:BulkNbTDGL}}) and for the two surface defect impurity models (Sec. \ref{sec:GBModel}, Sec. \ref{sec:GBHeight} and Sec. \ref{sec:TwoGBmodel})}
\label{tbl:TDGLparameters}
\end{center}
\end{table}

In TDGL, the relaxation dynamics of the order parameter $\psi$ is assumed to be described as follows \cite{tinkham2004introduction,gulian2020shortcut}

\begin{equation}
\begin{split}
-\gamma\left(\frac{\partial}{\partial t}+i\frac{e_{*}}{\hbar}\varPhi\right)\psi=\frac{\delta F_{GL}}{\delta\psi^{*}},
\end{split}
\end{equation}
with the Ginzburg-Landau free energy $F_{GL}$ given by
\begin{equation}
\begin{split}
F_{GL}= & \frac{-1}{2m_{*}}\left|\left(-i\hbar\nabla-e_{*}A\right)\psi\right|^{2}+\alpha\left|\psi\right|^{2}+\frac{\beta}{2}\left|\psi\right|^{4} \\
& +\frac{1}{2\mu_{0}}\left|\nabla\times A-B_{ext}\right|^{2}.
\end{split}
\end{equation}
Here $e_{*}=2e$ is the charge of the Cooper pair, $m_{*}=2m_{e}$ is the mass of the Cooper pair, $\varPhi$ is the electric potential, $\gamma=\frac{\pi \hbar \left|\alpha\right|}{8 k_{B} (T_{c}-T)}$ plays the role of a friction coefficient, $\psi$ is the order parameter, $A$ is the total vector potential (arising from both external and self-generated sources), $B_{ext}$ is the external magnetic field, and both $\alpha$ and $\beta$ are material-specific phenomenological parameters. 
The dynamics of $\psi$ and the electromagnetic field are described by Maxwell's equations and the following TDGL equations 

\begin{equation}
\begin{split}
\gamma\left(\frac{\partial}{\partial t}+i\frac{e_{*}}{\hbar}\varPhi\right)\psi= & \frac{-1}{2m_{*}}\left(-i\hbar\nabla-e_{*}A\right)^{2}\psi \\
& -\alpha\psi-\beta\left|\psi\right|^{2}\psi
\end{split}
\end{equation}
and
\begin{equation}
\begin{split}
\sigma\left(\nabla\varPhi+\frac{\partial A}{\partial t}\right)= & \frac{-i\hbar e_{*}}{2m_{*}}\left(\psi^{*}\nabla\psi-\psi\nabla\psi^{*}\right)-\frac{e_{*}^{2}}{m_{*}}\left|\psi\right|^{2}A \\
& -\frac{1}{\mu_{0}}\nabla\times\left(\nabla\times A-B_{ext}\right).
\end{split}
\end{equation}
Here $\sigma$ is the electric conductivity of the normal state.

The Ginzburg-Landau order parameter relaxation time $\tau_{GL}$ is given by
\begin{equation} \label{eq:TDGLtime}
\tau_{GL}=\frac{\pi\hbar}{8k_{B}\left(T_{c}-T\right)}=\left(3\cdot10^{-12}s\right)\left(\frac{K}{T_{c}-T}\right).
\end{equation}
The dynamics of a superconductor when driven by an RF field of a few GHz is in the adiabatic limit unless $T$ is extremely close to $T_c$ ($T_c-T<10^{-2}K$).

There is no external current inside the superconductor in our setup, and hence $\nabla\times B_{ext}=\mu_{0}J_{ext}=0$ inside the superconductor. In addition, we choose the gauge such that the scalar potential is zero, namely $\varPhi=0$. The two TDGL equations inside the superconductor become

\begin{equation} \label{eq:FirstTDGL}
\begin{split}
\gamma\frac{\partial\psi}{\partial t}=\frac{-1}{2m_{*}}\left(-i\hbar\nabla-e_{*}A\right)^{2}\psi-\alpha\psi-\beta\left|\psi\right|^{2}\psi
\end{split}
\end{equation}
and
\begin{equation} \label{eq:SecondTDGL}
\begin{split}
\sigma\frac{\partial A}{\partial t}= & \frac{-i\hbar e_{*}}{2m_{*}}\left(\psi^{*}\nabla\psi-\psi\nabla\psi^{*}\right)-\frac{e_{*}^{2}}{m_{*}}\left|\psi\right|^{2}A \\
& -\frac{1}{\mu_{0}}\nabla\times\nabla\times A.
\end{split}
\end{equation}
Note that $\alpha$ and $\beta$, namely the two material-specific phenomenological parameters of the Ginzburg-Landau free energy, can be related to the penetration depth $\lambda$ and the thermodynamic critical field $B_{c}$ by

\begin{equation} \label{eq:LambdaFormula}
\lambda=\sqrt{\frac{m_{*}\beta}{\mu_{0}e_{*}^{2}\left|\alpha\right|}}
\end{equation}
and
\begin{equation} \label{eq:BcFormula}
B_{c}=\frac{\sqrt{\mu_{0}}\left|\alpha\right|}{\sqrt{\beta}}.
\end{equation}

In the two surface defect models (Sec. \ref{sec:GBModel}, Sec. \ref{sec:GBHeight} and Sec. \ref{sec:TwoGBmodel}), the superconducting samples contain Nb and the low-$T_c$ impurity phase, and thus are inhomogeneous. Equations \ref{eq:FirstTDGL} and \ref{eq:SecondTDGL} are applied to both the Nb region and the low-$T_c$ impurity phase region, and the inhomogeneity is incorporated via the spatial variation of the five material-specific parameters $T_c$, $\alpha$, $\beta$, $\gamma$ and $\sigma$. In practice, the values of $\alpha$ and $\beta$ are obtained for a given choice of $\lambda$ and $B_c$ through Eqs. \ref{eq:LambdaFormula} and \ref{eq:BcFormula}.

Material parameters (penetration depth $\lambda$, Ginzburg-Landau parameter $\kappa$, etc.) of Nb films vary from one sample to another. For the Nb part in the TDGL simulations, we adopt the material parameters of bulk Nb as shown in the first row of Table 1 of Ref \cite{valente2016superconducting} for simplicity.

We now discuss the choice of the material parameters of the low-$T_c$ impurity phase in the two surface defect models. For simplicity, the normal state conductivity of the low-$T_c$ impurity phase is set to be the same as that of Nb. The transition temperature of the low-$T_c$ impurity phase is set to be 3 K. The penetration depth of the low-$T_c$ impurity phase is expected to be larger than that of Nb (40 nm) and is set to be 90 nm; the thermodynamic critical field of the low-$T_c$ impurity phase is expected to be smaller than that of Nb (200 mT) and is set to be 120 mT. Values of parameters used in TDGL simulations are summarized in Table \ref{tbl:TDGLparameters}. The ``impurity" sector specifies the material parameters of the low-$T_c$ impurity phase region in the two surface defect models (Sec. \ref{sec:GBModel}, Sec. \ref{sec:GBHeight} and Sec. \ref{sec:TwoGBmodel}).

To numerically simulate the superconducting domain, boundary conditions need to be specified. Any current passing through the boundary between a superconductor and a vacuum is unphysical, and hence on the boundary $\partial \Omega$ of a superconducting domain we expect \cite{oripov2020time}

\begin{equation} \label{eq:BCoriginal}
J \cdot \hat{n} = 0 \quad \textrm{on} \quad \partial \Omega.
\end{equation}
Here $\hat{n}$ is the unit vector normal to the boundary. The supercurrent $J_s$ is given by
\begin{equation} \label{eq:Js}
J_s = \frac{-i\hbar e_{*}}{2m_{*}}\left(\psi^{*}\nabla\psi-\psi\nabla\psi^{*}\right)-\frac{e_{*}^{2}}{m_{*}}\left|\psi\right|^{2}A.
\end{equation}
The boundary condition \ref{eq:BCoriginal} should hold even when $A=0$, and thus the boundary condition \ref{eq:BCoriginal} becomes
\begin{equation} \label{eq:BCfirst}
\nabla \psi \cdot \hat{n} = 0 \quad \textrm{on} \quad \partial \Omega.
\end{equation}
Combine Eq. \ref{eq:BCoriginal}, \ref{eq:Js}, and \ref{eq:BCfirst} and we obtain
\begin{equation} \label{eq:BCsecond}
A \cdot \hat{n} = 0 \quad \textrm{on} \quad \partial \Omega.
\end{equation}
Equations \ref{eq:BCfirst} and \ref{eq:BCsecond} are applied as the boundary conditions of the superconductor-vacuum interface.


\section{A top view for the grain boundary model in Sec. \ref{sec:GBModel}}
\label{sec:OneGBComplete}

\begin{figure}
\includegraphics[width=0.45\textwidth]{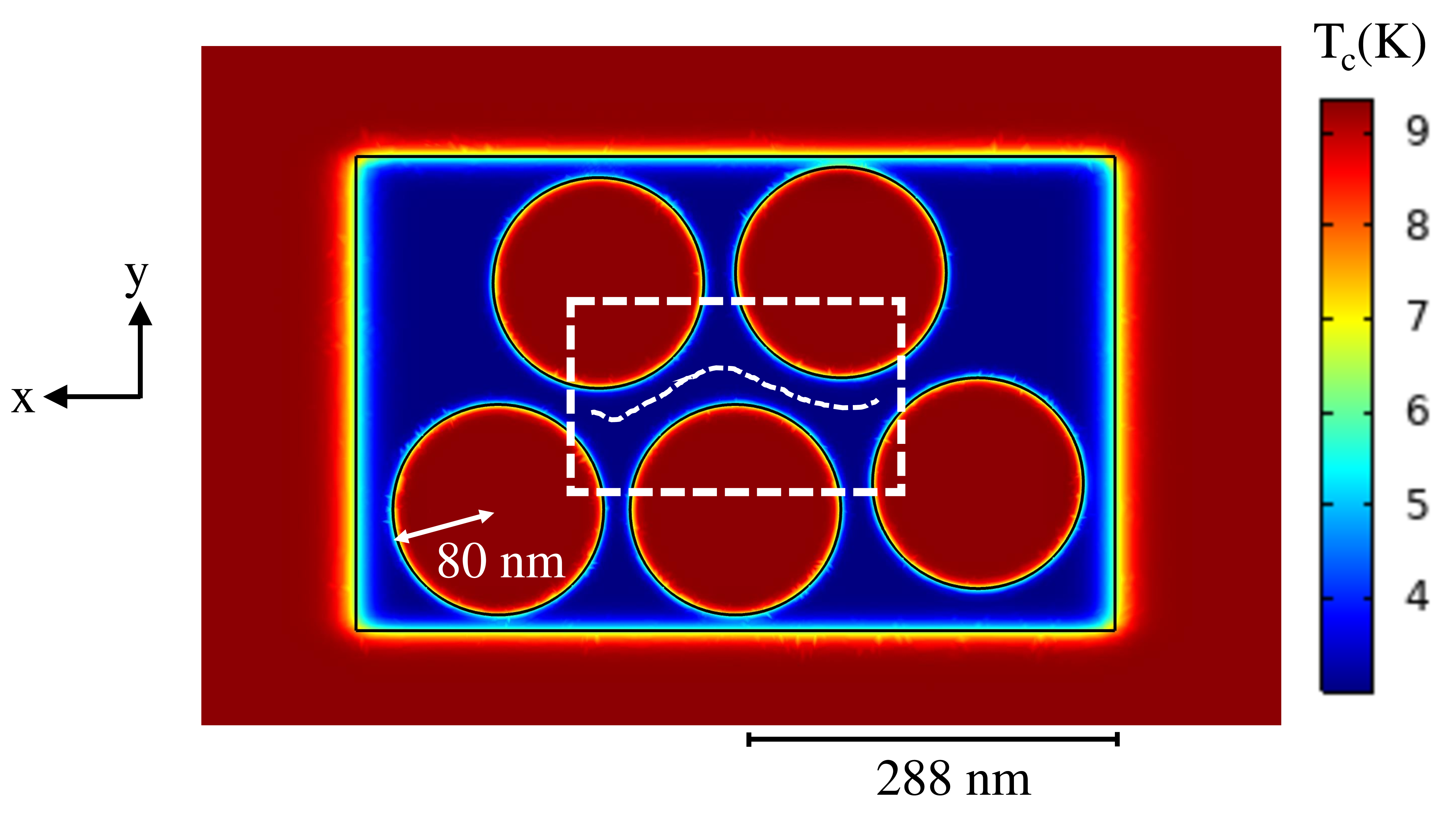}
\caption{\label{fig:OneGBComplete}A spatially extended view of the distribution of critical temperature on the XY plane for the grain boundary model discussed in Sec. \ref{sec:GBModel}. The region shown here corresponds to the region indicated by the green dashed line in Fig. \ref{fig:OneGBSideView}.}
\end{figure}

Fig. \ref{fig:OneGBTopView} (b) shows the region around the origin (which plays the dominant role in RF vortex nucleation and $P_{3f}$) for the grain boundary model discussed in Sec. \ref{sec:GBModel}. Compared to Fig. \ref{fig:OneGBTopView} (b), Fig. \ref{fig:OneGBComplete} shows the setup over a broader spatial extent as indicated by the green dashed line in Fig. \ref{fig:OneGBSideView}. Fig. \ref{fig:OneGBComplete} contains the entire defect region (the five Nb grains and the blue region) and part of the bulk Nb region as shown in Fig. \ref{fig:OneGBSideView}.


\section{Snapshots of a vortex in a grain boundary for the grain boundary model in Sec. \ref{sec:GBModel}}
\label{sec:VortexSnapshot}

\begin{figure}
\includegraphics[width=0.45\textwidth]{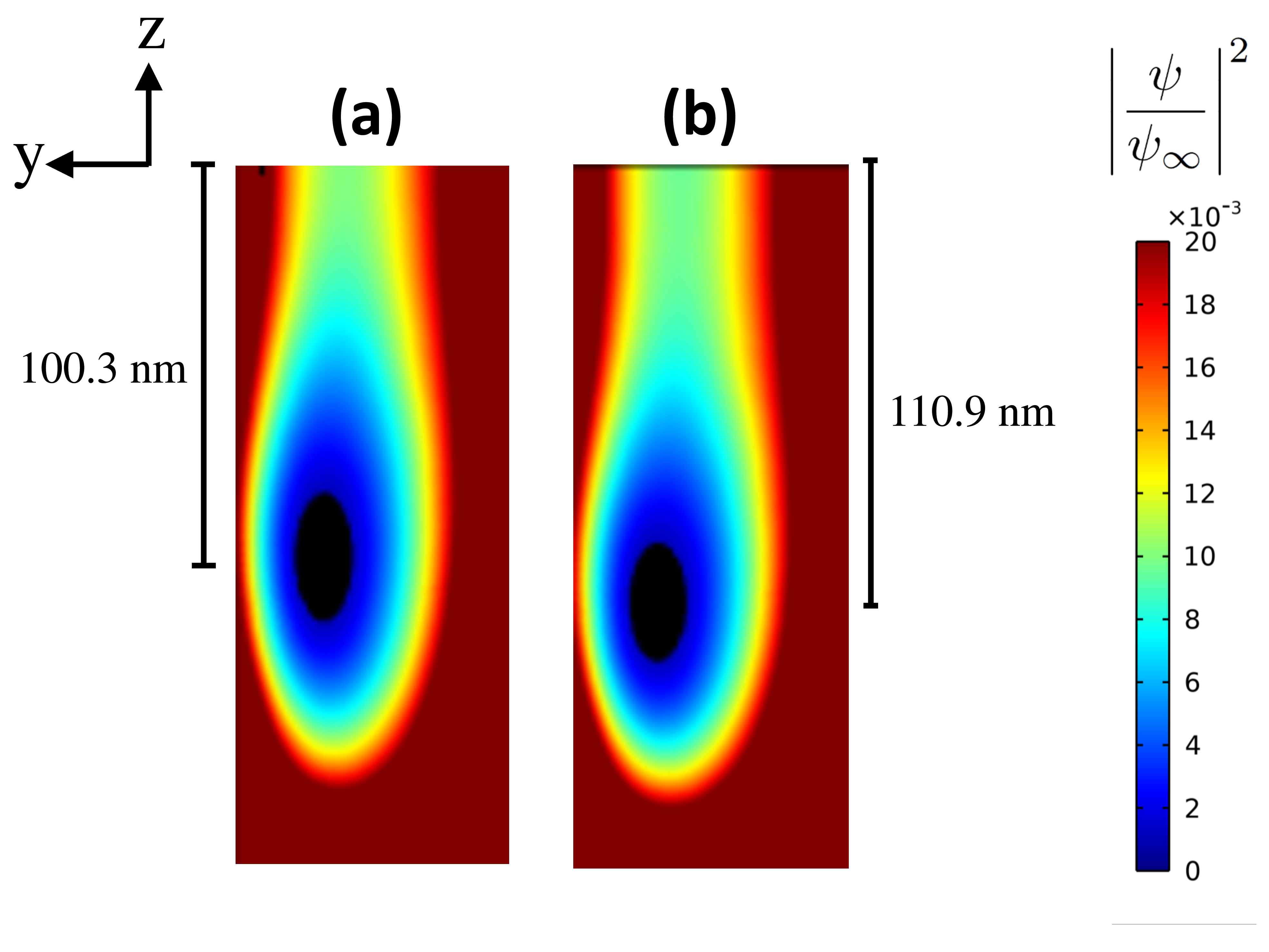}
\caption{\label{fig:TDGLVortexSnapshot}A snapshot of the vortex core obtained by the TDGL simulation for the grain boundary model in Sec. \ref{sec:GBModel} for $B_{pk} = 50.9$ mT (a) and for $B_{pk} = 56.6$ mT (b), with the temperature being 5.5 K and $h_\textrm{penetration}^\textrm{defect}=200$ nm. This snapshot is taken when the RF field reaches its maximum in an RF cycle ($\omega t= \pi /2$ and hence $B_{RF}sin(\omega t)=B_{RF}$). The snapshot shows the distribution of the square of the normalized order parameter on the YZ plane immediately below the dipole around the grain boundary that nucleates an RF vortex semi-loop. The black region is where $|\psi / \psi_{\infty}|^2 < 0.001$.}
\end{figure}

An RF vortex semi-loop is roughly parallel to the direction of the RF dipole, which points in the x direction, and hence the cross-section of the RF vortex semi-loop is on the YZ plane. Fig. \ref{fig:TDGLVortexSnapshot} visualizes an RF vortex semi-loop in the grain boundary marked by the white dashed curve in Fig. \ref{fig:OneGBTopView} (b), with the vortex core corresponding to the black region, where $|\psi / \psi_{\infty}|^2 < 0.001$. The RF vortex semi-loop penetrates the sample surface and the vortex core is around 100.3 nm deep for $B_{pk} = 50.9$ mT (Fig. \ref{fig:TDGLVortexSnapshot} (a)) and is around 110.9 nm deep for $B_{pk} = 56.6$ mT (Fig. \ref{fig:TDGLVortexSnapshot} (b)).


\bibliography{apssamp.bib}

\begin{thebibliography}{91}%
\makeatletter
\providecommand \@ifxundefined [1]{%
 \@ifx{#1\undefined}
}%
\providecommand \@ifnum [1]{%
 \ifnum #1\expandafter \@firstoftwo
 \else \expandafter \@secondoftwo
 \fi
}%
\providecommand \@ifx [1]{%
 \ifx #1\expandafter \@firstoftwo
 \else \expandafter \@secondoftwo
 \fi
}%
\providecommand \natexlab [1]{#1}%
\providecommand \enquote  [1]{``#1''}%
\providecommand \bibnamefont  [1]{#1}%
\providecommand \bibfnamefont [1]{#1}%
\providecommand \citenamefont [1]{#1}%
\providecommand \href@noop [0]{\@secondoftwo}%
\providecommand \href [0]{\begingroup \@sanitize@url \@href}%
\providecommand \@href[1]{\@@startlink{#1}\@@href}%
\providecommand \@@href[1]{\endgroup#1\@@endlink}%
\providecommand \@sanitize@url [0]{\catcode `\\12\catcode `\$12\catcode `\&12\catcode `\#12\catcode `\^12\catcode `\_12\catcode `\%12\relax}%
\providecommand \@@startlink[1]{}%
\providecommand \@@endlink[0]{}%
\providecommand \url  [0]{\begingroup\@sanitize@url \@url }%
\providecommand \@url [1]{\endgroup\@href {#1}{\urlprefix }}%
\providecommand \urlprefix  [0]{URL }%
\providecommand \Eprint [0]{\href }%
\providecommand \doibase [0]{http://dx.doi.org/}%
\providecommand \selectlanguage [0]{\@gobble}%
\providecommand \bibinfo  [0]{\@secondoftwo}%
\providecommand \bibfield  [0]{\@secondoftwo}%
\providecommand \translation [1]{[#1]}%
\providecommand \BibitemOpen [0]{}%
\providecommand \bibitemStop [0]{}%
\providecommand \bibitemNoStop [0]{.\EOS\space}%
\providecommand \EOS [0]{\spacefactor3000\relax}%
\providecommand \BibitemShut  [1]{\csname bibitem#1\endcsname}%
\let\auto@bib@innerbib\@empty
\bibitem [{\citenamefont {Padamsee}(2017)}]{padamsee201750}%
  \BibitemOpen
  \bibfield  {author} {\bibinfo {author} {\bibfnamefont {Hasan}\ \bibnamefont {Padamsee}},\ }\bibfield  {title} {\enquote {\bibinfo {title} {50 years of success for {SRF} accelerators—a review},}\ }\href {https://iopscience.iop.org/article/10.1088/1361-6668/aa6376/meta} {\bibfield  {journal} {\bibinfo  {journal} {Superconductor science and technology}\ }\textbf {\bibinfo {volume} {30}},\ \bibinfo {pages} {053003} (\bibinfo {year} {2017})}\BibitemShut {NoStop}%
\bibitem [{\citenamefont {Bambade}\ \emph {et~al.}(2019)\citenamefont {Bambade}, \citenamefont {Barklow}, \citenamefont {Behnke}, \citenamefont {Berggren}, \citenamefont {Brau}, \citenamefont {Burrows}, \citenamefont {Denisov}, \citenamefont {Faus-Golfe}, \citenamefont {Foster}, \citenamefont {Fujii} \emph {et~al.}}]{bambade2019international}%
  \BibitemOpen
  \bibfield  {author} {\bibinfo {author} {\bibfnamefont {Philip}\ \bibnamefont {Bambade}}, \bibinfo {author} {\bibfnamefont {Tim}\ \bibnamefont {Barklow}}, \bibinfo {author} {\bibfnamefont {Ties}\ \bibnamefont {Behnke}}, \bibinfo {author} {\bibfnamefont {Mikael}\ \bibnamefont {Berggren}}, \bibinfo {author} {\bibfnamefont {James}\ \bibnamefont {Brau}}, \bibinfo {author} {\bibfnamefont {Philip}\ \bibnamefont {Burrows}}, \bibinfo {author} {\bibfnamefont {Dmitri}\ \bibnamefont {Denisov}}, \bibinfo {author} {\bibfnamefont {Angeles}\ \bibnamefont {Faus-Golfe}}, \bibinfo {author} {\bibfnamefont {Brian}\ \bibnamefont {Foster}}, \bibinfo {author} {\bibfnamefont {Keisuke}\ \bibnamefont {Fujii}},  \emph {et~al.},\ }\bibfield  {title} {\enquote {\bibinfo {title} {The international linear collider: a global project},}\ }\href {https://arxiv.org/abs/1903.01629} {\bibfield  {journal} {\bibinfo  {journal} {arXiv preprint arXiv:1903.01629}\ } (\bibinfo {year} {2019})}\BibitemShut {NoStop}%
\bibitem [{\citenamefont {Ciovati}(2006)}]{ciovati2006review}%
  \BibitemOpen
  \bibfield  {author} {\bibinfo {author} {\bibfnamefont {Gianluigi}\ \bibnamefont {Ciovati}},\ }\bibfield  {title} {\enquote {\bibinfo {title} {Review of the frontier workshop and {Q}-slope results},}\ }\href {https://www.sciencedirect.com/science/article/pii/S0921453406001493?casa_token=bFpZ8zVlFEgAAAAA:T-oBJVsnwH5AU6w9jWov4RhlPymjk8JM4PHaSwLjMOBmrJEIot8ujOCPdV4raIzbngoMwCDjgHE} {\bibfield  {journal} {\bibinfo  {journal} {Physica C: Superconductivity}\ }\textbf {\bibinfo {volume} {441}},\ \bibinfo {pages} {44--50} (\bibinfo {year} {2006})}\BibitemShut {NoStop}%
\bibitem [{\citenamefont {Gurevich}(2006{\natexlab{a}})}]{gurevich2006multiscale}%
  \BibitemOpen
  \bibfield  {author} {\bibinfo {author} {\bibfnamefont {A.}~\bibnamefont {Gurevich}},\ }\bibfield  {title} {\enquote {\bibinfo {title} {Multiscale mechanisms of {SRF} breakdown},}\ }\href {https://www.sciencedirect.com/science/article/pii/S0921453406001481} {\bibfield  {journal} {\bibinfo  {journal} {Physica C: Superconductivity}\ }\textbf {\bibinfo {volume} {441}},\ \bibinfo {pages} {38--43} (\bibinfo {year} {2006}{\natexlab{a}})}\BibitemShut {NoStop}%
\bibitem [{\citenamefont {Gurevich}(2012)}]{gurevich2012superconducting}%
  \BibitemOpen
  \bibfield  {author} {\bibinfo {author} {\bibfnamefont {Alex}\ \bibnamefont {Gurevich}},\ }\bibfield  {title} {\enquote {\bibinfo {title} {Superconducting radio-frequency fundamentals for particle accelerators},}\ }\href {https://www.worldscientific.com/doi/abs/10.1142/S1793626812300058} {\bibfield  {journal} {\bibinfo  {journal} {Reviews of Accelerator Science and Technology}\ }\textbf {\bibinfo {volume} {5}},\ \bibinfo {pages} {119--146} (\bibinfo {year} {2012})}\BibitemShut {NoStop}%
\bibitem [{\citenamefont {Gurevich}(2017)}]{gurevich2017theory}%
  \BibitemOpen
  \bibfield  {author} {\bibinfo {author} {\bibfnamefont {Alex}\ \bibnamefont {Gurevich}},\ }\bibfield  {title} {\enquote {\bibinfo {title} {Theory of {RF} superconductivity for resonant cavities},}\ }\href {https://iopscience.iop.org/article/10.1088/1361-6668/30/3/034004/meta} {\bibfield  {journal} {\bibinfo  {journal} {Superconductor Science and Technology}\ }\textbf {\bibinfo {volume} {30}},\ \bibinfo {pages} {034004} (\bibinfo {year} {2017})}\BibitemShut {NoStop}%
\bibitem [{\citenamefont {Champion}\ \emph {et~al.}(2009)\citenamefont {Champion}, \citenamefont {Cooley}, \citenamefont {Ginsburg}, \citenamefont {Sergatskov}, \citenamefont {Geng}, \citenamefont {Hayano}, \citenamefont {Iwashita},\ and\ \citenamefont {Tajima}}]{champion2009quench}%
  \BibitemOpen
  \bibfield  {author} {\bibinfo {author} {\bibfnamefont {Mark~S.}\ \bibnamefont {Champion}}, \bibinfo {author} {\bibfnamefont {Lance~D.}\ \bibnamefont {Cooley}}, \bibinfo {author} {\bibfnamefont {Camille~M.}\ \bibnamefont {Ginsburg}}, \bibinfo {author} {\bibfnamefont {Dmitri~A.}\ \bibnamefont {Sergatskov}}, \bibinfo {author} {\bibfnamefont {Rongli~L.}\ \bibnamefont {Geng}}, \bibinfo {author} {\bibfnamefont {Hitoshi}\ \bibnamefont {Hayano}}, \bibinfo {author} {\bibfnamefont {Yoshihisa}\ \bibnamefont {Iwashita}}, \ and\ \bibinfo {author} {\bibfnamefont {Yujiro}\ \bibnamefont {Tajima}},\ }\bibfield  {title} {\enquote {\bibinfo {title} {Quench-limited {SRF} cavities: Failure at the heat-affected zone},}\ }\href {https://ieeexplore.ieee.org/abstract/document/5067195} {\bibfield  {journal} {\bibinfo  {journal} {IEEE transactions on applied superconductivity}\ }\textbf {\bibinfo {volume} {19}},\ \bibinfo {pages} {1384--1386} (\bibinfo {year} {2009})}\BibitemShut {NoStop}%
\bibitem [{\citenamefont {Bao}\ and\ \citenamefont {Guo}(2019)}]{bao2019quench}%
  \BibitemOpen
  \bibfield  {author} {\bibinfo {author} {\bibfnamefont {Shiran}\ \bibnamefont {Bao}}\ and\ \bibinfo {author} {\bibfnamefont {Wei}\ \bibnamefont {Guo}},\ }\bibfield  {title} {\enquote {\bibinfo {title} {Quench-spot detection for superconducting accelerator cavities via flow visualization in superfluid helium-4},}\ }\href {https://journals.aps.org/prapplied/abstract/10.1103/PhysRevApplied.11.044003} {\bibfield  {journal} {\bibinfo  {journal} {Physical Review Applied}\ }\textbf {\bibinfo {volume} {11}},\ \bibinfo {pages} {044003} (\bibinfo {year} {2019})}\BibitemShut {NoStop}%
\bibitem [{\citenamefont {Posen}\ \emph {et~al.}(2015{\natexlab{a}})\citenamefont {Posen}, \citenamefont {Valles},\ and\ \citenamefont {Liepe}}]{posen2015radio}%
  \BibitemOpen
  \bibfield  {author} {\bibinfo {author} {\bibfnamefont {S.}~\bibnamefont {Posen}}, \bibinfo {author} {\bibfnamefont {N.}~\bibnamefont {Valles}}, \ and\ \bibinfo {author} {\bibfnamefont {M.}~\bibnamefont {Liepe}},\ }\bibfield  {title} {\enquote {\bibinfo {title} {Radio frequency magnetic field limits of {N}b and {N}b3{S}n},}\ }\href {https://journals.aps.org/prl/abstract/10.1103/PhysRevLett.115.047001} {\bibfield  {journal} {\bibinfo  {journal} {Physical review letters}\ }\textbf {\bibinfo {volume} {115}},\ \bibinfo {pages} {047001} (\bibinfo {year} {2015}{\natexlab{a}})}\BibitemShut {NoStop}%
\bibitem [{\citenamefont {Kneisel}\ \emph {et~al.}(2015)\citenamefont {Kneisel}, \citenamefont {Ciovati}, \citenamefont {Dhakal}, \citenamefont {Saito}, \citenamefont {Singer}, \citenamefont {Singer},\ and\ \citenamefont {Myneni}}]{kneisel2015review}%
  \BibitemOpen
  \bibfield  {author} {\bibinfo {author} {\bibfnamefont {P.}~\bibnamefont {Kneisel}}, \bibinfo {author} {\bibfnamefont {G.}~\bibnamefont {Ciovati}}, \bibinfo {author} {\bibfnamefont {P.}~\bibnamefont {Dhakal}}, \bibinfo {author} {\bibfnamefont {K.}~\bibnamefont {Saito}}, \bibinfo {author} {\bibfnamefont {W.}~\bibnamefont {Singer}}, \bibinfo {author} {\bibfnamefont {Xenia}\ \bibnamefont {Singer}}, \ and\ \bibinfo {author} {\bibfnamefont {G.R.}\ \bibnamefont {Myneni}},\ }\bibfield  {title} {\enquote {\bibinfo {title} {Review of ingot niobium as a material for superconducting radiofrequency accelerating cavities},}\ }\href {https://www.sciencedirect.com/science/article/pii/S0168900214013977} {\bibfield  {journal} {\bibinfo  {journal} {Nuclear Instruments and Methods in Physics Research Section A: Accelerators, Spectrometers, Detectors and Associated Equipment}\ }\textbf {\bibinfo {volume} {774}},\ \bibinfo {pages} {133--150} (\bibinfo {year} {2015})}\BibitemShut {NoStop}%
\bibitem [{\citenamefont {Antoine}(2019)}]{antoine2019influence}%
  \BibitemOpen
  \bibfield  {author} {\bibinfo {author} {\bibfnamefont {C.Z.}\ \bibnamefont {Antoine}},\ }\bibfield  {title} {\enquote {\bibinfo {title} {Influence of crystalline structure on rf dissipation in superconducting niobium},}\ }\href {https://journals.aps.org/prab/abstract/10.1103/PhysRevAccelBeams.22.034801} {\bibfield  {journal} {\bibinfo  {journal} {Physical Review Accelerators and Beams}\ }\textbf {\bibinfo {volume} {22}},\ \bibinfo {pages} {034801} (\bibinfo {year} {2019})}\BibitemShut {NoStop}%
\bibitem [{\citenamefont {Weingarten}(2023)}]{weingarten2023field}%
  \BibitemOpen
  \bibfield  {author} {\bibinfo {author} {\bibfnamefont {Wolfgang}\ \bibnamefont {Weingarten}},\ }\bibfield  {title} {\enquote {\bibinfo {title} {Field-dependent surface resistance for superconducting niobium accelerating cavities-condensed overview of weak superconducting defect model},}\ }\href {https://ieeexplore.ieee.org/document/10040731} {\bibfield  {journal} {\bibinfo  {journal} {IEEE Transactions on Applied Superconductivity}\ } (\bibinfo {year} {2023})}\BibitemShut {NoStop}%
\bibitem [{\citenamefont {Yoon}\ \emph {et~al.}(2008)\citenamefont {Yoon}, \citenamefont {Seidman}, \citenamefont {Antoine},\ and\ \citenamefont {Bauer}}]{yoon2008atomic}%
  \BibitemOpen
  \bibfield  {author} {\bibinfo {author} {\bibfnamefont {Kevin~E.}\ \bibnamefont {Yoon}}, \bibinfo {author} {\bibfnamefont {David~N.}\ \bibnamefont {Seidman}}, \bibinfo {author} {\bibfnamefont {Claire}\ \bibnamefont {Antoine}}, \ and\ \bibinfo {author} {\bibfnamefont {Pierre}\ \bibnamefont {Bauer}},\ }\bibfield  {title} {\enquote {\bibinfo {title} {Atomic-scale chemical analyses of niobium oxide/niobium interfaces via atom-probe tomography},}\ }\href {https://aip.scitation.org/doi/full/10.1063/1.2987483} {\bibfield  {journal} {\bibinfo  {journal} {Applied Physics Letters}\ }\textbf {\bibinfo {volume} {93}},\ \bibinfo {pages} {132502} (\bibinfo {year} {2008})}\BibitemShut {NoStop}%
\bibitem [{\citenamefont {Proslier}\ \emph {et~al.}(2008)\citenamefont {Proslier}, \citenamefont {Zasadzinski}, \citenamefont {Cooley}, \citenamefont {Antoine}, \citenamefont {Moore}, \citenamefont {Norem}, \citenamefont {Pellin},\ and\ \citenamefont {Gray}}]{proslier2008tunneling}%
  \BibitemOpen
  \bibfield  {author} {\bibinfo {author} {\bibfnamefont {Th}~\bibnamefont {Proslier}}, \bibinfo {author} {\bibfnamefont {John~F.}\ \bibnamefont {Zasadzinski}}, \bibinfo {author} {\bibfnamefont {L.}~\bibnamefont {Cooley}}, \bibinfo {author} {\bibfnamefont {C.}~\bibnamefont {Antoine}}, \bibinfo {author} {\bibfnamefont {J.}~\bibnamefont {Moore}}, \bibinfo {author} {\bibfnamefont {J.}~\bibnamefont {Norem}}, \bibinfo {author} {\bibfnamefont {M.}~\bibnamefont {Pellin}}, \ and\ \bibinfo {author} {\bibfnamefont {K.E.}\ \bibnamefont {Gray}},\ }\bibfield  {title} {\enquote {\bibinfo {title} {Tunneling study of cavity grade {N}b: Possible magnetic scattering at the surface},}\ }\href {https://aip.scitation.org/doi/full/10.1063/1.2913764} {\bibfield  {journal} {\bibinfo  {journal} {Applied Physics Letters}\ }\textbf {\bibinfo {volume} {92}},\ \bibinfo {pages} {212505} (\bibinfo {year} {2008})}\BibitemShut {NoStop}%
\bibitem [{\citenamefont {Romanenko}\ and\ \citenamefont {Schuster}(2017)}]{romanenko2017understanding}%
  \BibitemOpen
  \bibfield  {author} {\bibinfo {author} {\bibfnamefont {A.}~\bibnamefont {Romanenko}}\ and\ \bibinfo {author} {\bibfnamefont {D.I.}\ \bibnamefont {Schuster}},\ }\bibfield  {title} {\enquote {\bibinfo {title} {Understanding quality factor degradation in superconducting niobium cavities at low microwave field amplitudes},}\ }\href {https://journals.aps.org/prl/abstract/10.1103/PhysRevLett.119.264801} {\bibfield  {journal} {\bibinfo  {journal} {Physical Review Letters}\ }\textbf {\bibinfo {volume} {119}},\ \bibinfo {pages} {264801} (\bibinfo {year} {2017})}\BibitemShut {NoStop}%
\bibitem [{\citenamefont {Semione}\ \emph {et~al.}(2019)\citenamefont {Semione}, \citenamefont {Pandey}, \citenamefont {Tober}, \citenamefont {Pfrommer}, \citenamefont {Poulain}, \citenamefont {Drnec}, \citenamefont {Sch{\"u}tz}, \citenamefont {Keller}, \citenamefont {Noei}, \citenamefont {Vonk} \emph {et~al.}}]{semione2019niobium}%
  \BibitemOpen
  \bibfield  {author} {\bibinfo {author} {\bibfnamefont {Guilherme Dalla~Lana}\ \bibnamefont {Semione}}, \bibinfo {author} {\bibfnamefont {A.~Dangwal}\ \bibnamefont {Pandey}}, \bibinfo {author} {\bibfnamefont {S.}~\bibnamefont {Tober}}, \bibinfo {author} {\bibfnamefont {J.}~\bibnamefont {Pfrommer}}, \bibinfo {author} {\bibfnamefont {A.}~\bibnamefont {Poulain}}, \bibinfo {author} {\bibfnamefont {J.}~\bibnamefont {Drnec}}, \bibinfo {author} {\bibfnamefont {G.}~\bibnamefont {Sch{\"u}tz}}, \bibinfo {author} {\bibfnamefont {T.F.}\ \bibnamefont {Keller}}, \bibinfo {author} {\bibfnamefont {H.}~\bibnamefont {Noei}}, \bibinfo {author} {\bibfnamefont {V.}~\bibnamefont {Vonk}},  \emph {et~al.},\ }\bibfield  {title} {\enquote {\bibinfo {title} {Niobium near-surface composition during nitrogen infusion relevant for superconducting radio-frequency cavities},}\ }\href {https://journals.aps.org/prab/abstract/10.1103/PhysRevAccelBeams.22.103102} {\bibfield  {journal} {\bibinfo  {journal} {Physical Review Accelerators and
  Beams}\ }\textbf {\bibinfo {volume} {22}},\ \bibinfo {pages} {103102} (\bibinfo {year} {2019})}\BibitemShut {NoStop}%
\bibitem [{\citenamefont {Semione}\ \emph {et~al.}(2021)\citenamefont {Semione}, \citenamefont {Vonk}, \citenamefont {Pandey}, \citenamefont {Gr{\aa}n{\"a}s}, \citenamefont {Arndt}, \citenamefont {Wenskat}, \citenamefont {Hillert}, \citenamefont {Noei},\ and\ \citenamefont {Stierle}}]{semione2021temperature}%
  \BibitemOpen
  \bibfield  {author} {\bibinfo {author} {\bibfnamefont {Guilherme Dalla~Lana}\ \bibnamefont {Semione}}, \bibinfo {author} {\bibfnamefont {Vedran}\ \bibnamefont {Vonk}}, \bibinfo {author} {\bibfnamefont {Arti~Dangwal}\ \bibnamefont {Pandey}}, \bibinfo {author} {\bibfnamefont {Elin}\ \bibnamefont {Gr{\aa}n{\"a}s}}, \bibinfo {author} {\bibfnamefont {Bj{\"o}rn}\ \bibnamefont {Arndt}}, \bibinfo {author} {\bibfnamefont {Marc}\ \bibnamefont {Wenskat}}, \bibinfo {author} {\bibfnamefont {Wolfgang}\ \bibnamefont {Hillert}}, \bibinfo {author} {\bibfnamefont {Heshmat}\ \bibnamefont {Noei}}, \ and\ \bibinfo {author} {\bibfnamefont {Andreas}\ \bibnamefont {Stierle}},\ }\bibfield  {title} {\enquote {\bibinfo {title} {Temperature-dependent near-surface interstitial segregation in niobium},}\ }\href {https://iopscience.iop.org/article/10.1088/1361-648X/abf9b7/meta} {\bibfield  {journal} {\bibinfo  {journal} {Journal of Physics: Condensed Matter}\ }\textbf {\bibinfo {volume} {33}},\ \bibinfo {pages} {265001} (\bibinfo {year}
  {2021})}\BibitemShut {NoStop}%
\bibitem [{\citenamefont {Russo}(2007)}]{russo2007quality}%
  \BibitemOpen
  \bibfield  {author} {\bibinfo {author} {\bibfnamefont {Roberto}\ \bibnamefont {Russo}},\ }\bibfield  {title} {\enquote {\bibinfo {title} {Quality measurement of niobium thin films for {N}b/{C}u superconducting {RF} cavities},}\ }\href {https://iopscience.iop.org/article/10.1088/0957-0233/18/8/003/meta} {\bibfield  {journal} {\bibinfo  {journal} {Measurement Science and Technology}\ }\textbf {\bibinfo {volume} {18}},\ \bibinfo {pages} {2299} (\bibinfo {year} {2007})}\BibitemShut {NoStop}%
\bibitem [{\citenamefont {Kharitonov}\ \emph {et~al.}(2012)\citenamefont {Kharitonov}, \citenamefont {Proslier}, \citenamefont {Glatz},\ and\ \citenamefont {Pellin}}]{kharitonov2012surface}%
  \BibitemOpen
  \bibfield  {author} {\bibinfo {author} {\bibfnamefont {Maxim}\ \bibnamefont {Kharitonov}}, \bibinfo {author} {\bibfnamefont {Thomas}\ \bibnamefont {Proslier}}, \bibinfo {author} {\bibfnamefont {Andreas}\ \bibnamefont {Glatz}}, \ and\ \bibinfo {author} {\bibfnamefont {Michael~J.}\ \bibnamefont {Pellin}},\ }\bibfield  {title} {\enquote {\bibinfo {title} {Surface impedance of superconductors with magnetic impurities},}\ }\href {https://journals.aps.org/prb/abstract/10.1103/PhysRevB.86.024514} {\bibfield  {journal} {\bibinfo  {journal} {Physical Review B}\ }\textbf {\bibinfo {volume} {86}},\ \bibinfo {pages} {024514} (\bibinfo {year} {2012})}\BibitemShut {NoStop}%
\bibitem [{\citenamefont {Carlson}\ \emph {et~al.}(2021)\citenamefont {Carlson}, \citenamefont {Pack}, \citenamefont {Transtrum}, \citenamefont {Lee}, \citenamefont {Seidman}, \citenamefont {Liarte}, \citenamefont {Sitaraman}, \citenamefont {Senanian}, \citenamefont {Kelley}, \citenamefont {Sethna} \emph {et~al.}}]{carlson2021analysis}%
  \BibitemOpen
  \bibfield  {author} {\bibinfo {author} {\bibfnamefont {Jared}\ \bibnamefont {Carlson}}, \bibinfo {author} {\bibfnamefont {Alden}\ \bibnamefont {Pack}}, \bibinfo {author} {\bibfnamefont {Mark~K.}\ \bibnamefont {Transtrum}}, \bibinfo {author} {\bibfnamefont {Jaeyel}\ \bibnamefont {Lee}}, \bibinfo {author} {\bibfnamefont {David~N.}\ \bibnamefont {Seidman}}, \bibinfo {author} {\bibfnamefont {Danilo~B.}\ \bibnamefont {Liarte}}, \bibinfo {author} {\bibfnamefont {Nathan~S.}\ \bibnamefont {Sitaraman}}, \bibinfo {author} {\bibfnamefont {Alen}\ \bibnamefont {Senanian}}, \bibinfo {author} {\bibfnamefont {Michelle~M.}\ \bibnamefont {Kelley}}, \bibinfo {author} {\bibfnamefont {James~P.}\ \bibnamefont {Sethna}},  \emph {et~al.},\ }\bibfield  {title} {\enquote {\bibinfo {title} {Analysis of magnetic vortex dissipation in {S}n-segregated boundaries in {N}b3{S}n superconducting {RF} cavities},}\ }\href {https://journals.aps.org/prb/abstract/10.1103/PhysRevB.103.024516} {\bibfield  {journal} {\bibinfo  {journal} {Physical
  Review B}\ }\textbf {\bibinfo {volume} {103}},\ \bibinfo {pages} {024516} (\bibinfo {year} {2021})}\BibitemShut {NoStop}%
\bibitem [{\citenamefont {Lee}\ \emph {et~al.}(2007)\citenamefont {Lee}, \citenamefont {Polyanskii}, \citenamefont {Sung}, \citenamefont {Larbalestier}, \citenamefont {Antoine}, \citenamefont {Bauer}, \citenamefont {Boffo},\ and\ \citenamefont {Edwards}}]{lee2007flux}%
  \BibitemOpen
  \bibfield  {author} {\bibinfo {author} {\bibfnamefont {P.J.}\ \bibnamefont {Lee}}, \bibinfo {author} {\bibfnamefont {A.A.}\ \bibnamefont {Polyanskii}}, \bibinfo {author} {\bibfnamefont {Zu-Hawn}\ \bibnamefont {Sung}}, \bibinfo {author} {\bibfnamefont {D.C.}\ \bibnamefont {Larbalestier}}, \bibinfo {author} {\bibfnamefont {C.}~\bibnamefont {Antoine}}, \bibinfo {author} {\bibfnamefont {P.C.}\ \bibnamefont {Bauer}}, \bibinfo {author} {\bibfnamefont {C.}~\bibnamefont {Boffo}}, \ and\ \bibinfo {author} {\bibfnamefont {H.T.}\ \bibnamefont {Edwards}},\ }\bibfield  {title} {\enquote {\bibinfo {title} {Flux penetration into grain boundaries large grain niobium sheet for {SRF} cavities: Angular sensitivity},}\ }in\ \href {https://aip.scitation.org/doi/abs/10.1063/1.2770684} {\emph {\bibinfo {booktitle} {AIP Conference Proceedings}}},\ Vol.\ \bibinfo {volume} {927}\ (\bibinfo {organization} {American Institute of Physics},\ \bibinfo {year} {2007})\ pp.\ \bibinfo {pages} {113--120}\BibitemShut {NoStop}%
\bibitem [{\citenamefont {Polyanskii}\ \emph {et~al.}(2011)\citenamefont {Polyanskii}, \citenamefont {Lee}, \citenamefont {Gurevich}, \citenamefont {Sung},\ and\ \citenamefont {Larbalestier}}]{polyanskii2011magneto}%
  \BibitemOpen
  \bibfield  {author} {\bibinfo {author} {\bibfnamefont {A.A.}\ \bibnamefont {Polyanskii}}, \bibinfo {author} {\bibfnamefont {P.J.}\ \bibnamefont {Lee}}, \bibinfo {author} {\bibfnamefont {A.}~\bibnamefont {Gurevich}}, \bibinfo {author} {\bibfnamefont {Zu-Hawn}\ \bibnamefont {Sung}}, \ and\ \bibinfo {author} {\bibfnamefont {D.C.}\ \bibnamefont {Larbalestier}},\ }\bibfield  {title} {\enquote {\bibinfo {title} {Magneto-{O}ptical study high-purity niobium for superconducting {RF} application},}\ }in\ \href {https://aip.scitation.org/doi/abs/10.1063/1.3579237} {\emph {\bibinfo {booktitle} {AIP Conference Proceedings}}},\ Vol.\ \bibinfo {volume} {1352}\ (\bibinfo {organization} {American Institute of Physics},\ \bibinfo {year} {2011})\ pp.\ \bibinfo {pages} {186--202}\BibitemShut {NoStop}%
\bibitem [{\citenamefont {K{\"o}szegi}\ \emph {et~al.}(2017)\citenamefont {K{\"o}szegi}, \citenamefont {Kugeler}, \citenamefont {Abou-Ras}, \citenamefont {Knobloch},\ and\ \citenamefont {Sch{\"a}fer}}]{koszegi2017magneto}%
  \BibitemOpen
  \bibfield  {author} {\bibinfo {author} {\bibfnamefont {J}~\bibnamefont {K{\"o}szegi}}, \bibinfo {author} {\bibfnamefont {O.}~\bibnamefont {Kugeler}}, \bibinfo {author} {\bibfnamefont {D.}~\bibnamefont {Abou-Ras}}, \bibinfo {author} {\bibfnamefont {J.}~\bibnamefont {Knobloch}}, \ and\ \bibinfo {author} {\bibfnamefont {R.}~\bibnamefont {Sch{\"a}fer}},\ }\bibfield  {title} {\enquote {\bibinfo {title} {A magneto-optical study on magnetic flux expulsion and pinning in high-purity niobium},}\ }\href {https://aip.scitation.org/doi/full/10.1063/1.4996113} {\bibfield  {journal} {\bibinfo  {journal} {Journal of Applied Physics}\ }\textbf {\bibinfo {volume} {122}},\ \bibinfo {pages} {173901} (\bibinfo {year} {2017})}\BibitemShut {NoStop}%
\bibitem [{\citenamefont {Wang}\ \emph {et~al.}(2018)\citenamefont {Wang}, \citenamefont {Balachandran}, \citenamefont {Bieler}, \citenamefont {Chetri}, \citenamefont {Compton}, \citenamefont {Lee}, \citenamefont {Polyanskii} \emph {et~al.}}]{wang2018investigation}%
  \BibitemOpen
  \bibfield  {author} {\bibinfo {author} {\bibfnamefont {Mingmin}\ \bibnamefont {Wang}}, \bibinfo {author} {\bibfnamefont {Shreyas}\ \bibnamefont {Balachandran}}, \bibinfo {author} {\bibfnamefont {Thomas}\ \bibnamefont {Bieler}}, \bibinfo {author} {\bibfnamefont {Santosh}\ \bibnamefont {Chetri}}, \bibinfo {author} {\bibfnamefont {Chris}\ \bibnamefont {Compton}}, \bibinfo {author} {\bibfnamefont {Peter}\ \bibnamefont {Lee}}, \bibinfo {author} {\bibfnamefont {Anatolii}\ \bibnamefont {Polyanskii}},  \emph {et~al.},\ }\bibfield  {title} {\enquote {\bibinfo {title} {Investigation of the effect of strategically selected grain boundaries on superconducting properties of {SRF} cavity niobium},}\ }in\ \href {https://accelconf.web.cern.ch/srf2017/papers/thpb026.pdf} {\emph {\bibinfo {booktitle} {18th international conference on rf superconductivity}}}\ (\bibinfo {year} {2018})\ p.\ \bibinfo {pages} {210}\BibitemShut {NoStop}%
\bibitem [{\citenamefont {Wang}\ \emph {et~al.}(2022{\natexlab{a}})\citenamefont {Wang}, \citenamefont {Polyanskii}, \citenamefont {Balachandran}, \citenamefont {Chetri}, \citenamefont {Crimp}, \citenamefont {Lee},\ and\ \citenamefont {Bieler}}]{wang2022investigation}%
  \BibitemOpen
  \bibfield  {author} {\bibinfo {author} {\bibfnamefont {Mingmin}\ \bibnamefont {Wang}}, \bibinfo {author} {\bibfnamefont {Anatolii}\ \bibnamefont {Polyanskii}}, \bibinfo {author} {\bibfnamefont {Shreyas}\ \bibnamefont {Balachandran}}, \bibinfo {author} {\bibfnamefont {Santosh}\ \bibnamefont {Chetri}}, \bibinfo {author} {\bibfnamefont {Martin~A.}\ \bibnamefont {Crimp}}, \bibinfo {author} {\bibfnamefont {Peter~J.}\ \bibnamefont {Lee}}, \ and\ \bibinfo {author} {\bibfnamefont {Thomas~R.}\ \bibnamefont {Bieler}},\ }\bibfield  {title} {\enquote {\bibinfo {title} {Investigation of the effect of structural defects from hydride precipitation on superconducting properties of high purity {SRF} cavity {N}b using magneto-optical and electron imaging methods},}\ }\href {https://iopscience.iop.org/article/10.1088/1361-6668/ac4f6a/meta} {\bibfield  {journal} {\bibinfo  {journal} {Superconductor Science and Technology}\ }\textbf {\bibinfo {volume} {35}},\ \bibinfo {pages} {045001} (\bibinfo {year}
  {2022}{\natexlab{a}})}\BibitemShut {NoStop}%
\bibitem [{\citenamefont {Lee}\ \emph {et~al.}(2020)\citenamefont {Lee}, \citenamefont {Mao}, \citenamefont {He}, \citenamefont {Spina}, \citenamefont {Baik}, \citenamefont {Hall}, \citenamefont {Liepe}, \citenamefont {Seidman}, \citenamefont {Posen} \emph {et~al.}}]{lee2020grain}%
  \BibitemOpen
  \bibfield  {author} {\bibinfo {author} {\bibfnamefont {Jaeyel}\ \bibnamefont {Lee}}, \bibinfo {author} {\bibfnamefont {Zugang}\ \bibnamefont {Mao}}, \bibinfo {author} {\bibfnamefont {Kai}\ \bibnamefont {He}}, \bibinfo {author} {\bibfnamefont {Tiziana}\ \bibnamefont {Spina}}, \bibinfo {author} {\bibfnamefont {Sung-Il}\ \bibnamefont {Baik}}, \bibinfo {author} {\bibfnamefont {Daniel~L.}\ \bibnamefont {Hall}}, \bibinfo {author} {\bibfnamefont {Matthias}\ \bibnamefont {Liepe}}, \bibinfo {author} {\bibfnamefont {David~N.}\ \bibnamefont {Seidman}}, \bibinfo {author} {\bibfnamefont {Sam}\ \bibnamefont {Posen}},  \emph {et~al.},\ }\bibfield  {title} {\enquote {\bibinfo {title} {Grain-boundary structure and segregation in {N}b3{S}n coatings on {N}b for high-performance superconducting radiofrequency cavity applications},}\ }\href {https://www.sciencedirect.com/science/article/pii/S135964542030080X} {\bibfield  {journal} {\bibinfo  {journal} {Acta Materialia}\ }\textbf {\bibinfo {volume} {188}},\ \bibinfo {pages}
  {155--165} (\bibinfo {year} {2020})}\BibitemShut {NoStop}%
\bibitem [{\citenamefont {Romanenko}\ and\ \citenamefont {Padamsee}(2010)}]{romanenko2010role}%
  \BibitemOpen
  \bibfield  {author} {\bibinfo {author} {\bibfnamefont {A.}~\bibnamefont {Romanenko}}\ and\ \bibinfo {author} {\bibfnamefont {H.}~\bibnamefont {Padamsee}},\ }\bibfield  {title} {\enquote {\bibinfo {title} {The role of near-surface dislocations in the high magnetic field performance of superconducting niobium cavities},}\ }\href {https://iopscience.iop.org/article/10.1088/0953-2048/23/4/045008/meta} {\bibfield  {journal} {\bibinfo  {journal} {Superconductor Science and Technology}\ }\textbf {\bibinfo {volume} {23}},\ \bibinfo {pages} {045008} (\bibinfo {year} {2010})}\BibitemShut {NoStop}%
\bibitem [{\citenamefont {Bieler}\ \emph {et~al.}(2010)\citenamefont {Bieler}, \citenamefont {Wright}, \citenamefont {Pourboghrat}, \citenamefont {Compton}, \citenamefont {Hartwig}, \citenamefont {Baars}, \citenamefont {Zamiri}, \citenamefont {Chandrasekaran}, \citenamefont {Darbandi}, \citenamefont {Jiang} \emph {et~al.}}]{bieler2010physical}%
  \BibitemOpen
  \bibfield  {author} {\bibinfo {author} {\bibfnamefont {T.R.}\ \bibnamefont {Bieler}}, \bibinfo {author} {\bibfnamefont {N.T.}\ \bibnamefont {Wright}}, \bibinfo {author} {\bibfnamefont {F.}~\bibnamefont {Pourboghrat}}, \bibinfo {author} {\bibfnamefont {C.}~\bibnamefont {Compton}}, \bibinfo {author} {\bibfnamefont {K.T.}\ \bibnamefont {Hartwig}}, \bibinfo {author} {\bibfnamefont {D.}~\bibnamefont {Baars}}, \bibinfo {author} {\bibfnamefont {A.}~\bibnamefont {Zamiri}}, \bibinfo {author} {\bibfnamefont {S.}~\bibnamefont {Chandrasekaran}}, \bibinfo {author} {\bibfnamefont {P.}~\bibnamefont {Darbandi}}, \bibinfo {author} {\bibfnamefont {H.}~\bibnamefont {Jiang}},  \emph {et~al.},\ }\bibfield  {title} {\enquote {\bibinfo {title} {Physical and mechanical metallurgy of high purity {N}b for accelerator cavities},}\ }\href {https://journals.aps.org/prab/abstract/10.1103/PhysRevSTAB.13.031002} {\bibfield  {journal} {\bibinfo  {journal} {Physical Review Special Topics-Accelerators and Beams}\ }\textbf {\bibinfo {volume}
  {13}},\ \bibinfo {pages} {031002} (\bibinfo {year} {2010})}\BibitemShut {NoStop}%
\bibitem [{\citenamefont {Wang}\ \emph {et~al.}(2022{\natexlab{b}})\citenamefont {Wang}, \citenamefont {Xue}, \citenamefont {Dong},\ and\ \citenamefont {Zhou}}]{wang2022effects}%
  \BibitemOpen
  \bibfield  {author} {\bibinfo {author} {\bibfnamefont {Qing-Yu}\ \bibnamefont {Wang}}, \bibinfo {author} {\bibfnamefont {Cun}\ \bibnamefont {Xue}}, \bibinfo {author} {\bibfnamefont {Chao}\ \bibnamefont {Dong}}, \ and\ \bibinfo {author} {\bibfnamefont {You-He}\ \bibnamefont {Zhou}},\ }\bibfield  {title} {\enquote {\bibinfo {title} {Effects of defects and surface roughness on the vortex penetration and vortex dynamics in superconductor--insulator--superconductor multilayer structures exposed to {RF} magnetic fields: numerical simulations within {TDGL} theory},}\ }\href {https://iopscience.iop.org/article/10.1088/1361-6668/ac4ad1/meta} {\bibfield  {journal} {\bibinfo  {journal} {Superconductor Science and Technology}\ }\textbf {\bibinfo {volume} {35}},\ \bibinfo {pages} {045004} (\bibinfo {year} {2022}{\natexlab{b}})}\BibitemShut {NoStop}%
\bibitem [{\citenamefont {Ries}\ \emph {et~al.}(2020)\citenamefont {Ries}, \citenamefont {Seiler}, \citenamefont {G{\"o}m{\"o}ry}, \citenamefont {Medvids}, \citenamefont {Pira},\ and\ \citenamefont {Malyshev}}]{ries2020superconducting}%
  \BibitemOpen
  \bibfield  {author} {\bibinfo {author} {\bibfnamefont {R.}~\bibnamefont {Ries}}, \bibinfo {author} {\bibfnamefont {E.}~\bibnamefont {Seiler}}, \bibinfo {author} {\bibfnamefont {F}~\bibnamefont {G{\"o}m{\"o}ry}}, \bibinfo {author} {\bibfnamefont {A.}~\bibnamefont {Medvids}}, \bibinfo {author} {\bibfnamefont {C.}~\bibnamefont {Pira}}, \ and\ \bibinfo {author} {\bibfnamefont {O.B.}\ \bibnamefont {Malyshev}},\ }\bibfield  {title} {\enquote {\bibinfo {title} {Superconducting properties and surface roughness of thin {N}b samples fabricated for {SRF} applications},}\ }in\ \href {https://iopscience.iop.org/article/10.1088/1742-6596/1559/1/012040/meta} {\emph {\bibinfo {booktitle} {Journal of Physics: Conference Series}}},\ Vol.\ \bibinfo {volume} {1559}\ (\bibinfo {organization} {IOP Publishing},\ \bibinfo {year} {2020})\ p.\ \bibinfo {pages} {012040}\BibitemShut {NoStop}%
\bibitem [{\citenamefont {Dhakal}\ \emph {et~al.}(2014)\citenamefont {Dhakal}, \citenamefont {Ciovati}, \citenamefont {Kneisel},\ and\ \citenamefont {Myneni}}]{dhakal2014enhancement}%
  \BibitemOpen
  \bibfield  {author} {\bibinfo {author} {\bibfnamefont {Pashupati}\ \bibnamefont {Dhakal}}, \bibinfo {author} {\bibfnamefont {Gianluigi}\ \bibnamefont {Ciovati}}, \bibinfo {author} {\bibfnamefont {Peter}\ \bibnamefont {Kneisel}}, \ and\ \bibinfo {author} {\bibfnamefont {Ganapati~Rao}\ \bibnamefont {Myneni}},\ }\bibfield  {title} {\enquote {\bibinfo {title} {Enhancement in quality factor of {SRF} niobium cavities by material diffusion},}\ }\href {https://ieeexplore.ieee.org/abstract/document/6906233} {\bibfield  {journal} {\bibinfo  {journal} {IEEE Transactions on Applied Superconductivity}\ }\textbf {\bibinfo {volume} {25}},\ \bibinfo {pages} {1--4} (\bibinfo {year} {2014})}\BibitemShut {NoStop}%
\bibitem [{\citenamefont {Gonnella}\ \emph {et~al.}(2014)\citenamefont {Gonnella}, \citenamefont {Liepe} \emph {et~al.}}]{gonnella2014cool}%
  \BibitemOpen
  \bibfield  {author} {\bibinfo {author} {\bibfnamefont {Dan}\ \bibnamefont {Gonnella}}, \bibinfo {author} {\bibfnamefont {Matthias}\ \bibnamefont {Liepe}},  \emph {et~al.},\ }\bibfield  {title} {\enquote {\bibinfo {title} {Cool down and flux trapping studies on {SRF} cavities},}\ }\href {https://epaper.kek.jp/LINAC2014/papers/mopp017.pdf} {\bibfield  {journal} {\bibinfo  {journal} {Proceedings of LINAC 2014}\ } (\bibinfo {year} {2014})}\BibitemShut {NoStop}%
\bibitem [{\citenamefont {Kleindienst}\ \emph {et~al.}(2015)\citenamefont {Kleindienst}, \citenamefont {Burrill}, \citenamefont {Kugeler},\ and\ \citenamefont {Knobloch}}]{kleindienst2015commissioning}%
  \BibitemOpen
  \bibfield  {author} {\bibinfo {author} {\bibfnamefont {Raphael}\ \bibnamefont {Kleindienst}}, \bibinfo {author} {\bibfnamefont {Andrew}\ \bibnamefont {Burrill}}, \bibinfo {author} {\bibfnamefont {Oliver}\ \bibnamefont {Kugeler}}, \ and\ \bibinfo {author} {\bibfnamefont {Jens}\ \bibnamefont {Knobloch}},\ }\href {https://cds.cern.ch/record/2058158} {\emph {\bibinfo {title} {Commissioning results of the {HZB} quadrupole resonator}}},\ \bibinfo {type} {Tech. Rep.}\ (\bibinfo {year} {2015})\BibitemShut {NoStop}%
\bibitem [{\citenamefont {Keckert}\ \emph {et~al.}(2021{\natexlab{a}})\citenamefont {Keckert}, \citenamefont {Kleindienst}, \citenamefont {Kugeler}, \citenamefont {Tikhonov},\ and\ \citenamefont {Knobloch}}]{keckert2021characterizing}%
  \BibitemOpen
  \bibfield  {author} {\bibinfo {author} {\bibfnamefont {S.}~\bibnamefont {Keckert}}, \bibinfo {author} {\bibfnamefont {R.}~\bibnamefont {Kleindienst}}, \bibinfo {author} {\bibfnamefont {O.}~\bibnamefont {Kugeler}}, \bibinfo {author} {\bibfnamefont {D.}~\bibnamefont {Tikhonov}}, \ and\ \bibinfo {author} {\bibfnamefont {J.}~\bibnamefont {Knobloch}},\ }\bibfield  {title} {\enquote {\bibinfo {title} {Characterizing materials for superconducting radiofrequency applications—a comprehensive overview of the quadrupole resonator design and measurement capabilities},}\ }\href {https://aip.scitation.org/doi/full/10.1063/5.0046971} {\bibfield  {journal} {\bibinfo  {journal} {Review of Scientific Instruments}\ }\textbf {\bibinfo {volume} {92}},\ \bibinfo {pages} {064710} (\bibinfo {year} {2021}{\natexlab{a}})}\BibitemShut {NoStop}%
\bibitem [{\citenamefont {Arzeo}\ \emph {et~al.}(2022)\citenamefont {Arzeo}, \citenamefont {Avino}, \citenamefont {Pfeiffer}, \citenamefont {Rosaz}, \citenamefont {Taborelli}, \citenamefont {Vega-Cid},\ and\ \citenamefont {Venturini-Delsolaro}}]{arzeo2022enhanced}%
  \BibitemOpen
  \bibfield  {author} {\bibinfo {author} {\bibfnamefont {Marco}\ \bibnamefont {Arzeo}}, \bibinfo {author} {\bibfnamefont {F.}~\bibnamefont {Avino}}, \bibinfo {author} {\bibfnamefont {S.}~\bibnamefont {Pfeiffer}}, \bibinfo {author} {\bibfnamefont {G.}~\bibnamefont {Rosaz}}, \bibinfo {author} {\bibfnamefont {M.}~\bibnamefont {Taborelli}}, \bibinfo {author} {\bibfnamefont {L.}~\bibnamefont {Vega-Cid}}, \ and\ \bibinfo {author} {\bibfnamefont {W.}~\bibnamefont {Venturini-Delsolaro}},\ }\bibfield  {title} {\enquote {\bibinfo {title} {Enhanced radio-frequency performance of niobium films on copper substrates deposited by high power impulse magnetron sputtering},}\ }\href {https://iopscience.iop.org/article/10.1088/1361-6668/ac5646/meta} {\bibfield  {journal} {\bibinfo  {journal} {Superconductor Science and Technology}\ }\textbf {\bibinfo {volume} {35}},\ \bibinfo {pages} {054008} (\bibinfo {year} {2022})}\BibitemShut {NoStop}%
\bibitem [{\citenamefont {Keckert}\ \emph {et~al.}(2021{\natexlab{b}})\citenamefont {Keckert}, \citenamefont {Ackermann}, \citenamefont {De~Gersem}, \citenamefont {Jiang}, \citenamefont {Sezgin}, \citenamefont {Vogel}, \citenamefont {Wenskat}, \citenamefont {Kleindienst}, \citenamefont {Knobloch}, \citenamefont {Kugeler} \emph {et~al.}}]{keckert2021mitigation}%
  \BibitemOpen
  \bibfield  {author} {\bibinfo {author} {\bibfnamefont {S.}~\bibnamefont {Keckert}}, \bibinfo {author} {\bibfnamefont {W.}~\bibnamefont {Ackermann}}, \bibinfo {author} {\bibfnamefont {H.}~\bibnamefont {De~Gersem}}, \bibinfo {author} {\bibfnamefont {X.}~\bibnamefont {Jiang}}, \bibinfo {author} {\bibfnamefont {A{\"O}}~\bibnamefont {Sezgin}}, \bibinfo {author} {\bibfnamefont {M.}~\bibnamefont {Vogel}}, \bibinfo {author} {\bibfnamefont {M.}~\bibnamefont {Wenskat}}, \bibinfo {author} {\bibfnamefont {R.}~\bibnamefont {Kleindienst}}, \bibinfo {author} {\bibfnamefont {J.}~\bibnamefont {Knobloch}}, \bibinfo {author} {\bibfnamefont {O.}~\bibnamefont {Kugeler}},  \emph {et~al.},\ }\bibfield  {title} {\enquote {\bibinfo {title} {Mitigation of parasitic losses in the quadrupole resonator enabling direct measurements of low residual resistances of {SRF} samples},}\ }\href {https://aip.scitation.org/doi/full/10.1063/5.0076715} {\bibfield  {journal} {\bibinfo  {journal} {Aip Advances}\ }\textbf {\bibinfo {volume} {11}},\
  \bibinfo {pages} {125326} (\bibinfo {year} {2021}{\natexlab{b}})}\BibitemShut {NoStop}%
\bibitem [{\citenamefont {James}\ \emph {et~al.}(2012)\citenamefont {James}, \citenamefont {Krishnan}, \citenamefont {Bures}, \citenamefont {Tajima}, \citenamefont {Civale}, \citenamefont {Edwards}, \citenamefont {Spradlin},\ and\ \citenamefont {Inoue}}]{james2012superconducting}%
  \BibitemOpen
  \bibfield  {author} {\bibinfo {author} {\bibfnamefont {Colt}\ \bibnamefont {James}}, \bibinfo {author} {\bibfnamefont {Mahadevan}\ \bibnamefont {Krishnan}}, \bibinfo {author} {\bibfnamefont {Brian}\ \bibnamefont {Bures}}, \bibinfo {author} {\bibfnamefont {Tsuyoshi}\ \bibnamefont {Tajima}}, \bibinfo {author} {\bibfnamefont {Leonardo}\ \bibnamefont {Civale}}, \bibinfo {author} {\bibfnamefont {Randy}\ \bibnamefont {Edwards}}, \bibinfo {author} {\bibfnamefont {Josh}\ \bibnamefont {Spradlin}}, \ and\ \bibinfo {author} {\bibfnamefont {Hitoshi}\ \bibnamefont {Inoue}},\ }\bibfield  {title} {\enquote {\bibinfo {title} {Superconducting {N}b thin films on {C}u for applications in {SRF} accelerators},}\ }\href {https://ieeexplore.ieee.org/abstract/document/6389725} {\bibfield  {journal} {\bibinfo  {journal} {IEEE transactions on applied superconductivity}\ }\textbf {\bibinfo {volume} {23}},\ \bibinfo {pages} {3500205--3500205} (\bibinfo {year} {2012})}\BibitemShut {NoStop}%
\bibitem [{\citenamefont {Antoine}\ \emph {et~al.}(2019)\citenamefont {Antoine}, \citenamefont {Aburas}, \citenamefont {Four}, \citenamefont {Weiss}, \citenamefont {Iwashita}, \citenamefont {Hayano}, \citenamefont {Kato}, \citenamefont {Kubo},\ and\ \citenamefont {Saeki}}]{antoine2019optimization}%
  \BibitemOpen
  \bibfield  {author} {\bibinfo {author} {\bibfnamefont {Claire~Z.}\ \bibnamefont {Antoine}}, \bibinfo {author} {\bibfnamefont {M.}~\bibnamefont {Aburas}}, \bibinfo {author} {\bibfnamefont {A.}~\bibnamefont {Four}}, \bibinfo {author} {\bibfnamefont {F.}~\bibnamefont {Weiss}}, \bibinfo {author} {\bibfnamefont {Y.}~\bibnamefont {Iwashita}}, \bibinfo {author} {\bibfnamefont {H.}~\bibnamefont {Hayano}}, \bibinfo {author} {\bibfnamefont {S.}~\bibnamefont {Kato}}, \bibinfo {author} {\bibfnamefont {Takayuki}\ \bibnamefont {Kubo}}, \ and\ \bibinfo {author} {\bibfnamefont {T.}~\bibnamefont {Saeki}},\ }\bibfield  {title} {\enquote {\bibinfo {title} {Optimization of tailored multilayer superconductors for {RF} application and protection against premature vortex penetration},}\ }\href {https://iopscience.iop.org/article/10.1088/1361-6668/ab1bf1/meta} {\bibfield  {journal} {\bibinfo  {journal} {Superconductor Science and Technology}\ }\textbf {\bibinfo {volume} {32}},\ \bibinfo {pages} {085005} (\bibinfo {year}
  {2019})}\BibitemShut {NoStop}%
\bibitem [{\citenamefont {Lamura}\ \emph {et~al.}(2009)\citenamefont {Lamura}, \citenamefont {Aurino}, \citenamefont {Andreone},\ and\ \citenamefont {Vill{\'e}gier}}]{lamura2009first}%
  \BibitemOpen
  \bibfield  {author} {\bibinfo {author} {\bibfnamefont {G.}~\bibnamefont {Lamura}}, \bibinfo {author} {\bibfnamefont {M.}~\bibnamefont {Aurino}}, \bibinfo {author} {\bibfnamefont {A.}~\bibnamefont {Andreone}}, \ and\ \bibinfo {author} {\bibfnamefont {J-C}\ \bibnamefont {Vill{\'e}gier}},\ }\bibfield  {title} {\enquote {\bibinfo {title} {First critical field measurements of superconducting films by third harmonic analysis},}\ }\href {https://aip.scitation.org/doi/full/10.1063/1.3211321} {\bibfield  {journal} {\bibinfo  {journal} {Journal of Applied Physics}\ }\textbf {\bibinfo {volume} {106}},\ \bibinfo {pages} {053903} (\bibinfo {year} {2009})}\BibitemShut {NoStop}%
\bibitem [{\citenamefont {Antoine}\ \emph {et~al.}(2010)\citenamefont {Antoine}, \citenamefont {Berry}, \citenamefont {Bouat}, \citenamefont {Jacquot}, \citenamefont {Villegier}, \citenamefont {Lamura},\ and\ \citenamefont {Gurevich}}]{antoine2010characterization}%
  \BibitemOpen
  \bibfield  {author} {\bibinfo {author} {\bibfnamefont {C.Z.}\ \bibnamefont {Antoine}}, \bibinfo {author} {\bibfnamefont {S.}~\bibnamefont {Berry}}, \bibinfo {author} {\bibfnamefont {S.}~\bibnamefont {Bouat}}, \bibinfo {author} {\bibfnamefont {J.F.}\ \bibnamefont {Jacquot}}, \bibinfo {author} {\bibfnamefont {J.C.}\ \bibnamefont {Villegier}}, \bibinfo {author} {\bibfnamefont {G.}~\bibnamefont {Lamura}}, \ and\ \bibinfo {author} {\bibfnamefont {A.}~\bibnamefont {Gurevich}},\ }\bibfield  {title} {\enquote {\bibinfo {title} {Characterization of superconducting nanometric multilayer samples for superconducting rf applications: First evidence of magnetic screening effect},}\ }\href {https://journals.aps.org/prab/abstract/10.1103/PhysRevSTAB.13.121001} {\bibfield  {journal} {\bibinfo  {journal} {Physical Review Special Topics-Accelerators and Beams}\ }\textbf {\bibinfo {volume} {13}},\ \bibinfo {pages} {121001} (\bibinfo {year} {2010})}\BibitemShut {NoStop}%
\bibitem [{\citenamefont {Antoine}\ \emph {et~al.}(2011)\citenamefont {Antoine}, \citenamefont {Berry}, \citenamefont {Aurino}, \citenamefont {Jacquot}, \citenamefont {Villegier}, \citenamefont {Lamura},\ and\ \citenamefont {Andreone}}]{antoine2011characterization}%
  \BibitemOpen
  \bibfield  {author} {\bibinfo {author} {\bibfnamefont {C.Z.}\ \bibnamefont {Antoine}}, \bibinfo {author} {\bibfnamefont {S.}~\bibnamefont {Berry}}, \bibinfo {author} {\bibfnamefont {M.}~\bibnamefont {Aurino}}, \bibinfo {author} {\bibfnamefont {J-F}\ \bibnamefont {Jacquot}}, \bibinfo {author} {\bibfnamefont {J-C}\ \bibnamefont {Villegier}}, \bibinfo {author} {\bibfnamefont {G.}~\bibnamefont {Lamura}}, \ and\ \bibinfo {author} {\bibfnamefont {A.}~\bibnamefont {Andreone}},\ }\bibfield  {title} {\enquote {\bibinfo {title} {Characterization of field penetration in superconducting multilayers samples},}\ }\href {https://ieeexplore.ieee.org/abstract/document/5699933} {\bibfield  {journal} {\bibinfo  {journal} {IEEE Transactions on Applied Superconductivity}\ }\textbf {\bibinfo {volume} {21}},\ \bibinfo {pages} {2601--2604} (\bibinfo {year} {2011})}\BibitemShut {NoStop}%
\bibitem [{\citenamefont {Antoine}\ \emph {et~al.}(2013)\citenamefont {Antoine}, \citenamefont {Villegier},\ and\ \citenamefont {Martinet}}]{antoine2013study}%
  \BibitemOpen
  \bibfield  {author} {\bibinfo {author} {\bibfnamefont {C.Z.}\ \bibnamefont {Antoine}}, \bibinfo {author} {\bibfnamefont {J-C}\ \bibnamefont {Villegier}}, \ and\ \bibinfo {author} {\bibfnamefont {G.}~\bibnamefont {Martinet}},\ }\bibfield  {title} {\enquote {\bibinfo {title} {Study of nanometric superconducting multilayers for {RF} field screening applications},}\ }\href {https://aip.scitation.org/doi/full/10.1063/1.4794938} {\bibfield  {journal} {\bibinfo  {journal} {Applied Physics Letters}\ }\textbf {\bibinfo {volume} {102}},\ \bibinfo {pages} {102603} (\bibinfo {year} {2013})}\BibitemShut {NoStop}%
\bibitem [{\citenamefont {Katyan}\ and\ \citenamefont {Antoine}(2015)}]{katyan2015characterization}%
  \BibitemOpen
  \bibfield  {author} {\bibinfo {author} {\bibfnamefont {N.}~\bibnamefont {Katyan}}\ and\ \bibinfo {author} {\bibfnamefont {C.Z.}\ \bibnamefont {Antoine}},\ }\href {https://cds.cern.ch/record/2162710} {\emph {\bibinfo {title} {Characterization of thin films using local magneometer}}},\ \bibinfo {type} {Tech. Rep.}\ (\bibinfo {year} {2015})\BibitemShut {NoStop}%
\bibitem [{\citenamefont {Aburas}\ \emph {et~al.}(2017)\citenamefont {Aburas}, \citenamefont {Antoine}, \citenamefont {Four} \emph {et~al.}}]{aburas2017local}%
  \BibitemOpen
  \bibfield  {author} {\bibinfo {author} {\bibfnamefont {Muhammad}\ \bibnamefont {Aburas}}, \bibinfo {author} {\bibfnamefont {C.Z.}\ \bibnamefont {Antoine}}, \bibinfo {author} {\bibfnamefont {Aurelien}\ \bibnamefont {Four}},  \emph {et~al.},\ }\bibfield  {title} {\enquote {\bibinfo {title} {Local magnetometer: first critical field measurement of multilayer superconductors},}\ }in\ \href {https://inspirehep.net/files/354c3b7e9609d9af52acce28e0d48965} {\emph {\bibinfo {booktitle} {Proc. 18th Int. Conf. RF Superconductivity (SRF'17)}}}\ (\bibinfo {year} {2017})\ pp.\ \bibinfo {pages} {830--834}\BibitemShut {NoStop}%
\bibitem [{\citenamefont {Ito}\ \emph {et~al.}(2019)\citenamefont {Ito}, \citenamefont {Hayano}, \citenamefont {Kubo}, \citenamefont {Saeki}, \citenamefont {Iwashita}, \citenamefont {Katayama}, \citenamefont {Tongu}, \citenamefont {ICR}, \citenamefont {Ito}, \citenamefont {Nagata} \emph {et~al.}}]{ito2019lower}%
  \BibitemOpen
  \bibfield  {author} {\bibinfo {author} {\bibfnamefont {Hayato}\ \bibnamefont {Ito}}, \bibinfo {author} {\bibfnamefont {H.}~\bibnamefont {Hayano}}, \bibinfo {author} {\bibfnamefont {T.}~\bibnamefont {Kubo}}, \bibinfo {author} {\bibfnamefont {T.}~\bibnamefont {Saeki}}, \bibinfo {author} {\bibfnamefont {Y.}~\bibnamefont {Iwashita}}, \bibinfo {author} {\bibfnamefont {R.}~\bibnamefont {Katayama}}, \bibinfo {author} {\bibfnamefont {H.}~\bibnamefont {Tongu}}, \bibinfo {author} {\bibfnamefont {Kyoto}\ \bibnamefont {ICR}}, \bibinfo {author} {\bibfnamefont {R.}~\bibnamefont {Ito}}, \bibinfo {author} {\bibfnamefont {T.}~\bibnamefont {Nagata}},  \emph {et~al.},\ }\bibfield  {title} {\enquote {\bibinfo {title} {Lower critical field measurement of thin film superconductor},}\ }in\ \href {https://accelconf.web.cern.ch/linac2018/papers/tupo066.pdf} {\emph {\bibinfo {booktitle} {29$\backslash$textsuperscript $\{$th$\}$ Linear Accelerator Conf.(LINAC'18), Beijing, China, 16-21 September 2018}}}\ (\bibinfo {organization}
  {JACOW Publishing, Geneva, Switzerland},\ \bibinfo {year} {2019})\ pp.\ \bibinfo {pages} {484--487}\BibitemShut {NoStop}%
\bibitem [{\citenamefont {Ito}\ \emph {et~al.}(2020)\citenamefont {Ito}, \citenamefont {Hayano}, \citenamefont {Kubo},\ and\ \citenamefont {Saeki}}]{ito2020vortex}%
  \BibitemOpen
  \bibfield  {author} {\bibinfo {author} {\bibfnamefont {Hayato}\ \bibnamefont {Ito}}, \bibinfo {author} {\bibfnamefont {Hitoshi}\ \bibnamefont {Hayano}}, \bibinfo {author} {\bibfnamefont {Takayuki}\ \bibnamefont {Kubo}}, \ and\ \bibinfo {author} {\bibfnamefont {Takayuki}\ \bibnamefont {Saeki}},\ }\bibfield  {title} {\enquote {\bibinfo {title} {Vortex penetration field measurement system based on third-harmonic method for superconducting {RF} materials},}\ }\href {https://www.sciencedirect.com/science/article/pii/S0168900219315414} {\bibfield  {journal} {\bibinfo  {journal} {Nuclear Instruments and Methods in Physics Research Section A: Accelerators, Spectrometers, Detectors and Associated Equipment}\ }\textbf {\bibinfo {volume} {955}},\ \bibinfo {pages} {163284} (\bibinfo {year} {2020})}\BibitemShut {NoStop}%
\bibitem [{\citenamefont {Lee}\ \emph {et~al.}(2000)\citenamefont {Lee}, \citenamefont {Vlahacos}, \citenamefont {Feenstra}, \citenamefont {Schwartz}, \citenamefont {Steinhauer}, \citenamefont {Wellstood},\ and\ \citenamefont {Anlage}}]{lee2000magnetic}%
  \BibitemOpen
  \bibfield  {author} {\bibinfo {author} {\bibfnamefont {Sheng-Chiang}\ \bibnamefont {Lee}}, \bibinfo {author} {\bibfnamefont {C.P.}\ \bibnamefont {Vlahacos}}, \bibinfo {author} {\bibfnamefont {B.J.}\ \bibnamefont {Feenstra}}, \bibinfo {author} {\bibfnamefont {Andrew}\ \bibnamefont {Schwartz}}, \bibinfo {author} {\bibfnamefont {D.E.}\ \bibnamefont {Steinhauer}}, \bibinfo {author} {\bibfnamefont {F.C.}\ \bibnamefont {Wellstood}}, \ and\ \bibinfo {author} {\bibfnamefont {Steven~M.}\ \bibnamefont {Anlage}},\ }\bibfield  {title} {\enquote {\bibinfo {title} {Magnetic permeability imaging of metals with a scanning near-field microwave microscope},}\ }\href {https://aip.scitation.org/doi/10.1063/1.1332978} {\bibfield  {journal} {\bibinfo  {journal} {Applied Physics Letters}\ }\textbf {\bibinfo {volume} {77}},\ \bibinfo {pages} {4404--4406} (\bibinfo {year} {2000})}\BibitemShut {NoStop}%
\bibitem [{\citenamefont {Lee}\ and\ \citenamefont {Anlage}(2003)}]{lee2003spatially}%
  \BibitemOpen
  \bibfield  {author} {\bibinfo {author} {\bibfnamefont {Sheng-Chiang}\ \bibnamefont {Lee}}\ and\ \bibinfo {author} {\bibfnamefont {Steven~M.}\ \bibnamefont {Anlage}},\ }\bibfield  {title} {\enquote {\bibinfo {title} {Spatially-resolved nonlinearity measurements of {Y}{B}a2{C}u3{O}7$-\delta$ bicrystal grain boundaries},}\ }\href {https://aip.scitation.org/doi/10.1063/1.1561152} {\bibfield  {journal} {\bibinfo  {journal} {Applied physics letters}\ }\textbf {\bibinfo {volume} {82}},\ \bibinfo {pages} {1893--1895} (\bibinfo {year} {2003})}\BibitemShut {NoStop}%
\bibitem [{\citenamefont {Lee}\ \emph {et~al.}(2005{\natexlab{a}})\citenamefont {Lee}, \citenamefont {Sullivan}, \citenamefont {Ruchti}, \citenamefont {Anlage}, \citenamefont {Palmer}, \citenamefont {Maiorov},\ and\ \citenamefont {Osquiguil}}]{lee2005doping}%
  \BibitemOpen
  \bibfield  {author} {\bibinfo {author} {\bibfnamefont {Sheng-Chiang}\ \bibnamefont {Lee}}, \bibinfo {author} {\bibfnamefont {Mathew}\ \bibnamefont {Sullivan}}, \bibinfo {author} {\bibfnamefont {Gregory~R.}\ \bibnamefont {Ruchti}}, \bibinfo {author} {\bibfnamefont {Steven~M.}\ \bibnamefont {Anlage}}, \bibinfo {author} {\bibfnamefont {Benjamin~S.}\ \bibnamefont {Palmer}}, \bibinfo {author} {\bibfnamefont {B.}~\bibnamefont {Maiorov}}, \ and\ \bibinfo {author} {\bibfnamefont {E.}~\bibnamefont {Osquiguil}},\ }\bibfield  {title} {\enquote {\bibinfo {title} {Doping-dependent nonlinear {M}eissner effect and spontaneous currents in high-{T}c superconductors},}\ }\href {https://journals.aps.org/prb/abstract/10.1103/PhysRevB.71.014507} {\bibfield  {journal} {\bibinfo  {journal} {Physical Review B}\ }\textbf {\bibinfo {volume} {71}},\ \bibinfo {pages} {014507} (\bibinfo {year} {2005}{\natexlab{a}})}\BibitemShut {NoStop}%
\bibitem [{\citenamefont {Lee}\ \emph {et~al.}(2005{\natexlab{b}})\citenamefont {Lee}, \citenamefont {Lee},\ and\ \citenamefont {Anlage}}]{lee2005microwave}%
  \BibitemOpen
  \bibfield  {author} {\bibinfo {author} {\bibfnamefont {Sheng-Chiang}\ \bibnamefont {Lee}}, \bibinfo {author} {\bibfnamefont {Su-Young}\ \bibnamefont {Lee}}, \ and\ \bibinfo {author} {\bibfnamefont {Steven~M.}\ \bibnamefont {Anlage}},\ }\bibfield  {title} {\enquote {\bibinfo {title} {Microwave nonlinearities of an isolated long {Y}{B}a2{C}u3{O}7$-\delta$ bicrystal grain boundary},}\ }\href {https://journals.aps.org/prb/abstract/10.1103/PhysRevB.72.024527} {\bibfield  {journal} {\bibinfo  {journal} {Physical Review B}\ }\textbf {\bibinfo {volume} {72}},\ \bibinfo {pages} {024527} (\bibinfo {year} {2005}{\natexlab{b}})}\BibitemShut {NoStop}%
\bibitem [{\citenamefont {Mircea}\ \emph {et~al.}(2009)\citenamefont {Mircea}, \citenamefont {Xu},\ and\ \citenamefont {Anlage}}]{mircea2009phase}%
  \BibitemOpen
  \bibfield  {author} {\bibinfo {author} {\bibfnamefont {Dragos~I.}\ \bibnamefont {Mircea}}, \bibinfo {author} {\bibfnamefont {Hua}\ \bibnamefont {Xu}}, \ and\ \bibinfo {author} {\bibfnamefont {Steven~M.}\ \bibnamefont {Anlage}},\ }\bibfield  {title} {\enquote {\bibinfo {title} {Phase-sensitive harmonic measurements of microwave nonlinearities in cuprate thin films},}\ }\href {https://journals.aps.org/prb/abstract/10.1103/PhysRevB.80.144505} {\bibfield  {journal} {\bibinfo  {journal} {Physical Review B}\ }\textbf {\bibinfo {volume} {80}},\ \bibinfo {pages} {144505} (\bibinfo {year} {2009})}\BibitemShut {NoStop}%
\bibitem [{\citenamefont {Tai}\ \emph {et~al.}(2011)\citenamefont {Tai}, \citenamefont {Xi}, \citenamefont {Zhuang}, \citenamefont {Mircea},\ and\ \citenamefont {Anlage}}]{tai2011nonlinear}%
  \BibitemOpen
  \bibfield  {author} {\bibinfo {author} {\bibfnamefont {Tamin}\ \bibnamefont {Tai}}, \bibinfo {author} {\bibfnamefont {X.~X.}\ \bibnamefont {Xi}}, \bibinfo {author} {\bibfnamefont {C.G.}\ \bibnamefont {Zhuang}}, \bibinfo {author} {\bibfnamefont {Dragos~I.}\ \bibnamefont {Mircea}}, \ and\ \bibinfo {author} {\bibfnamefont {Steven~M.}\ \bibnamefont {Anlage}},\ }\bibfield  {title} {\enquote {\bibinfo {title} {Nonlinear near-field microwave microscope for {RF} defect localization in superconductors},}\ }\href {https://ieeexplore.ieee.org/abstract/document/5682057} {\bibfield  {journal} {\bibinfo  {journal} {IEEE transactions on applied superconductivity}\ }\textbf {\bibinfo {volume} {21}},\ \bibinfo {pages} {2615--2618} (\bibinfo {year} {2011})}\BibitemShut {NoStop}%
\bibitem [{\citenamefont {Tai}\ \emph {et~al.}(2012)\citenamefont {Tai}, \citenamefont {Ghamsari},\ and\ \citenamefont {Anlage}}]{tai2012nanoscale}%
  \BibitemOpen
  \bibfield  {author} {\bibinfo {author} {\bibfnamefont {Tamin}\ \bibnamefont {Tai}}, \bibinfo {author} {\bibfnamefont {Behnood~G.}\ \bibnamefont {Ghamsari}}, \ and\ \bibinfo {author} {\bibfnamefont {Steven~M.}\ \bibnamefont {Anlage}},\ }\bibfield  {title} {\enquote {\bibinfo {title} {Nanoscale electrodynamic response of {N}b superconductors},}\ }\href {https://ieeexplore.ieee.org/abstract/document/6353538} {\bibfield  {journal} {\bibinfo  {journal} {IEEE Transactions on Applied Superconductivity}\ }\textbf {\bibinfo {volume} {23}},\ \bibinfo {pages} {7100104--7100104} (\bibinfo {year} {2012})}\BibitemShut {NoStop}%
\bibitem [{\citenamefont {Tai}\ \emph {et~al.}(2014{\natexlab{a}})\citenamefont {Tai}, \citenamefont {Ghamsari},\ and\ \citenamefont {Anlage}}]{tai2014modeling}%
  \BibitemOpen
  \bibfield  {author} {\bibinfo {author} {\bibfnamefont {Tamin}\ \bibnamefont {Tai}}, \bibinfo {author} {\bibfnamefont {B.G.}\ \bibnamefont {Ghamsari}}, \ and\ \bibinfo {author} {\bibfnamefont {Steven~M.}\ \bibnamefont {Anlage}},\ }\bibfield  {title} {\enquote {\bibinfo {title} {Modeling the nanoscale linear response of superconducting thin films measured by a scanning probe microwave microscope},}\ }\href {https://aip.scitation.org/doi/full/10.1063/1.4878937} {\bibfield  {journal} {\bibinfo  {journal} {Journal of Applied Physics}\ }\textbf {\bibinfo {volume} {115}},\ \bibinfo {pages} {203908} (\bibinfo {year} {2014}{\natexlab{a}})}\BibitemShut {NoStop}%
\bibitem [{\citenamefont {Tai}\ \emph {et~al.}(2014{\natexlab{b}})\citenamefont {Tai}, \citenamefont {Ghamsari}, \citenamefont {Bieler}, \citenamefont {Tan}, \citenamefont {Xi},\ and\ \citenamefont {Anlage}}]{tai2014near}%
  \BibitemOpen
  \bibfield  {author} {\bibinfo {author} {\bibfnamefont {Tamin}\ \bibnamefont {Tai}}, \bibinfo {author} {\bibfnamefont {Behnood~G.}\ \bibnamefont {Ghamsari}}, \bibinfo {author} {\bibfnamefont {Thomas~R.}\ \bibnamefont {Bieler}}, \bibinfo {author} {\bibfnamefont {Teng}\ \bibnamefont {Tan}}, \bibinfo {author} {\bibfnamefont {X.~X.}\ \bibnamefont {Xi}}, \ and\ \bibinfo {author} {\bibfnamefont {Steven~M.}\ \bibnamefont {Anlage}},\ }\bibfield  {title} {\enquote {\bibinfo {title} {Near-field microwave magnetic nanoscopy of superconducting radio frequency cavity materials},}\ }\href {https://aip.scitation.org/doi/full/10.1063/1.4881880} {\bibfield  {journal} {\bibinfo  {journal} {Applied Physics Letters}\ }\textbf {\bibinfo {volume} {104}},\ \bibinfo {pages} {232603} (\bibinfo {year} {2014}{\natexlab{b}})}\BibitemShut {NoStop}%
\bibitem [{\citenamefont {Tai}\ \emph {et~al.}(2015)\citenamefont {Tai}, \citenamefont {Ghamsari}, \citenamefont {Bieler},\ and\ \citenamefont {Anlage}}]{tai2015nanoscale}%
  \BibitemOpen
  \bibfield  {author} {\bibinfo {author} {\bibfnamefont {Tamin}\ \bibnamefont {Tai}}, \bibinfo {author} {\bibfnamefont {Behnood~G.}\ \bibnamefont {Ghamsari}}, \bibinfo {author} {\bibfnamefont {Tom}\ \bibnamefont {Bieler}}, \ and\ \bibinfo {author} {\bibfnamefont {Steven~M.}\ \bibnamefont {Anlage}},\ }\bibfield  {title} {\enquote {\bibinfo {title} {Nanoscale nonlinear radio frequency properties of bulk {N}b: Origins of extrinsic nonlinear effects},}\ }\href {https://journals.aps.org/prb/abstract/10.1103/PhysRevB.92.134513} {\bibfield  {journal} {\bibinfo  {journal} {Physical Review B}\ }\textbf {\bibinfo {volume} {92}},\ \bibinfo {pages} {134513} (\bibinfo {year} {2015})}\BibitemShut {NoStop}%
\bibitem [{\citenamefont {Oripov}\ \emph {et~al.}(2019)\citenamefont {Oripov}, \citenamefont {Bieler}, \citenamefont {Ciovati}, \citenamefont {Calatroni}, \citenamefont {Dhakal}, \citenamefont {Junginger}, \citenamefont {Malyshev}, \citenamefont {Terenziani}, \citenamefont {Valente-Feliciano}, \citenamefont {Valizadeh} \emph {et~al.}}]{oripov2019high}%
  \BibitemOpen
  \bibfield  {author} {\bibinfo {author} {\bibfnamefont {Bakhrom}\ \bibnamefont {Oripov}}, \bibinfo {author} {\bibfnamefont {Thomas}\ \bibnamefont {Bieler}}, \bibinfo {author} {\bibfnamefont {Gianluigi}\ \bibnamefont {Ciovati}}, \bibinfo {author} {\bibfnamefont {Sergio}\ \bibnamefont {Calatroni}}, \bibinfo {author} {\bibfnamefont {Pashupati}\ \bibnamefont {Dhakal}}, \bibinfo {author} {\bibfnamefont {Tobias}\ \bibnamefont {Junginger}}, \bibinfo {author} {\bibfnamefont {Oleg~B.}\ \bibnamefont {Malyshev}}, \bibinfo {author} {\bibfnamefont {Giovanni}\ \bibnamefont {Terenziani}}, \bibinfo {author} {\bibfnamefont {Anne-Marie}\ \bibnamefont {Valente-Feliciano}}, \bibinfo {author} {\bibfnamefont {Reza}\ \bibnamefont {Valizadeh}},  \emph {et~al.},\ }\bibfield  {title} {\enquote {\bibinfo {title} {High-frequency nonlinear response of superconducting cavity-grade {N}b surfaces},}\ }\href {https://journals.aps.org/prapplied/abstract/10.1103/PhysRevApplied.11.064030} {\bibfield  {journal} {\bibinfo  {journal} {Physical
  Review Applied}\ }\textbf {\bibinfo {volume} {11}},\ \bibinfo {pages} {064030} (\bibinfo {year} {2019})}\BibitemShut {NoStop}%
\bibitem [{\citenamefont {Valente-Feliciano}(2016)}]{valente2016superconducting}%
  \BibitemOpen
  \bibfield  {author} {\bibinfo {author} {\bibfnamefont {Anne-Marie}\ \bibnamefont {Valente-Feliciano}},\ }\bibfield  {title} {\enquote {\bibinfo {title} {Superconducting {RF} materials other than bulk niobium: a review},}\ }\href {https://iopscience.iop.org/article/10.1088/0953-2048/29/11/113002/meta} {\bibfield  {journal} {\bibinfo  {journal} {Superconductor Science and Technology}\ }\textbf {\bibinfo {volume} {29}},\ \bibinfo {pages} {113002} (\bibinfo {year} {2016})}\BibitemShut {NoStop}%
\bibitem [{\citenamefont {Arbet-Engels}\ \emph {et~al.}(2001)\citenamefont {Arbet-Engels}, \citenamefont {Benvenuti}, \citenamefont {Calatroni}, \citenamefont {Darriulat}, \citenamefont {Peck}, \citenamefont {Valente},\ and\ \citenamefont {Van’t~Hof}}]{arbet2001superconducting}%
  \BibitemOpen
  \bibfield  {author} {\bibinfo {author} {\bibfnamefont {V.}~\bibnamefont {Arbet-Engels}}, \bibinfo {author} {\bibfnamefont {Cristoforo}\ \bibnamefont {Benvenuti}}, \bibinfo {author} {\bibfnamefont {S.}~\bibnamefont {Calatroni}}, \bibinfo {author} {\bibfnamefont {Pierre}\ \bibnamefont {Darriulat}}, \bibinfo {author} {\bibfnamefont {M.A.}\ \bibnamefont {Peck}}, \bibinfo {author} {\bibfnamefont {A-M}\ \bibnamefont {Valente}}, \ and\ \bibinfo {author} {\bibfnamefont {C.A.}\ \bibnamefont {Van’t~Hof}},\ }\bibfield  {title} {\enquote {\bibinfo {title} {Superconducting niobium cavities, a case for the film technology},}\ }\href {https://www.sciencedirect.com/science/article/pii/S0168900201001656?casa_token=sb1goArN98AAAAAA:a_kYD-Eagf6w_DuvjDwXdTtiBnUv59-42i4IRcKYAzMSsbsF96P3n_-OC192CvF0gBfa75r59FQ} {\bibfield  {journal} {\bibinfo  {journal} {Nuclear Instruments and Methods in Physics Research Section A: Accelerators, Spectrometers, Detectors and Associated Equipment}\ }\textbf {\bibinfo {volume} {463}},\ \bibinfo
  {pages} {1--8} (\bibinfo {year} {2001})}\BibitemShut {NoStop}%
\bibitem [{\citenamefont {Sublet}\ \emph {et~al.}(2015)\citenamefont {Sublet}, \citenamefont {Venturini~Delsolaro}, \citenamefont {Therasse}, \citenamefont {Richard}, \citenamefont {Rosaz}, \citenamefont {Aull}, \citenamefont {Zhang}, \citenamefont {B{\'a}rtov{\'a}}, \citenamefont {Calatroni},\ and\ \citenamefont {Taborelli}}]{sublet2015developments}%
  \BibitemOpen
  \bibfield  {author} {\bibinfo {author} {\bibfnamefont {Alban}\ \bibnamefont {Sublet}}, \bibinfo {author} {\bibfnamefont {Walter}\ \bibnamefont {Venturini~Delsolaro}}, \bibinfo {author} {\bibfnamefont {Mathieu}\ \bibnamefont {Therasse}}, \bibinfo {author} {\bibfnamefont {Thibaut}\ \bibnamefont {Richard}}, \bibinfo {author} {\bibfnamefont {Guillaume}\ \bibnamefont {Rosaz}}, \bibinfo {author} {\bibfnamefont {Sarah}\ \bibnamefont {Aull}}, \bibinfo {author} {\bibfnamefont {Pei}\ \bibnamefont {Zhang}}, \bibinfo {author} {\bibfnamefont {Barbora}\ \bibnamefont {B{\'a}rtov{\'a}}}, \bibinfo {author} {\bibfnamefont {Sergio}\ \bibnamefont {Calatroni}}, \ and\ \bibinfo {author} {\bibfnamefont {Mauro}\ \bibnamefont {Taborelli}},\ }\bibfield  {title} {\enquote {\bibinfo {title} {Developments on {SRF} coatings at {CERN}},}\ }\href {https://cds.cern.ch/record/2288263} {\bibfield  {journal} {\bibinfo  {journal} {17th International Conference on RF Superconductivity}\ }\textbf {\bibinfo {volume} {TUPB027}} (\bibinfo {year}
  {2015})}\BibitemShut {NoStop}%
\bibitem [{\citenamefont {Roach}\ \emph {et~al.}(2012)\citenamefont {Roach}, \citenamefont {Beringer}, \citenamefont {Skuza}, \citenamefont {Oliver}, \citenamefont {Clavero}, \citenamefont {Reece},\ and\ \citenamefont {Lukaszew}}]{roach2012niobium}%
  \BibitemOpen
  \bibfield  {author} {\bibinfo {author} {\bibfnamefont {W.M.}\ \bibnamefont {Roach}}, \bibinfo {author} {\bibfnamefont {D.B.}\ \bibnamefont {Beringer}}, \bibinfo {author} {\bibfnamefont {J.R.}\ \bibnamefont {Skuza}}, \bibinfo {author} {\bibfnamefont {W.A.}\ \bibnamefont {Oliver}}, \bibinfo {author} {\bibfnamefont {C.}~\bibnamefont {Clavero}}, \bibinfo {author} {\bibfnamefont {C.E.}\ \bibnamefont {Reece}}, \ and\ \bibinfo {author} {\bibfnamefont {R.A.}\ \bibnamefont {Lukaszew}},\ }\bibfield  {title} {\enquote {\bibinfo {title} {Niobium thin film deposition studies on copper surfaces for superconducting radio frequency cavity applications},}\ }\href {https://journals.aps.org/prab/abstract/10.1103/PhysRevSTAB.15.062002} {\bibfield  {journal} {\bibinfo  {journal} {Physical Review Special Topics-Accelerators and Beams}\ }\textbf {\bibinfo {volume} {15}},\ \bibinfo {pages} {062002} (\bibinfo {year} {2012})}\BibitemShut {NoStop}%
\bibitem [{\citenamefont {Rosaz}\ \emph {et~al.}(2022)\citenamefont {Rosaz}, \citenamefont {Bartkowska}, \citenamefont {Carlos}, \citenamefont {Richard},\ and\ \citenamefont {Taborelli}}]{rosaz2022niobium}%
  \BibitemOpen
  \bibfield  {author} {\bibinfo {author} {\bibfnamefont {Guillaume}\ \bibnamefont {Rosaz}}, \bibinfo {author} {\bibfnamefont {Aleksandra}\ \bibnamefont {Bartkowska}}, \bibinfo {author} {\bibfnamefont {Carlota~P.A.}\ \bibnamefont {Carlos}}, \bibinfo {author} {\bibfnamefont {Thibaut}\ \bibnamefont {Richard}}, \ and\ \bibinfo {author} {\bibfnamefont {Mauro}\ \bibnamefont {Taborelli}},\ }\bibfield  {title} {\enquote {\bibinfo {title} {Niobium thin film thickness profile tailoring on complex shape substrates using unbalanced biased {H}igh {P}ower {I}mpulse {M}agnetron {S}puttering},}\ }\href {https://doi.org/10.1016/j.surfcoat.2022.128306} {\bibfield  {journal} {\bibinfo  {journal} {Surface and Coatings Technology}\ }\textbf {\bibinfo {volume} {436}},\ \bibinfo {pages} {128306} (\bibinfo {year} {2022})}\BibitemShut {NoStop}%
\bibitem [{\citenamefont {Ghaemi}\ \emph {et~al.}(2024)\citenamefont {Ghaemi}, \citenamefont {Lopez-Cazalilla}, \citenamefont {Sarakinos}, \citenamefont {Rosaz}, \citenamefont {Carlos}, \citenamefont {Leith}, \citenamefont {Calatroni}, \citenamefont {Himmerlich},\ and\ \citenamefont {Djurabekova}}]{ghaemi2023growth}%
  \BibitemOpen
  \bibfield  {author} {\bibinfo {author} {\bibfnamefont {M.}~\bibnamefont {Ghaemi}}, \bibinfo {author} {\bibfnamefont {A.}~\bibnamefont {Lopez-Cazalilla}}, \bibinfo {author} {\bibfnamefont {K.}~\bibnamefont {Sarakinos}}, \bibinfo {author} {\bibfnamefont {G.J.}\ \bibnamefont {Rosaz}}, \bibinfo {author} {\bibfnamefont {C.P.A.}\ \bibnamefont {Carlos}}, \bibinfo {author} {\bibfnamefont {S.}~\bibnamefont {Leith}}, \bibinfo {author} {\bibfnamefont {S.}~\bibnamefont {Calatroni}}, \bibinfo {author} {\bibfnamefont {M.}~\bibnamefont {Himmerlich}}, \ and\ \bibinfo {author} {\bibfnamefont {F.}~\bibnamefont {Djurabekova}},\ }\bibfield  {title} {\enquote {\bibinfo {title} {Growth of {N}b films on {C}u for superconducting radio frequency cavities by direct current and high power impulse magnetron sputtering: A molecular dynamics and experimental study},}\ }\href {https://www.sciencedirect.com/science/article/pii/S025789722300974X} {\bibfield  {journal} {\bibinfo  {journal} {Surface and Coatings Technology}\ ,\ \bibinfo
  {pages} {130199}} (\bibinfo {year} {2024})}\BibitemShut {NoStop}%
\bibitem [{\citenamefont {Posen}\ \emph {et~al.}(2015{\natexlab{b}})\citenamefont {Posen}, \citenamefont {Liepe},\ and\ \citenamefont {Hall}}]{posen2015proof}%
  \BibitemOpen
  \bibfield  {author} {\bibinfo {author} {\bibfnamefont {S.}~\bibnamefont {Posen}}, \bibinfo {author} {\bibfnamefont {M.}~\bibnamefont {Liepe}}, \ and\ \bibinfo {author} {\bibfnamefont {D.L.}\ \bibnamefont {Hall}},\ }\bibfield  {title} {\enquote {\bibinfo {title} {Proof-of-principle demonstration of {N}b3{S}n superconducting radiofrequency cavities for high {Q} applications},}\ }\href {https://aip.scitation.org/doi/full/10.1063/1.4913247} {\bibfield  {journal} {\bibinfo  {journal} {Applied Physics Letters}\ }\textbf {\bibinfo {volume} {106}},\ \bibinfo {pages} {082601} (\bibinfo {year} {2015}{\natexlab{b}})}\BibitemShut {NoStop}%
\bibitem [{\citenamefont {Posen}\ and\ \citenamefont {Hall}(2017)}]{posen2017nb3sn}%
  \BibitemOpen
  \bibfield  {author} {\bibinfo {author} {\bibfnamefont {Sam}\ \bibnamefont {Posen}}\ and\ \bibinfo {author} {\bibfnamefont {Daniel~Leslie}\ \bibnamefont {Hall}},\ }\bibfield  {title} {\enquote {\bibinfo {title} {{N}b3{S}n superconducting radiofrequency cavities: fabrication, results, properties, and prospects},}\ }\href {https://iopscience.iop.org/article/10.1088/1361-6668/30/3/033004/meta} {\bibfield  {journal} {\bibinfo  {journal} {Superconductor Science and Technology}\ }\textbf {\bibinfo {volume} {30}},\ \bibinfo {pages} {033004} (\bibinfo {year} {2017})}\BibitemShut {NoStop}%
\bibitem [{\citenamefont {Trenikhina}\ \emph {et~al.}(2017)\citenamefont {Trenikhina}, \citenamefont {Posen}, \citenamefont {Romanenko}, \citenamefont {Sardela}, \citenamefont {Zuo}, \citenamefont {Hall},\ and\ \citenamefont {Liepe}}]{trenikhina2017performance}%
  \BibitemOpen
  \bibfield  {author} {\bibinfo {author} {\bibfnamefont {Y.}~\bibnamefont {Trenikhina}}, \bibinfo {author} {\bibfnamefont {S.}~\bibnamefont {Posen}}, \bibinfo {author} {\bibfnamefont {A.}~\bibnamefont {Romanenko}}, \bibinfo {author} {\bibfnamefont {M.}~\bibnamefont {Sardela}}, \bibinfo {author} {\bibfnamefont {J.M.}\ \bibnamefont {Zuo}}, \bibinfo {author} {\bibfnamefont {D.L.}\ \bibnamefont {Hall}}, \ and\ \bibinfo {author} {\bibfnamefont {M.}~\bibnamefont {Liepe}},\ }\bibfield  {title} {\enquote {\bibinfo {title} {Performance-defining properties of {N}b3{S}n coating in {SRF} cavities},}\ }\href {https://iopscience.iop.org/article/10.1088/1361-6668/aa9694/meta} {\bibfield  {journal} {\bibinfo  {journal} {Superconductor Science and Technology}\ }\textbf {\bibinfo {volume} {31}},\ \bibinfo {pages} {015004} (\bibinfo {year} {2017})}\BibitemShut {NoStop}%
\bibitem [{\citenamefont {Ilyina}\ \emph {et~al.}(2019)\citenamefont {Ilyina}, \citenamefont {Rosaz}, \citenamefont {Descarrega}, \citenamefont {Vollenberg}, \citenamefont {Lunt}, \citenamefont {Leaux}, \citenamefont {Calatroni}, \citenamefont {Venturini-Delsolaro},\ and\ \citenamefont {Taborelli}}]{ilyina2019development}%
  \BibitemOpen
  \bibfield  {author} {\bibinfo {author} {\bibfnamefont {E.A.}\ \bibnamefont {Ilyina}}, \bibinfo {author} {\bibfnamefont {Guillaume}\ \bibnamefont {Rosaz}}, \bibinfo {author} {\bibfnamefont {Josep~Busom}\ \bibnamefont {Descarrega}}, \bibinfo {author} {\bibfnamefont {Wilhelmus}\ \bibnamefont {Vollenberg}}, \bibinfo {author} {\bibfnamefont {AJG}\ \bibnamefont {Lunt}}, \bibinfo {author} {\bibfnamefont {Floriane}\ \bibnamefont {Leaux}}, \bibinfo {author} {\bibfnamefont {Sergio}\ \bibnamefont {Calatroni}}, \bibinfo {author} {\bibfnamefont {W.}~\bibnamefont {Venturini-Delsolaro}}, \ and\ \bibinfo {author} {\bibfnamefont {Mauro}\ \bibnamefont {Taborelli}},\ }\bibfield  {title} {\enquote {\bibinfo {title} {Development of sputtered {N}b3{S}n films on copper substrates for superconducting radiofrequency applications},}\ }\href {https://iopscience.iop.org/article/10.1088/1361-6668/aaf61f/meta} {\bibfield  {journal} {\bibinfo  {journal} {Superconductor Science and Technology}\ }\textbf {\bibinfo {volume} {32}},\ \bibinfo
  {pages} {035002} (\bibinfo {year} {2019})}\BibitemShut {NoStop}%
\bibitem [{\citenamefont {Gurevich}(2006{\natexlab{b}})}]{gurevich2006enhancement}%
  \BibitemOpen
  \bibfield  {author} {\bibinfo {author} {\bibfnamefont {Alexander}\ \bibnamefont {Gurevich}},\ }\bibfield  {title} {\enquote {\bibinfo {title} {Enhancement of rf breakdown field of superconductors by multilayer coating},}\ }\href {https://aip.scitation.org/doi/full/10.1063/1.2162264} {\bibfield  {journal} {\bibinfo  {journal} {Applied Physics Letters}\ }\textbf {\bibinfo {volume} {88}},\ \bibinfo {pages} {012511} (\bibinfo {year} {2006}{\natexlab{b}})}\BibitemShut {NoStop}%
\bibitem [{\citenamefont {Kubo}\ \emph {et~al.}(2014)\citenamefont {Kubo}, \citenamefont {Iwashita},\ and\ \citenamefont {Saeki}}]{kubo2014radio}%
  \BibitemOpen
  \bibfield  {author} {\bibinfo {author} {\bibfnamefont {Takayuki}\ \bibnamefont {Kubo}}, \bibinfo {author} {\bibfnamefont {Yoshihisa}\ \bibnamefont {Iwashita}}, \ and\ \bibinfo {author} {\bibfnamefont {Takayuki}\ \bibnamefont {Saeki}},\ }\bibfield  {title} {\enquote {\bibinfo {title} {Radio-frequency electromagnetic field and vortex penetration in multilayered superconductors},}\ }\href {https://aip.scitation.org/doi/full/10.1063/1.4862892} {\bibfield  {journal} {\bibinfo  {journal} {Applied Physics Letters}\ }\textbf {\bibinfo {volume} {104}},\ \bibinfo {pages} {032603} (\bibinfo {year} {2014})}\BibitemShut {NoStop}%
\bibitem [{\citenamefont {Kubo}(2016)}]{kubo2016multilayer}%
  \BibitemOpen
  \bibfield  {author} {\bibinfo {author} {\bibfnamefont {Takayuki}\ \bibnamefont {Kubo}},\ }\bibfield  {title} {\enquote {\bibinfo {title} {Multilayer coating for higher accelerating fields in superconducting radio-frequency cavities: a review of theoretical aspects},}\ }\href {https://iopscience.iop.org/article/10.1088/1361-6668/30/2/023001/meta} {\bibfield  {journal} {\bibinfo  {journal} {Superconductor Science and Technology}\ }\textbf {\bibinfo {volume} {30}},\ \bibinfo {pages} {023001} (\bibinfo {year} {2016})}\BibitemShut {NoStop}%
\bibitem [{\citenamefont {Calatroni}(2006)}]{calatroni200620}%
  \BibitemOpen
  \bibfield  {author} {\bibinfo {author} {\bibfnamefont {S.}~\bibnamefont {Calatroni}},\ }\bibfield  {title} {\enquote {\bibinfo {title} {20 years of experience with the {N}b/{C}u technology for superconducting cavities and perspectives for future developments},}\ }\href {https://www.sciencedirect.com/science/article/pii/S0921453406001596} {\bibfield  {journal} {\bibinfo  {journal} {Physica C: Superconductivity}\ }\textbf {\bibinfo {volume} {441}},\ \bibinfo {pages} {95--101} (\bibinfo {year} {2006})}\BibitemShut {NoStop}%
\bibitem [{\citenamefont {Benvenuti}\ \emph {et~al.}(1984)\citenamefont {Benvenuti}, \citenamefont {Circelli},\ and\ \citenamefont {Hauer}}]{benvenuti1984niobium}%
  \BibitemOpen
  \bibfield  {author} {\bibinfo {author} {\bibfnamefont {Cristoforo}\ \bibnamefont {Benvenuti}}, \bibinfo {author} {\bibfnamefont {N.}~\bibnamefont {Circelli}}, \ and\ \bibinfo {author} {\bibfnamefont {M.}~\bibnamefont {Hauer}},\ }\bibfield  {title} {\enquote {\bibinfo {title} {Niobium films for superconducting accelerating cavities},}\ }\href {https://aip.scitation.org/doi/abs/10.1063/1.95289} {\bibfield  {journal} {\bibinfo  {journal} {Applied Physics Letters}\ }\textbf {\bibinfo {volume} {45}},\ \bibinfo {pages} {583--584} (\bibinfo {year} {1984})}\BibitemShut {NoStop}%
\bibitem [{\citenamefont {Aull}\ \emph {et~al.}(2015)\citenamefont {Aull}, \citenamefont {Venturini~Delsolaro}, \citenamefont {Junginger}, \citenamefont {Valente-Feliciano}, \citenamefont {Knobloch}, \citenamefont {Sublet},\ and\ \citenamefont {Zhang}}]{aull2015understanding}%
  \BibitemOpen
  \bibfield  {author} {\bibinfo {author} {\bibfnamefont {Sarah}\ \bibnamefont {Aull}}, \bibinfo {author} {\bibfnamefont {Walter}\ \bibnamefont {Venturini~Delsolaro}}, \bibinfo {author} {\bibfnamefont {Tobias}\ \bibnamefont {Junginger}}, \bibinfo {author} {\bibfnamefont {Anne-Marie}\ \bibnamefont {Valente-Feliciano}}, \bibinfo {author} {\bibfnamefont {Jens}\ \bibnamefont {Knobloch}}, \bibinfo {author} {\bibfnamefont {Alban}\ \bibnamefont {Sublet}}, \ and\ \bibinfo {author} {\bibfnamefont {Pei}\ \bibnamefont {Zhang}},\ }\bibfield  {title} {\enquote {\bibinfo {title} {On the understanding of {Q}-slope of niobium thin films},}\ }\href {https://cds.cern.ch/record/2288262} {\  (\bibinfo {year} {2015})}\BibitemShut {NoStop}%
\bibitem [{\citenamefont {Palmieri}\ and\ \citenamefont {Vaglio}(2015)}]{palmieri2015thermal}%
  \BibitemOpen
  \bibfield  {author} {\bibinfo {author} {\bibfnamefont {V.}~\bibnamefont {Palmieri}}\ and\ \bibinfo {author} {\bibfnamefont {R.}~\bibnamefont {Vaglio}},\ }\bibfield  {title} {\enquote {\bibinfo {title} {Thermal contact resistance at the {N}b/{C}u interface as a limiting factor for sputtered thin film {RF} superconducting cavities},}\ }\href {https://iopscience.iop.org/article/10.1088/0953-2048/29/1/015004/meta} {\bibfield  {journal} {\bibinfo  {journal} {Superconductor Science and Technology}\ }\textbf {\bibinfo {volume} {29}},\ \bibinfo {pages} {015004} (\bibinfo {year} {2015})}\BibitemShut {NoStop}%
\bibitem [{\citenamefont {Groll}\ \emph {et~al.}(2010)\citenamefont {Groll}, \citenamefont {Gurevich},\ and\ \citenamefont {Chiorescu}}]{groll2010measurement}%
  \BibitemOpen
  \bibfield  {author} {\bibinfo {author} {\bibfnamefont {Nickolas}\ \bibnamefont {Groll}}, \bibinfo {author} {\bibfnamefont {Alexander}\ \bibnamefont {Gurevich}}, \ and\ \bibinfo {author} {\bibfnamefont {Irinel}\ \bibnamefont {Chiorescu}},\ }\bibfield  {title} {\enquote {\bibinfo {title} {Measurement of the nonlinear {M}eissner effect in superconducting {N}b films using a resonant microwave cavity: A probe of unconventional pairing symmetries},}\ }\href {https://journals.aps.org/prb/abstract/10.1103/PhysRevB.81.020504} {\bibfield  {journal} {\bibinfo  {journal} {Physical Review B}\ }\textbf {\bibinfo {volume} {81}},\ \bibinfo {pages} {020504} (\bibinfo {year} {2010})}\BibitemShut {NoStop}%
\bibitem [{\citenamefont {Makita}\ \emph {et~al.}(2022)\citenamefont {Makita}, \citenamefont {Sundahl}, \citenamefont {Ciovati}, \citenamefont {Eom},\ and\ \citenamefont {Gurevich}}]{makita2022nonlinear}%
  \BibitemOpen
  \bibfield  {author} {\bibinfo {author} {\bibfnamefont {Junki}\ \bibnamefont {Makita}}, \bibinfo {author} {\bibfnamefont {C.}~\bibnamefont {Sundahl}}, \bibinfo {author} {\bibfnamefont {Gianluigi}\ \bibnamefont {Ciovati}}, \bibinfo {author} {\bibfnamefont {C.B.}\ \bibnamefont {Eom}}, \ and\ \bibinfo {author} {\bibfnamefont {Alex}\ \bibnamefont {Gurevich}},\ }\bibfield  {title} {\enquote {\bibinfo {title} {Nonlinear {M}eissner effect in {N}b3{S}n coplanar resonators},}\ }\href {https://journals.aps.org/prresearch/abstract/10.1103/PhysRevResearch.4.013156} {\bibfield  {journal} {\bibinfo  {journal} {Physical Review Research}\ }\textbf {\bibinfo {volume} {4}},\ \bibinfo {pages} {013156} (\bibinfo {year} {2022})}\BibitemShut {NoStop}%
\bibitem [{\citenamefont {Gurevich}\ and\ \citenamefont {Ciovati}(2008)}]{gurevich2008dynamics}%
  \BibitemOpen
  \bibfield  {author} {\bibinfo {author} {\bibfnamefont {Alexander}\ \bibnamefont {Gurevich}}\ and\ \bibinfo {author} {\bibfnamefont {Gianluigi}\ \bibnamefont {Ciovati}},\ }\bibfield  {title} {\enquote {\bibinfo {title} {Dynamics of vortex penetration, jumpwise instabilities, and nonlinear surface resistance of type-{II} superconductors in strong rf fields},}\ }\href {https://journals.aps.org/prb/abstract/10.1103/PhysRevB.77.104501} {\bibfield  {journal} {\bibinfo  {journal} {Physical Review B}\ }\textbf {\bibinfo {volume} {77}},\ \bibinfo {pages} {104501} (\bibinfo {year} {2008})}\BibitemShut {NoStop}%
\bibitem [{\citenamefont {Cid}\ \emph {et~al.}(2023)\citenamefont {Cid}, \citenamefont {Bellini}, \citenamefont {Bianchi}, \citenamefont {Ferreira}, \citenamefont {Leith}, \citenamefont {Carlos}, \citenamefont {Rosaz},\ and\ \citenamefont {Delsolaro}}]{vegacid:srf2023-weixa02}%
  \BibitemOpen
  \bibfield  {author} {\bibinfo {author} {\bibfnamefont {L.~Vega}\ \bibnamefont {Cid}}, \bibinfo {author} {\bibfnamefont {G.}~\bibnamefont {Bellini}}, \bibinfo {author} {\bibfnamefont {A.}~\bibnamefont {Bianchi}}, \bibinfo {author} {\bibfnamefont {L.M.A.}\ \bibnamefont {Ferreira}}, \bibinfo {author} {\bibfnamefont {S.B.}\ \bibnamefont {Leith}}, \bibinfo {author} {\bibfnamefont {C.~Pereira}\ \bibnamefont {Carlos}}, \bibinfo {author} {\bibfnamefont {G.J.}\ \bibnamefont {Rosaz}}, \ and\ \bibinfo {author} {\bibfnamefont {W.~Venturini}\ \bibnamefont {Delsolaro}},\ }\bibfield  {title} {\enquote {\bibinfo {title} {{Results of the R\&D RF Testing Campaign of 1.3 GHz Nb/Cu Cavities}},}\ }in\ \href {\doibase 10.18429/JACoW-SRF2023-WEIXA02} {\emph {\bibinfo {booktitle} {Proc. 21th Int. Conf. RF Supercond. (SRF'23)}}},\ \bibinfo {series and number} {\bibinfo {series} {International Conference on RF Superconductivity}\ No.~\bibinfo {number} {21}}\ (\bibinfo  {publisher} {JACoW Publishing, Geneva, Switzerland},\ \bibinfo
  {year} {2023})\ pp.\ \bibinfo {pages} {621--626}\BibitemShut {NoStop}%
\bibitem [{\citenamefont {Oripov}\ and\ \citenamefont {Anlage}(2020)}]{oripov2020time}%
  \BibitemOpen
  \bibfield  {author} {\bibinfo {author} {\bibfnamefont {Bakhrom}\ \bibnamefont {Oripov}}\ and\ \bibinfo {author} {\bibfnamefont {Steven~M.}\ \bibnamefont {Anlage}},\ }\bibfield  {title} {\enquote {\bibinfo {title} {Time-dependent {G}inzburg-{L}andau treatment of rf magnetic vortices in superconductors: Vortex semiloops in a spatially nonuniform magnetic field},}\ }\href {https://journals.aps.org/pre/abstract/10.1103/PhysRevE.101.033306} {\bibfield  {journal} {\bibinfo  {journal} {Physical Review E}\ }\textbf {\bibinfo {volume} {101}},\ \bibinfo {pages} {033306} (\bibinfo {year} {2020})}\BibitemShut {NoStop}%
\bibitem [{\citenamefont {Pack}\ \emph {et~al.}(2020)\citenamefont {Pack}, \citenamefont {Carlson}, \citenamefont {Wadsworth},\ and\ \citenamefont {Transtrum}}]{pack2020vortex}%
  \BibitemOpen
  \bibfield  {author} {\bibinfo {author} {\bibfnamefont {Alden~R.}\ \bibnamefont {Pack}}, \bibinfo {author} {\bibfnamefont {Jared}\ \bibnamefont {Carlson}}, \bibinfo {author} {\bibfnamefont {Spencer}\ \bibnamefont {Wadsworth}}, \ and\ \bibinfo {author} {\bibfnamefont {Mark~K.}\ \bibnamefont {Transtrum}},\ }\bibfield  {title} {\enquote {\bibinfo {title} {Vortex nucleation in superconductors within time-dependent {G}inzburg-{L}andau theory in two and three dimensions: role of surface defects and material inhomogeneities},}\ }\href {https://journals.aps.org/prb/abstract/10.1103/PhysRevB.101.144504} {\bibfield  {journal} {\bibinfo  {journal} {Physical Review B}\ }\textbf {\bibinfo {volume} {101}},\ \bibinfo {pages} {144504} (\bibinfo {year} {2020})}\BibitemShut {NoStop}%
\bibitem [{\citenamefont {Kato}(1999)}]{kato1999charging}%
  \BibitemOpen
  \bibfield  {author} {\bibinfo {author} {\bibfnamefont {Yusuke}\ \bibnamefont {Kato}},\ }\bibfield  {title} {\enquote {\bibinfo {title} {Charging effect on the {H}all conductivity of single vortex in type {II} superconductors},}\ }\href {https://journals.jps.jp/doi/10.1143/JPSJ.68.3798} {\bibfield  {journal} {\bibinfo  {journal} {Journal of the Physical Society of Japan}\ }\textbf {\bibinfo {volume} {68}},\ \bibinfo {pages} {3798--3801} (\bibinfo {year} {1999})}\BibitemShut {NoStop}%
\bibitem [{\citenamefont {Lara}\ \emph {et~al.}(2015)\citenamefont {Lara}, \citenamefont {Aliev}, \citenamefont {Silhanek},\ and\ \citenamefont {Moshchalkov}}]{lara2015microwave}%
  \BibitemOpen
  \bibfield  {author} {\bibinfo {author} {\bibfnamefont {Antonio}\ \bibnamefont {Lara}}, \bibinfo {author} {\bibfnamefont {Farkhad~G.}\ \bibnamefont {Aliev}}, \bibinfo {author} {\bibfnamefont {Alejandro~V.}\ \bibnamefont {Silhanek}}, \ and\ \bibinfo {author} {\bibfnamefont {Victor~V.}\ \bibnamefont {Moshchalkov}},\ }\bibfield  {title} {\enquote {\bibinfo {title} {Microwave-stimulated superconductivity due to presence of vortices},}\ }\href {https://www.nature.com/articles/srep09187} {\bibfield  {journal} {\bibinfo  {journal} {Scientific reports}\ }\textbf {\bibinfo {volume} {5}},\ \bibinfo {pages} {9187} (\bibinfo {year} {2015})}\BibitemShut {NoStop}%
\bibitem [{\citenamefont {Dobrovolskiy}\ \emph {et~al.}(2020)\citenamefont {Dobrovolskiy}, \citenamefont {Gonz{\'a}lez-Ruano}, \citenamefont {Lara}, \citenamefont {Sachser}, \citenamefont {Bevz}, \citenamefont {Shklovskij}, \citenamefont {Bezuglyj}, \citenamefont {Vovk}, \citenamefont {Huth},\ and\ \citenamefont {Aliev}}]{dobrovolskiy2020moving}%
  \BibitemOpen
  \bibfield  {author} {\bibinfo {author} {\bibfnamefont {O.V.}\ \bibnamefont {Dobrovolskiy}}, \bibinfo {author} {\bibfnamefont {C.}~\bibnamefont {Gonz{\'a}lez-Ruano}}, \bibinfo {author} {\bibfnamefont {A.}~\bibnamefont {Lara}}, \bibinfo {author} {\bibfnamefont {R.}~\bibnamefont {Sachser}}, \bibinfo {author} {\bibfnamefont {V.M.}\ \bibnamefont {Bevz}}, \bibinfo {author} {\bibfnamefont {V.A.}\ \bibnamefont {Shklovskij}}, \bibinfo {author} {\bibfnamefont {A.I.}\ \bibnamefont {Bezuglyj}}, \bibinfo {author} {\bibfnamefont {R.V.}\ \bibnamefont {Vovk}}, \bibinfo {author} {\bibfnamefont {M.}~\bibnamefont {Huth}}, \ and\ \bibinfo {author} {\bibfnamefont {F.G.}\ \bibnamefont {Aliev}},\ }\bibfield  {title} {\enquote {\bibinfo {title} {Moving flux quanta cool superconductors by a microwave breath},}\ }\href {https://www.nature.com/articles/s42005-020-0329-z} {\bibfield  {journal} {\bibinfo  {journal} {Communications Physics}\ }\textbf {\bibinfo {volume} {3}},\ \bibinfo {pages} {64} (\bibinfo {year} {2020})}\BibitemShut
  {NoStop}%
\bibitem [{\citenamefont {Hern{\'a}ndez}\ and\ \citenamefont {Dom{\'\i}nguez}(2008)}]{hernandez2008dissipation}%
  \BibitemOpen
  \bibfield  {author} {\bibinfo {author} {\bibfnamefont {Alexander~D.}\ \bibnamefont {Hern{\'a}ndez}}\ and\ \bibinfo {author} {\bibfnamefont {Daniel}\ \bibnamefont {Dom{\'\i}nguez}},\ }\bibfield  {title} {\enquote {\bibinfo {title} {Dissipation spots generated by vortex nucleation points in mesoscopic superconductors driven by microwave magnetic fields},}\ }\href {https://journals.aps.org/prb/abstract/10.1103/PhysRevB.77.224505} {\bibfield  {journal} {\bibinfo  {journal} {Physical Review B}\ }\textbf {\bibinfo {volume} {77}},\ \bibinfo {pages} {224505} (\bibinfo {year} {2008})}\BibitemShut {NoStop}%
\bibitem [{\citenamefont {Tinkham}(2004)}]{tinkham2004introduction}%
  \BibitemOpen
  \bibfield  {author} {\bibinfo {author} {\bibfnamefont {Michael}\ \bibnamefont {Tinkham}},\ }\href@noop {} {\emph {\bibinfo {title} {Introduction to superconductivity}}}\ (\bibinfo  {publisher} {Courier Corporation},\ \bibinfo {year} {2004})\BibitemShut {NoStop}%
\bibitem [{\citenamefont {Halbritter}(1987)}]{halbritter1987oxidation}%
  \BibitemOpen
  \bibfield  {author} {\bibinfo {author} {\bibfnamefont {J.}~\bibnamefont {Halbritter}},\ }\bibfield  {title} {\enquote {\bibinfo {title} {On the oxidation and on the superconductivity of niobium},}\ }\href {https://link.springer.com/article/10.1007/BF00615201} {\bibfield  {journal} {\bibinfo  {journal} {Applied Physics A}\ }\textbf {\bibinfo {volume} {43}},\ \bibinfo {pages} {1--28} (\bibinfo {year} {1987})}\BibitemShut {NoStop}%
\bibitem [{\citenamefont {DeSorbo}(1963)}]{desorbo1963effect}%
  \BibitemOpen
  \bibfield  {author} {\bibinfo {author} {\bibfnamefont {Warren}\ \bibnamefont {DeSorbo}},\ }\bibfield  {title} {\enquote {\bibinfo {title} {Effect of dissolved gases on some superconducting properties of niobium},}\ }\href {https://journals.aps.org/pr/abstract/10.1103/PhysRev.132.107} {\bibfield  {journal} {\bibinfo  {journal} {Physical Review}\ }\textbf {\bibinfo {volume} {132}},\ \bibinfo {pages} {107} (\bibinfo {year} {1963})}\BibitemShut {NoStop}%
\bibitem [{\citenamefont {Koch}\ \emph {et~al.}(1974)\citenamefont {Koch}, \citenamefont {Scarbrough},\ and\ \citenamefont {Kroeger}}]{koch1974effects}%
  \BibitemOpen
  \bibfield  {author} {\bibinfo {author} {\bibfnamefont {C.C.}\ \bibnamefont {Koch}}, \bibinfo {author} {\bibfnamefont {J.O.}\ \bibnamefont {Scarbrough}}, \ and\ \bibinfo {author} {\bibfnamefont {D.M.}\ \bibnamefont {Kroeger}},\ }\bibfield  {title} {\enquote {\bibinfo {title} {Effects of interstitial oxygen on the superconductivity of niobium},}\ }\href {https://journals.aps.org/prb/abstract/10.1103/PhysRevB.9.888} {\bibfield  {journal} {\bibinfo  {journal} {Physical Review B}\ }\textbf {\bibinfo {volume} {9}},\ \bibinfo {pages} {888} (\bibinfo {year} {1974})}\BibitemShut {NoStop}%
\bibitem [{\citenamefont {Ford}\ \emph {et~al.}(2013)\citenamefont {Ford}, \citenamefont {Cooley},\ and\ \citenamefont {Seidman}}]{ford2013first}%
  \BibitemOpen
  \bibfield  {author} {\bibinfo {author} {\bibfnamefont {Denise~C}\ \bibnamefont {Ford}}, \bibinfo {author} {\bibfnamefont {Lance~D}\ \bibnamefont {Cooley}}, \ and\ \bibinfo {author} {\bibfnamefont {David~N}\ \bibnamefont {Seidman}},\ }\bibfield  {title} {\enquote {\bibinfo {title} {First-principles calculations of niobium hydride formation in superconducting radio-frequency cavities},}\ }\href {https://iopscience.iop.org/article/10.1088/0953-2048/26/9/095002} {\bibfield  {journal} {\bibinfo  {journal} {Superconductor Science and Technology}\ }\textbf {\bibinfo {volume} {26}},\ \bibinfo {pages} {095002} (\bibinfo {year} {2013})}\BibitemShut {NoStop}%
\bibitem [{\citenamefont {Isagawa}(1980)}]{isagawa1980hydrogen}%
  \BibitemOpen
  \bibfield  {author} {\bibinfo {author} {\bibfnamefont {S}~\bibnamefont {Isagawa}},\ }\bibfield  {title} {\enquote {\bibinfo {title} {Hydrogen absorption and its effect on low-temperature electric properties of niobium},}\ }\href {https://pubs.aip.org/aip/jap/article-abstract/51/8/4460/10202} {\bibfield  {journal} {\bibinfo  {journal} {Journal of Applied Physics}\ }\textbf {\bibinfo {volume} {51}},\ \bibinfo {pages} {4460--4470} (\bibinfo {year} {1980})}\BibitemShut {NoStop}%
\bibitem [{\citenamefont {Gulian}(2020)}]{gulian2020shortcut}%
  \BibitemOpen
  \bibfield  {author} {\bibinfo {author} {\bibfnamefont {Armen}\ \bibnamefont {Gulian}},\ }\href@noop {} {\emph {\bibinfo {title} {Shortcut to superconductivity}}}\ (\bibinfo  {publisher} {Springer},\ \bibinfo {year} {2020})\BibitemShut {NoStop}%
\end{thebibliography}%

\end{document}